\documentclass[12pt]{iopart}

\usepackage{graphicx}
\usepackage[geometry]{ifsym}
\usepackage{iopams}

\newcommand {\ra}{\rightarrow}

\begin{document}
\title{Superregular solitonic solutions: a novel scenario of the nonlinear stage of Modulation Instability}
\author{V. E. Zakharov$^{1,2,3}$ and A. A. Gelash$^{2}$}
\address{$^{1}$University of Arizona, Tucson, AZ, 857201, USA}
\address{$^{2}$Novosibirsk State University, Novosibirsk, 630090, Russia}
\address{$^{3}$Lebedev Physical Institute, Russian Academy of Sciences, Moscow, 119991, Russia}
\eads{\mailto{zakharov@math.arizona.edu}, \mailto{gelash@srd.nsu.ru}}
\begin{abstract}
We describe a general $N$-solitonic solution of the focusing NLSE in the presence of a condensate by using the dressing method. We give the explicit form of one- and two- solitonic solutions and study them in detail. We distinguish a special class of solutions that we call regular solitonic solutions. Regular solitonic solutions do not disturb phases of the condensate at infinity by coordinate. All of them can be treated as localized perturbations of the condensate. We find a broad class of superregular solitonic solutions which are small perturbations at certain a moment of time. Superregular solitonic solutions are generated by pairs of poles located on opposite sides of the cut. They describe the nonlinear stage of the modulation instability of the condensate and play an important role in the theory of freak waves.
\end{abstract}

\maketitle
\section{Introduction}
This research was motivated by the intention to develop an analytic theory of freak (or rogue) waves in ocean and optic fibers. In recent time the simplest and most universal model for description of these waves is the focusing NLSE (Nonlinear Schr\"{o}dinger Equation). In application to the theory of ocean waves this equation has been used since 1968 ~\cite{Zakharov1968}. In nonlinear optics it was known even earlier ~\cite{Townes1964}.

The focusing NLSE is a first approximation model. For the surface of fluid this model describes the essentially weakly nonlinear quasimonochromatic wave trains with maximal steepness not more than $0.15$  ~\cite{Dyachenko2008}. In nonlinear optics its application is also limited to the case of waves of small amplitudes (see, for instance ~\cite{Kivshar-Agraval2003}). Nowadays numerous models generalizing the NLSE have been developed. For the surface waves they are Dysthe equations ~\cite{Dysthe-Trulsen1999,Zakharov-Dyachenko2010}, for the waves in optic fibers equations include the third time derivatives and more complex forms of nonlinearity (see for instance ~\cite{Zakharov-Kuznetsov1998,Balakin2007}). Also, freak waves in the ocean were studied by numerical modeling of exact Euler equations for potential flow with free boundary ~\cite{Zakharov-Dyachenko-Prokofiev2006,Chalikov-Sheinin2005}. The behavior of freak waves studied by NLSE and by more sophisticated models shows considerable quantitative difference. Nevertheless, advanced improvement of NLSE does not lead to any qualitatively new effects. That means that a careful and detailed study of NLSE
solutions is still very important problem.

There is a point of common agreement that extreme waves on deep water appear as a result of modulational instability of quasimonochromatic weakly nonlinear stationarity waves ~\cite{Zakharov-Dyachenko-Prokofiev2006,Zakharov-Dyachenko-Vasilyev2006,Pelinovsky-Harif2008}. In terms of the NLSE model it means that we should study instability of the condensate in the focusing version of this equation.

It has been known since 1971 that the NLSE is a system that completely integrable by the Inverse Scattering Method (ISM) ~\cite{Zakharov-Shabat1972}. Since this time hundreds of papers and several books have been written on this subject (see for instance the monographs ~\cite{Faddeev-Takhtajan2007,Sulem-Sulem1999,Kharif-Pelinovsky2009,Zakharov-Manakov-Novikov-Pitaevskii1984}. In this sea of literature one can find some articles devoted to development of the ISM for NLSE with focusing in the presence of condensate (see the literature cited below). The application of the ISM technique to the NLSE is not a novelty.

However, the central question appearing in this theory what is long-time consequence of the modulational instability has not been answered so far. In this article we only study the evolution of a special class of localized initial data presented by exact $N$-solitonic solutions of NLSE. Solitonic solutions in the presence of an unstable condensate have a long history.

In the nonsolitonic case a solution of the auxillary linear Zakharov-Shabat system (wave function) is analytic in the right half-plane of the spectral parameter $\lambda=\lambda_R+i \lambda_I$ with the exception of a cut on the real axis $0<\lambda_R<A$, where A is the condensate amplitude. Each soliton adds a simple pole to some point of the right half plane, including possibly the cut.

The simplest solution was found by E. Kuznetsov in 1977 ~\cite{Kuznetsov1977}. Later on this solution was rediscovered by other authors~\cite{Ma1979,Kawata1978}. In this case the pole is located on the real axis outside the cut $\lambda_R>A$. The Kuznetsov solution is a localized bump oscillating in time. The oscillation period grows as $\lambda_R \ra A$ and become infinite in the limit $\lambda_R = A$. In this limit Kuznetsov's solution turns into the instanton found in 1983 by D. Peregrine ~\cite{Peregrine1983}. This is a purely homoclinic solution. It starts at $t \ra -\infty$ from the condensate and returns at $t \ra +\infty$ to the condensate with the same phase. The importance of these solutions to the development of the theory of freak waves is stressed in papers ~\cite{Shrira-Geogjaev2010},~\cite{Akhmediev2009_PRA}. Peregrine's solution was rediscovered by A. Its, A. Rybin and M. Sall in 1988 ~\cite{Its-Rybin-Sall1988}. In their article it was called an "exulton".  In 1985 a second order Peregrine solution was found ~\cite{Akhmediev-Elonskii-Kulagin1985}. Today "multi-Peregrine" solutions are actively studied by different groups (see for instance ~\cite{Akhmediev-PRE2009}, ~\cite{Matveev2010}, ~\cite{Matveev2011}).

In 1985 N. Akhmediev, V. Eleonskii and N. Kulagin discovered a solitonic solution periodic in space but localized in time (the so-called Akhmediev breather) ~\cite{Akhmediev-Elonskii-Kulagin1985}. This solution is almost homoclinic. It starts from a condensate and returns to the condensate with the same amplitude and a different phase. In the Akhmediev case the pole is located inside the cut $0 <\lambda_R < A$.

In the general case the pole is located in any point on the right half-plane. A generic solution moves and oscillates. At $x \ra \pm \infty$ it goes to two condensates with the same amplitude and different phases. In the explicit form it was found in the paper of A. Its, A. Rybin and M. Sall ~\cite{Its-Rybin-Sall1988}. Later this solution was obtained by other methods and discussed in the frame of freak wave theory by A. Slunyaev at al. ~\cite{Slunyaev2002,Slunyaev2006} and N. Akhmediev at al. ~\cite{Akhmediev2009_PRA,Akhmediev2009_PLA}. In 2011 a general one-solitonic solution was obtained by the authors of this article using the $\overline{\partial}$ problem \cite{Zakharov-Gelash2011}. General two-solitonic and N-solitonic solution were suggested in 1998 by M. Tajiri and Y. Watanabe \cite{Tajiri-Watanabe1998}. Some types of two-solitonic solutions and their degenerations were studied in the paper \cite{Akhmediev2012}.

In this article we describe a general N-solitonic solution of the focusing NLSE in the presence of condensate. We do not insist on our priority in this point but we believe that our method for its construction is the most simple and economic. But this is not a major point of our article. This article is more "practically oriented". We give a partial answer to a major question  - what is the nonlinear stage of the modulational instability. When we speak about this subject we must clearly separate development of two types of initial perturbations. One can consider periodic or quasiperiodic in space perturbations. Or one can study development of perturbation localized in space. The time-behavior of these two types of perturbations is different even in the linear theory. The nonlinearity amplifiers this difference enormously. In this article we will speak only on nonlinear behavior of spatially localized perturbations. We will show that a certain class of small perturbations can be described by 2N - solitonic solutions.

A general N-solitonic solution tends at $x \ra \pm \infty$ to the condensate with different phases. In this article we distinguish a class of regular solitonic solutions that do not disturb the phases of the condensate at infinity. All regular solitonic solutions can be treated as localized perturbations of the condensate. In general case they are never small. However we are able to construct an ample class of solutions that we call superregular solitonic which is a certain moment of time (suppose at $t=0$) are small perturbations of condensate. In fact they are pairs of "quasi-Akhmediev" breathers.

\section{NLSE via dressing method}

We study solutions of the following NLSE
\begin{equation}
\rmi\varphi_{t}-\frac{1}{2}\varphi_{xx}-(|\varphi|^{2}-|A|^{2})\varphi=0. \label{NLSE}
\end{equation}
with nonvanishing boundary conditions $|\varphi|^2\rightarrow |A|^{2}$ at $x\rightarrow\pm\infty$. Without loss of generality we assume that $A$ is a real constant. Equation (\ref{NLSE}) is the compatibility condition for the following overdetermined linear system for a matrix function $\bPsi$ ~\cite{Zakharov-Shabat1972}
\begin{eqnarray}
\label{lax system 1}
\frac{\partial\bPsi}{\partial x}=\widehat{\bi{U}}\bPsi,
\\
\label{lax system 2}
\rmi\frac{\partial\bPsi}{\partial t}=(\lambda\widehat{\bi{U}}+\widehat{\bi{W}})\bPsi.
\end{eqnarray}
Here
\begin{eqnarray}
\widehat{\bi{U}}=\bi{I}\lambda+u,
&\qquad&
\widehat{\bi{W}}=\frac{1}{2}
\left(
  \begin{array}{cc}
    |\varphi|^{2}-A^{2} & \varphi_{x} \\
    \varphi^*_{x} & -|\varphi|^{2}+A^{2} \\
  \end{array}
\right),
\nonumber\\
\bi{I}=
\left(
  \begin{array}{cc}
    1 & 0 \\
    0 & -1 \\
  \end{array}
\right),
&\qquad&
\bi{u}=
\left(
  \begin{array}{cc}
    0 & \varphi \\
    -\varphi^* & 0 \\
  \end{array}
\right).
\label{U and W def}
\end{eqnarray}
From (\ref{lax system 1}),(\ref{lax system 2}) we get
\begin{eqnarray}
\frac{\partial\bPsi^{-1}}{\partial x}=-\bPsi^{-1}\widehat{\bi{U}},
\nonumber\\
\rmi\frac{\partial\bPsi^{-1}}{\partial t}=-\bPsi^{-1}(\lambda \widehat{\bi{U}}+\widehat{\bi{W}}).
\label{lax system for Psi-}
\end{eqnarray}
and
\begin{eqnarray}
\frac{\partial\bPsi^{+}}{\partial x}=\bPsi^{+}\widehat{\bi{U}}^{+},
\nonumber\\
\rmi\frac{\partial\bPsi^{+}}{\partial t}=-\bPsi^{+}(\lambda^*\widehat{\bi{U}}^{+}+\widehat{\bi{W}}^{+}).
\label{lax system for Psi+}
\end{eqnarray}
We consider equations (\ref{lax system for Psi+}) in the point $\lambda=-\lambda^*$. We see that
\begin{eqnarray}
\widehat{\bi{U}}^{+}(-\lambda^*)=-\widehat{\bi{U}}(\lambda),
\nonumber\\
\widehat{\bi{W}}^+=\widehat{\bi{W}}.
\label{U, W and U^+, W^+}
\end{eqnarray}
Hence systems (\ref{lax system for Psi-}), (\ref{lax system for Psi+}) coincide. This means that they have a class of solutions $\bPsi$ satisfying the constraint
\begin{equation}
\bPsi^{+}(-\lambda^*)=\bPsi^{-1}(\lambda). \label{Psi^-1 and Psi^+ eq.}
\end{equation}
In what follows we assume that this condition (\ref{Psi^-1 and Psi^+ eq.}) is satisfied.

The idea of the dressing method is the following ~\cite{Zakharov-Shabat1979}. Suppose that we know some solution $\varphi_0$ of the NLSE (\ref{NLSE}) together with a fundamental solution $\bPsi_0$ as a matrix function on $x,t,\lambda$, satisfying the corresponding linear system
\begin{eqnarray}
\frac{\partial\bPsi_0}{\partial x}=\widehat{\bi{U}}_0\bPsi_0,
\nonumber\\
\rmi\frac{\partial\bPsi_0}{\partial t}=(\lambda\widehat{\bi{U}}_0+\widehat{\bi{W}}_0)\bPsi_0 . \label{lax system for Psi_0}
\end{eqnarray}
Here $\widehat{\bi{U}}_0$ and $\widehat{\bi{W}}_0$ are obtained by replacing $\varphi \ra \varphi_0$ in (\ref{U and W def}). Then we introduce the dressing function
\begin{equation}
\bchi=\bPsi\bPsi_{0}^{-1}.
\label{bchi definition}
\end{equation}
(notice that $\bPsi$ is still unknown!). We require that $\bchi$ is regular at infinity
\begin{eqnarray}
\bchi(\lambda) \ra \bi{E} + \frac{\widetilde{\bchi}}{\lambda}+ \cdots, &\qquad& |\lambda| \ra \infty
\nonumber\\
\bi{E}=
\left(
  \begin{array}{cc}
    1 & 0 \\
    0 & 1 \\
  \end{array}
\right).
\label{bchi asymptotic}
\end{eqnarray}
Evidently
\begin{equation}
\bchi^+(-\lambda^*)=\bchi^{-1}(\lambda).
\label{bchi^-1 and bchi^+ eq.}
\end{equation}
The dressing function $\bchi$ satisfies an overdeterminated system of linear equations. The first equation is
\begin{eqnarray}
\frac{\partial \bchi}{\partial x}=\widehat{\bi{U}}\bchi-\bchi \widehat{\bi{U}}_{0}, \label{d(bchi)dx 1}
\\
\frac{\partial \bchi^{-1}}{\partial x}=-\bchi^{-1}\widehat{\bi{U}}+\widehat{\bi{U}}_{0}\bchi^{-1}. \label{d(bchi)dx 2}
\end{eqnarray}
Equation (\ref{d(bchi)dx 2}) can be rewritten as follows
\begin{equation}
\widehat{\bi{U}}=-\bchi\biggl(\frac{\partial}{\partial x}-\widehat{\bi{U}}_0\biggr)\bchi^{-1}.
\label{U for extra poles}
\end{equation}
Now if choose the dressing function $\bchi$ such that $\widehat{\bi{U}}$ defined from (\ref{U for extra poles}) has no singularities on the $\lambda$-plane we construct new solution of equation (\ref{lax system 1}).

According to the Louiville theorem in this case the  function $\widehat{U}$ must be completely defined by its asymptotics at $\lambda \ra \infty$. By plugging (\ref{bchi asymptotic}) to (\ref{d(bchi)dx 1}) we find the so-called dressing formula
\begin{equation}
\bi{u}=\bi{u}_0+[\widetilde{\bchi},\bi{I}],
\end{equation}
or
\begin{equation}
\varphi=\varphi_0-2\widetilde{\chi}_{12}.
\label{general_NLSE solution and bchi asymptotic}
\end{equation}
Here $[\;,\;]$ is the commutator. Until now we perform dressing of equation (\ref{lax system 1}) only. From equation (\ref{lax system 2}) we can derive following relation
\begin{equation}
\lambda \widehat{\bi{U}}+\widehat{\bi{W}} = -\bchi
\biggl(
\rmi\frac{\partial}{\partial t}-\lambda \widehat{\bi{U}}_0-\widehat{\bi{W}}_0
\biggl)\bchi^{-1}.
\end{equation}
If we now demand that $\widehat{\bi{U}},\widehat{\bi{W}}$ have no singularities on the entire $\lambda$ - plane, including infinity (in other words, we require that $\widehat{\bi{W}}$ does not depend on $\lambda$), we realize dressing of equations (\ref{lax system 1}), (\ref{lax system 2}) either. After performing the dressing procedure and determining $\bchi$, the function
\begin{equation}
\bPsi=\bchi\bPsi_0,
\end{equation}
satisfies equations (\ref{lax system 1}), (\ref{lax system 2}) where $\varphi$ is given by (\ref{general_NLSE solution and bchi asymptotic}). This is a new solution of equation (\ref{NLSE}).

There several methods for constructing the dressing function $\bchi$. The method based on the Riemann--Hilbert problem was described in article of Zakharov and Shabat published in 1979 \cite{Zakharov-Shabat1979}. A more advanced approach is based on the use of the local $\overline{\partial}$-problem. This will be a subject of our next paper. In this article we use a poor man's version of the dressing method to construct multisolitonic solutions of NLSE on an arbitrary background.

\section{General N-solitonic solution}
In this chapter we construct solutions of the NLSE following the method developed by V. Zakharov and A. Mikhailov ~\cite{Zakharov-Mikhailov1978}. Let us assume that $\bPsi_0(x,t,\lambda)$ is known and assume that $\bchi$ is a rational function of $\lambda$
\begin{equation}
\bchi=\bi{E}+\sum_m \frac{\bi{U}_m}{\lambda-\lambda_m}.
\label{bchi sum}
\end{equation}
Without loss of generality we can assume $Re(\lambda_n) > 0$. As $\bchi$ satisfies the condition (\ref{bchi^-1 and bchi^+ eq.}), $\bchi^{-1}$ can be written as:
\begin{equation}
\bchi^{-1}=\bi{E}-\sum_m \frac{\bi{U}^+_m}{\lambda+\lambda^*_m}.
\label{bchi^-1 sum}
\end{equation}
Let us denote
\begin{equation}
\bchi_n = \bchi \biggl|_{\lambda=-\lambda^*_n}=\bi{E}-\sum\frac{\bi{U}_m}{\lambda^*_m+\lambda_n}
\end{equation}
and consider the identity
\begin{equation}
\bchi \bchi^{-1} = E \label{bchi bchi^-1 = E}.
\end{equation}
At the point $\lambda=-\lambda^*_n$ the identity (\ref{bchi bchi^-1 = E}) implies that
\begin{equation}
\bchi_n \bi{U}^+_n=0 \label{bchi_n U^+_n=0}.
\end{equation}
Hence $\bi{U}_n,\bi{U}_n^+$ are degenerate matrices, $\bchi_n, \bchi_n^+$ are degenerate also. Now we introduce two sets of complex valued vectors $p_{n,\alpha}$, $q_{n,\alpha}$ and set
\begin{eqnarray}
U_{n,\alpha\beta}=p_{n,\alpha}q_{n,\beta},
&\qquad&
U^{+}_{n,\alpha \beta}=q^*_{n,\alpha}p^*_{n,\beta}.
\nonumber
\end{eqnarray}
Condition (\ref{bchi_n U^+_n=0}) means that
\begin{equation}
\bchi_n q^*_{n,\alpha}=0
\label{bchi_n q_nalpha=0}
\end{equation}
Now
\begin{eqnarray}
\chi_{\alpha\beta}=\delta_{\alpha\beta}+\sum_n \frac{p_{n,\alpha}q_{n,\beta}}{\lambda-\lambda_n},
\nonumber\\
\chi^{-1}_{\alpha\beta}=\delta_{\alpha\beta}-\sum_n \frac{q^*_{n,\alpha}p^*_{n,\beta}}{\lambda+\lambda^*_n}.
\label{bchi_alpha beta}
\end{eqnarray}
Let us plug (\ref{bchi_alpha beta}) into equation (\ref{U for extra poles}). In a general case the function $\widehat{\bi{U}}$ acquires poles at points $\lambda=\lambda_n$ and $\lambda=-\lambda^*_n$. To perform dressing we must cancel residues in all poles. We consider equation (\ref{U for extra poles}) in a neighborhood of the point $\lambda=-\lambda_n^*$. To kill the residue at this point we demand
\begin{equation}
\bchi_n
\biggl(
\frac{\partial}{\partial x}-\widehat{U}_0(-\lambda^*_n)
\biggr)
q^{*}_{n,\alpha}p^{*}_{n,\beta}=0,
\end{equation}
or
\begin{equation}
\bchi_n
\biggl(
\frac{\partial q^*_n}{\partial x}-\widehat{U}_0 (-\lambda^*_n)q^*_{n,\alpha}+q^*_{n,\alpha} \frac{\partial p^*_{n,\beta}}{\partial x}
\biggr)=0.
\label{bchi_n condition}
\end{equation}
However in virtue of (\ref{bchi_n q_nalpha=0}) the last term in (\ref{bchi_n condition}) is cancelled. Hence it is enough to demand
\begin{equation}
\frac{\partial q^*_{n,\alpha}}{\partial x}-\widehat{U}_{0,\alpha\beta}(-\lambda^*_n)q^*_{n,\beta}=0.
\end{equation}
This equality is resolved as follows
\begin{equation}
q^*_{n,\alpha}=\Psi_{0,\alpha\beta}(-\lambda^*_n)\xi_{n,\beta}.
\label{q for onesolitonic solution.0}
\end{equation}
Here $\bxi_{\alpha}$ is an arbitrary complex-valued vector. In what follows we use the notation
\begin{equation}
F_{n,\alpha\beta}=\Psi_{0,\alpha\beta}(-\lambda^*_n).
\label{F_n defenition}
\end{equation}
From this moment we assume that the set of vectors $\bi{q}_n$, $\bi{q}^*_n$ is known. To find the second $\bi{p}_n$, $\bi{p}^*_n$ we need to solve relation (\ref{bchi_n q_nalpha=0}) which is equivalent a system of linear algebraic equations.
\begin{equation}
\sum_{m} \frac{(\bi{q}_{n}\cdot \bi{q}^*_{m})}{\lambda_{n}+\lambda^*_{m}}\bi{p}^*_m=\bi{q}_n
\label{p and q system}
\end{equation}
Here $(\bi{q}_{n}\cdot \bi{q}^*_{m})=q_{n,1} q^*_{m,1}+q_{n,2} q^*_{m,2}$ is a scalar product of $\bi{q}_{n}$ and $\bi{q}_{m}$ vectors. Denote
\begin{equation}
M_{nm}=\frac{(\bi{q}_{n}\cdot \bi{q}^*_{m})}{\lambda_{n}+\lambda^*_m},
\nonumber
\end{equation}
and $M=det(M_{nm})$. $M$ is a Hermitian matrix:
\begin{equation}
M^{*}_{nm}=M_{mn}=M^{T}_{nm}.
\nonumber\\
\end{equation}
Now (\ref{q for onesolitonic solution.0}) can be rewritten as
\begin{equation}
\sum_{m}M^{T}_{nm}\bi{p}_{m}=\bi{q}^{*}_{n}.
\label{p and q* system}
\end{equation}
We need to find $\widetilde{\bchi}$ from the asymptotic expansion of $\bchi$ (\ref{bchi asymptotic}), which can be represented as
\begin{equation}
\widetilde{\chi}_{\alpha\beta}=\sum_{n} p_{n,\alpha}q_{n,\beta}.
\nonumber
\end{equation}
This sum can be calculated as a determinant ratio:
\begin{equation}
\widetilde{\chi}_{\alpha\beta}=-\frac{\widetilde{M}_{\alpha\beta}}{M}.
\label{bchi1}
\end{equation}
Here $\widetilde{M}_{\alpha\beta}$ is the extended matrix:
\begin{equation}
\widetilde{M}_{\alpha\beta}=
\left(\begin{array}{cc}
        0 & \begin{array}{ccc}
              q_{1,\beta} & \cdots & q_{n,\beta}
            \end{array}
         \\
        \begin{array}{c}
          q^*_{1,\alpha} \\
          \vdots \\
          q^*_{n,\alpha}
        \end{array}
         &  \begin{array}{c}
              M^{T}_{nm}
            \end{array}
      \end{array}\right).
\label{M1}
\end{equation}
We find the solution of NLSE (\ref{NLSE}) from condition (\ref{general_NLSE solution and bchi asymptotic}) as:
\begin{equation}
\varphi=
\varphi_0+2 \frac{\widetilde{M}_{12}}{M}.
\label{N-solitoniic solution}
\end{equation}
Note that formulas (\ref{bchi1}), (\ref{M1}) in the simplest case of dressing on a zero background were found by Faddeev and Takhtajan ~\cite{Faddeev-Takhtajan2007}. Note also that function $\chi_{\alpha\beta}$ also can be presented as the ratio of two determinants
\begin{equation}
\bchi_{\alpha\beta}=-\frac{\widehat{M}_{\alpha\beta}}{M}.
\end{equation}
Here
\begin{equation}
\widehat{M}_{\alpha\beta}=
\left(\begin{array}{cc}
        \delta_{\alpha\beta} & \begin{array}{ccc}
              \frac{q_{1,\beta}}{\lambda-\lambda_1} & \cdots & \frac{q_{n,\beta}}{\lambda-\lambda_n}
            \end{array}
         \\
        \begin{array}{c}
          q^*_{1,\alpha} \\
          \vdots \\
          q^*_{n,\alpha}
        \end{array}
         &  \begin{array}{c}
              M^{T}_{nm}
            \end{array}
      \end{array}\right).
\end{equation}
The dressing formula (\ref{general_NLSE solution and bchi asymptotic}) can be written in the explicit form. Let
\begin{equation}
\bPsi_0=
\left(
  \begin{array}{cc}
    \Psi_{011} & \Psi_{012} \\
    \Psi_{021} & \Psi_{022} \\
  \end{array}
\right),
\end{equation}
then
\begin{equation}
\bPsi=\frac{1}{M}
\left(
  \begin{array}{cc}
    \widehat{M}_{11}\Psi_{011}+\widehat{M}_{12}\Psi_{021} & \widehat{M}_{11}\Psi_{012}+\widehat{M}_{12}\Psi_{022} \\
    \widehat{M}_{21}\Psi_{011}+\widehat{M}_{22}\Psi_{021} & \widehat{M}_{21}\Psi_{012}+\widehat{M}_{22}\Psi_{022} \\
  \end{array}
\right).
\end{equation}
Now mention that transformation
\begin{eqnarray}
\bi{q}_n \ra a_n \bi{q}_n,
&\qquad&
\bi{p}_n \ra \frac{1}{a_n}\bi{p}_n,
\label{q_p_invariance}
\end{eqnarray}
where $a_n$ are arbitrary complex numbers that do not change the result of dressing. Thus one can put
\begin{equation}
\bxi_n = \left(
          \begin{array}{c}
            1 \\
            C_n \\
          \end{array}
        \right).
\end{equation}
The finally constructed $N$-solitonic solution depends on $2N$ complex numbers $\lambda_n, C_n$ or on $4N$ real parameters. In what follows we assume $Re(\lambda) > 0$. In fact, it is enough to enumerate all possible solitonic solutions. We will prove this fact in a separate paper. So far we are sure that $\lambda_n+\lambda_m^* \ne 0$, thus equations (\ref{p and q system}) are always solvable.

Now we present one-solitonic solution. Function $\bchi$ has only one pole at $\lambda=\eta$, while $\bchi^{-1}$ has a pole at $\lambda=-\eta^*$. They can be presented in the following form (see ~\cite{Zakharov-Shabat1972})
\begin{eqnarray}
\bchi=\bi{E}+\frac{\bi{U}}{\lambda-\eta},
&\qquad&
\bchi^{-1}=\bi{E}-\frac{\bi{U}^+}{\lambda+\eta^*}.
\end{eqnarray}
As before
\begin{equation}
U_{\alpha\beta}=p_{\alpha}q_{\beta}.
\end{equation}
Vectors $\bi{p}$ and $\bi{q}$ are connected by the relation
\begin{equation}
p_{\alpha}=\frac{(\eta+\eta^*)q^*_{\alpha}}{|q_1|^2+|q_2|^2}.
\end{equation}
As a result $\bchi$ and $\bchi^{-1}$ are
\begin{eqnarray}
\bchi=\bi{E}+\frac{(\eta+\eta^*)\bi{P}}{\lambda-\eta},
&\qquad&
\bchi^{-1}=\bi{E}-\frac{(\eta+\eta^*)\bi{P}^+}{\lambda+\eta^{*}}.
\end{eqnarray}
Here
\begin{equation}
P_{\alpha\beta}=
\frac{q^*_{\alpha}q_{\beta}}{|q_1|^2+|q_2|^2}.
\end{equation}
$\bi{P}^2=\bi{P}$, thus $P$ is a projection matrix. As before we assume that the seed matrix $\bPsi_0(x,t,\lambda)$ is known. According to our definition (\ref{F_n defenition})
\begin{equation}
F_{\alpha\beta}=\Psi_{0,\alpha\beta}(-\eta^*).
\end{equation}
Now
\begin{eqnarray}
q^*_1=F_{11}+C F_{12},
\nonumber\\
q^*_2=F_{21}+C F_{22}.
\end{eqnarray}
Here $C$ is arbitrary complex constant. Finally a new solution of NLSE is
\begin{equation}
\varphi=
\varphi_0
-
\frac
{
2(\eta+\eta^*)q^*_1 q_2
}
{
|q_1|^2+|q_2|^2
}.
\label{Onesolitonic solution.form0}
\end{equation}
This formula presents the solitonic solution on an arbitrary background. It was established first in 1979 by Zakharov and Shabat ~\cite{Zakharov-Shabat1979} and reobtained in 1988 by Its, Rybin and Sall ~\cite{Its-Rybin-Sall1988} (see also ~\cite{Matveev-Salli1991}).

All results of this chapter can be extended to the much more general class of nonlinear wave systems which can be presented as compatibility conditions for overdetermined linear system
\begin{eqnarray}
\bPsi_x=\widehat{\bi{U}}(\lambda)\bPsi,
\nonumber\\
\bPsi_t=\widehat{\bi{V}}(\lambda)\bPsi.
\end{eqnarray}
Where $\bi{U}$ and $\bi{V}$ are rational matrix $2\times2$ functions on $\lambda$ satisfying the follows condition
\begin{eqnarray}
\widehat{\bi{U}}^+(-\lambda^*)=-\widehat{\bi{U}}(\lambda),
\nonumber\\
\widehat{\bi{V}}^+(-\lambda^*)=-\widehat{\bi{V}}(\lambda).
\end{eqnarray}
In particular they can be extended to all higher members of the NLSE hierarchy. Only a minor generalization is needed to extend this procedure to the case of $n\times n$ matrix systems.

\section{$N$-solitonic solution on the condensate}
\begin{figure}[h]
\centering
\includegraphics[width=2in]{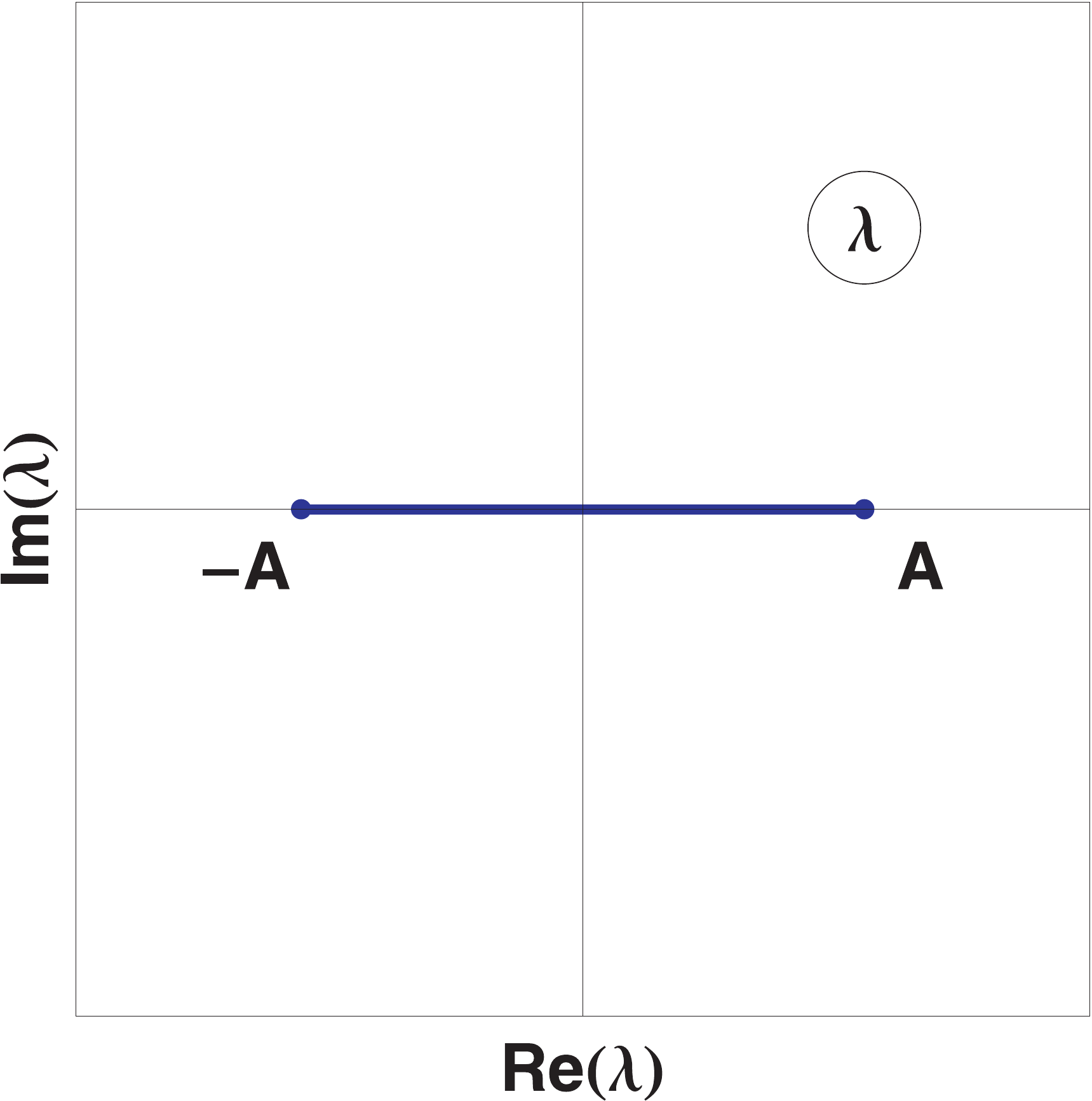}
\includegraphics[width=2in]{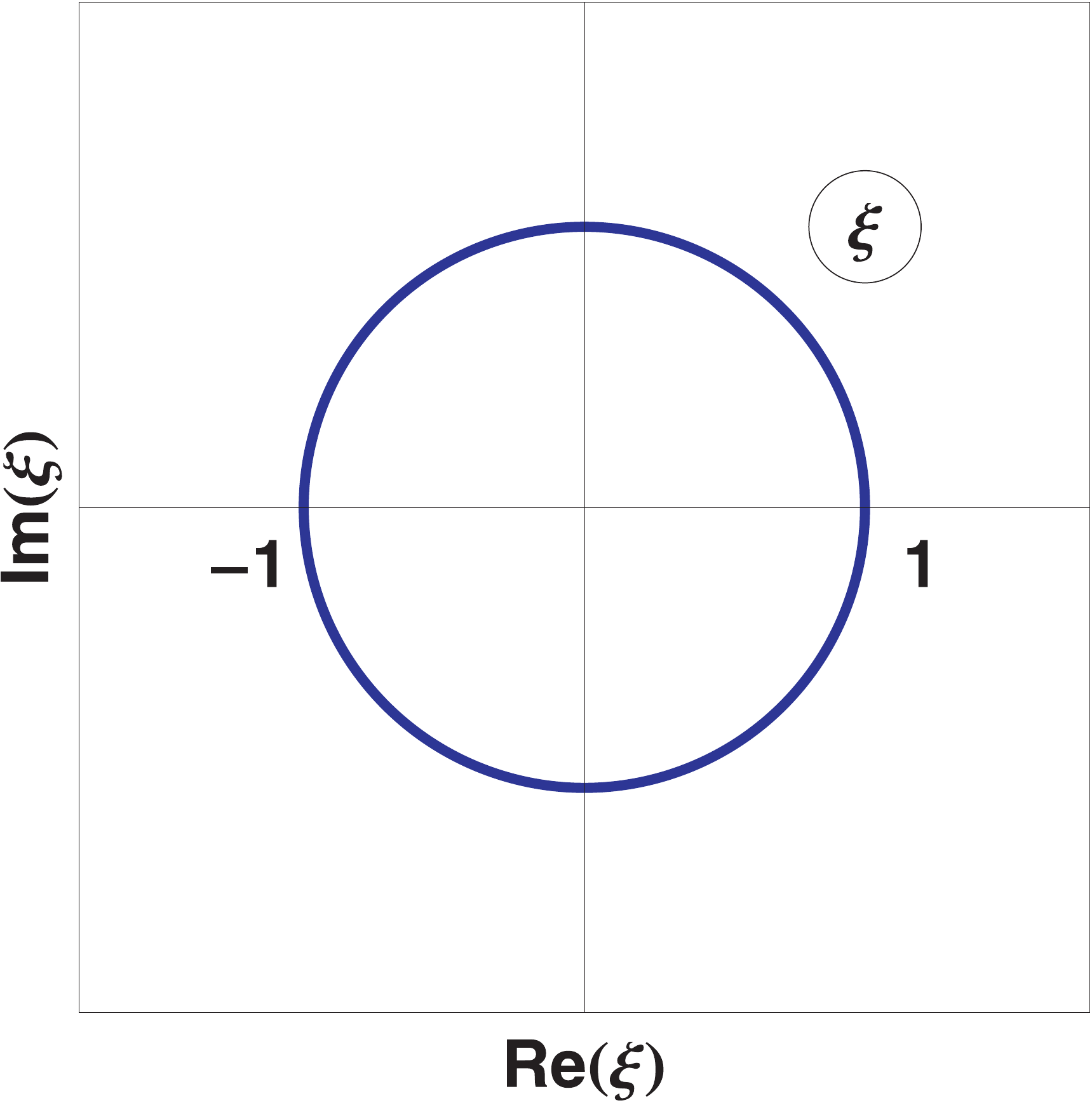}
\caption{\label{Uniformization}
Uniformization of the plane of spectral parameter with the help of Jukovsky transform.
}
\end{figure}

In what follows we study dressing only on a condensate background. Now $\varphi_0=A$, then
\begin{eqnarray}
\bi{\widehat{U}}_0=
\left(
  \begin{array}{cc}
    \lambda & A \\
    -A & -\lambda \\
  \end{array}
\right),
&\qquad&
\widehat{\bi{W}}_0=0.
\end{eqnarray}
$\bPsi_0$ can be found as
\begin{equation}
\fl \bPsi_{0}(x,t,\lambda)=\frac{1}{\sqrt{1-s^{2}(\lambda)}}
\left(
  \begin{array}{cc}
    \exp(\phi(x,t,\lambda)) & s(\lambda) \exp(-\phi(x,t,\lambda)) \\
    s(\lambda) \exp(\phi(x,t,\lambda)) & \exp(-\phi(x,t,\lambda)) \\
  \end{array}
\right).
\label{condensat lax solution}
\end{equation}
Here
\begin{eqnarray}
\phi=kx+\Omega t, &\qquad& k^{2}=\lambda^{2}-A^{2},
\nonumber\\
\Omega=-\rmi\lambda k, &\qquad& s=-\frac{A}{\lambda+k}.
\nonumber
\end{eqnarray}
In what follows we assume that function $k(\lambda)=\sqrt{\lambda^2-A^2}$ has a cut at $-A< Re(\lambda) < A$. Thus $k(\lambda) \ra \lambda$ at $\lambda \ra \infty$. Then
\begin{equation}
\fl \bPsi_{0}^{-1}(x,t,\lambda)=\frac{1}{\sqrt{1-s^{2}(\lambda)}}
\left(
  \begin{array}{cc}
    \exp(-\phi(x,t,\lambda)) & -s(\lambda)\exp(-\phi(x,t,\lambda)) \\
    -s(\lambda)\exp(\phi(x,t,\lambda)) & \exp(\phi(x,t,\lambda)) \\
  \end{array}
\right).
\label{condensat lax solution^-1}
\end{equation}
Notice that
\begin{equation}
k^*(-\lambda^*)=-k(\lambda), $\qquad$ s^*(-\lambda^*)=-s(\lambda), $\qquad$ \phi^*(-\lambda^*)=-\phi(\lambda).
\label{k s phi properties}
\end{equation}
One can check that
\begin{equation}
\bPsi_{0}^{-1}(-\lambda^*)=\bPsi_0^+(\lambda).
\end{equation}
We denote for simplicity
\begin{eqnarray}
\phi_n=\phi_n(\lambda_n),
&\qquad&
s_n=s(\lambda_n),
\end{eqnarray}
and by virtue of (\ref{k s phi properties})
\begin{eqnarray}
\phi_n(-\lambda_n^*)=-\phi_n^*,
&\qquad&
s_n(-\lambda_n^*)=-s_n^*.
\end{eqnarray}
Then
\begin{eqnarray}
\fl \bi{F}_n=\bPsi_0(-\lambda^*_n)=
\left(
  \begin{array}{cc}
    \exp(-\phi_n^*) & -s_n^* \exp(\phi_n^*) \\
    -s_n^* \exp(-\phi_n^*) & \exp(\phi_n^*) \\
  \end{array}
\right),
&\qquad&
\bi{q}_n^*=\bi{F}_n
\left(
    \begin{array}{c}
       1 \\
       C_n \\
     \end{array}
\right),
\label{condensat lax solution}
\end{eqnarray}
(we can omit the factor $(1-s_n^2)^{-1/2}$ because, as we mention before, it does not change the result of dressing) and
\begin{eqnarray}
\fl q_{n1}=\exp(-\phi_n)-C^*_n s_n \exp(\phi_n), &\qquad& q_{n2}=-s_n \exp(-\phi_n)+C^*_n \exp(\phi_n).
\end{eqnarray}
So far we have assumed that $\bchi$ is a rational function on the $\lambda$-plane with a cut at $-A<Re(\lambda)<A$. Now we perform the Jukowsky transform and map this plane onto the outer part of the circle of unit radius (see figure \ref{Uniformization}).
\begin{equation}
\lambda=\frac{A}{2}(\xi+\xi^{-1}), $\qquad$ k=\frac{A}{2}(\xi-\xi^{-1}), $\qquad$ s=-\xi^{-1}.
\label{Uniformizing variable}
\end{equation}
The $\lambda$-plane is mapped onto exterior of the circle of radius $R$. If the pole is located at $\lambda=\lambda_n$, then $\lambda_n=\frac{A}{2}(\xi_n+\xi^{-1}_n)$. In new variables
\begin{equation}
\bi{F}_n=
\left(
  \begin{array}{cc}
    \exp(-\phi_n^*) & \xi^{*^{-1}}_n \exp(\phi_n^*) \\
    \xi^{*^{-1}}_n \exp(-\phi_n^*) & \exp(\phi_n^*) \\
  \end{array}
\right),
\end{equation}
and
\begin{eqnarray}
\fl q_{n1}=\exp(-\phi_n)+\xi^{-1}_n C^*_n \exp(\phi_n), &\qquad& q_{n2}=\xi^{-1}_n \exp(-\phi_n)+C^*_n \exp(\phi_n).
\end{eqnarray}
Finally we use parametrization
\begin{eqnarray}
\xi_n=R_n \exp(\rmi\alpha_n), &\qquad& C_n=\exp(\rmi\theta_n+\mu_n),
\end{eqnarray}
and denote $R_n=\exp(z_n)$. Now
\begin{eqnarray}
\lambda_n=
&\frac{A}{2}( R_n+R^{-1}_n) \cos\alpha_n+\frac{\rmi A}{2}( R_n-R_n^{-1}) \sin\alpha_n
\nonumber\\
&=A( \cosh z_n\cos\alpha_n+\rmi \sinh z_n\sin\alpha_n ).
\end{eqnarray}
We also denote
\begin{eqnarray}
w_n=-\rmi\alpha_n-z_n.
\end{eqnarray}
Now
\begin{eqnarray}
F_n=
\left(
  \begin{array}{cc}
    \exp(-\phi^*_n) & \exp(w^*_n+\phi^*_n) \\
    \exp(w^*_n-\phi^*_n) & \exp(\phi^*_n) \\
  \end{array}
\right).
\end{eqnarray}
Expressions for $\bi{q}_n$ can be simplified by redefining the phase factor $\phi_n$. One can put
\begin{eqnarray}
\fl q_{n1}=\exp(-\phi_n)+\exp(w_n+\phi_n), &\qquad& q_{n2}=\exp(w_n-\phi_n)+\exp(\phi_n).
\label{vectors q_n}
\end{eqnarray}
Now
\begin{eqnarray}
\phi_n=u_n+\rmi v_n,
\nonumber\\
u_n=\ae_n x-\gamma_n t + \mu_n/2,
\nonumber\\
v_n=k_n x-\omega_n t - \theta_n/2,
\nonumber\\
\ae_n=\frac{A}{2}(R_n-R_n^{-1}) \cos\alpha_n = A\sinh z_n\cos\alpha_n,
\nonumber\\
k_n=\frac{A}{2} (R_n+R_n^{-1}) \sin\alpha_n = A\cosh z_n\sin\alpha_n,
\nonumber\\
\gamma_n=-\frac{A^2}{4} (R_n^2+R_n^{-2}) \sin 2\alpha_n=-\frac{A^2}{2}\cosh 2z_n \sin 2\alpha_n,
\nonumber\\
\omega_n=\frac{A^2}{4} (R_n^2-R_n^{-2}) \cos 2\alpha_n=\frac{A^2}{2}\sinh 2z_n\cos 2\alpha_n.
\label{phi_n}
\end{eqnarray}
We write following expressions for quadratic combinations of $\bi{q}_n$ which will be useful in the next paragraphs:
\begin{eqnarray}
q^*_{n1}q_{n2}=2\rme^{-z_n}\bigl[
\cos\alpha_n\cosh 2u_n+
\nonumber\\
\cosh z_n\cos 2v_n+\rmi(\sin\alpha_n\sinh 2u_n+\sinh z_n\sin 2v_n)
\bigr],
\nonumber\\
|q_{n1}|^2-|q_{n2}|^2=4\rme^{-z_n} ( \sin\alpha_n\sin 2v_n-\sinh z_n\sinh 2u_n ),
\nonumber\\
|\bi{q}_n|^2=4\rme^{-z_n} ( \cos\alpha_n\cos 2v_n+\cosh z_n\cosh 2u_n ).
\label{f(q)_a}
\end{eqnarray}
The $N$ - solitonic solution is invariant with respect to shifts in time and space. If we replace
\begin{eqnarray}
x \ra x-x_0, &\qquad& t \ra t-t_0.
\end{eqnarray}
Then
\begin{eqnarray}
\mu_n \ra \mu_n+2(\ae_n x_0 + \gamma_n t_0), &\qquad& \theta_n \ra \theta_n-2(k_n x_0 - \omega_n t_0).
\end{eqnarray}
This means that a space-time shift leads to renormalization of constants $C_n$:
\begin{eqnarray}
C_n \ra C_n \exp[2(\ae_n x_0 + \gamma_n t_0)-2\rmi(k_n x_0 - \omega_n t_0)]
\end{eqnarray}
$N$-solitonic solution can be considered as a nonlinear superposition of N separate solitons. Each of them is characterized by the group velocity
\begin{eqnarray}
V_{Gr_n}=\frac{\gamma_n}{\ae_n}=-\frac{A\cosh 2z_n \sin \alpha_n}{\sinh z_n},
\end{eqnarray}
and phase velocity
\begin{eqnarray}
V_{Ph_n}=\frac{\omega_n}{k_n}=\frac{A\sinh z_n\cos 2\alpha_n}{\sin\alpha_n}.
\end{eqnarray}
When all $z_n > 0$ the solution does not contain "Akhmediev components". If all group velocities are different, the $N$ - solitonic solution separates asymptotically at $t \ra \pm \infty$ into a superposition of individual solitons remote from one another.  This makes it possible to determine asymptotic properties at $x \ra \pm \infty$ of an $N$-solitonic solution. For an $N$-solitonic solution,
\begin{eqnarray}
\varphi \ra A\exp(\rmi\alpha^{\pm}), &\qquad& x \ra \pm \infty.
\end{eqnarray}
The phases $\alpha^{\pm}$ are constants in time. In the next paragraph we show that the phase of a one-solitonic solution in the case $z \ne 0$ is
\begin{eqnarray}
\varphi \ra A\exp(\pm 2\i\alpha), &\qquad& x \ra \pm \infty.
\end{eqnarray}
Then for far separated solitons
\begin{eqnarray}
\alpha^+=2(\alpha_1+ \cdots +\alpha_n), &\qquad& \alpha^-=-2(\alpha_1+ \cdots +\alpha_n).
\label{asymptotic phases}
\end{eqnarray}
This fact holds even in a general case when some group velocities coincide.

If we are interested in $N$-solitonic solutions localized in a finite spatial domain and not perturbing the remote condensate we must demand
\begin{eqnarray}
\alpha^+=\alpha^-.
\end{eqnarray}
We call this solution a regular solitonic solution and in what follows study only this case. If we assume that the modulation instability develops from a localized perturbation, only a regular solution can be used as a model for its nonlinear behavior. Looking at (\ref{asymptotic phases}) we conclude that an N-solitonic solution is regular when
\begin{eqnarray}
\alpha_1+ \cdots +\alpha_n=0,\; \pm \frac{\pi}{2}.
\end{eqnarray}
Among one-solitonic solutions only the Kuznetsov and Peregrine solutions are regular. In two-solitonic case we can construct a broad class of regular solutions.

When all $z_n=0$ the solution is the $N$-Akhmediev breather, which is periodic in space and localized in time. In this case we should study asymptotics at $t \ra \pm \infty$. The result is analogous to the previous case but now the phase of the one-solitonic solution (Akhmediev breather) is
\begin{eqnarray}
\varphi \ra A\exp(\pm 2\rmi|\alpha|), &\qquad& t \ra \pm \infty.
\end{eqnarray}
The modulus sign appears because the sign of $\alpha$ is not important for the Akhmediev breather (see the next paragraph). Then for the $N$-Akhmediev breather
\begin{eqnarray}
\varphi \ra A\exp(\rmi\alpha^{\pm}), &\qquad& t \ra \pm \infty
\end{eqnarray}
\begin{eqnarray}
\alpha^+=2(|\alpha_1|+ \cdots +|\alpha_n|), &\qquad& \alpha^-=-2(|\alpha_1|+ \cdots +|\alpha_n|).
\label{asymptotic phases_AB}
\end{eqnarray}
We developed an analytical code making it possible to calculate $N$-solitonic solutions using Mathematica and checked relations (\ref{asymptotic phases}) and (\ref{asymptotic phases_AB}) directly for $n=2,3$.

In the case of the N-Akhmediev breather a condition of equal phases at $t \ra \pm \infty$ is following
\begin{eqnarray}
|\alpha_1|+ \cdots +|\alpha_n|=\pm \frac{\pi}{2}.
\end{eqnarray}
This is a pure homoclinic N-Akhmediev breather.

\section{One-solitonic solution on the condensate}
\begin{figure}[h]
\centering
\includegraphics[width=2in]{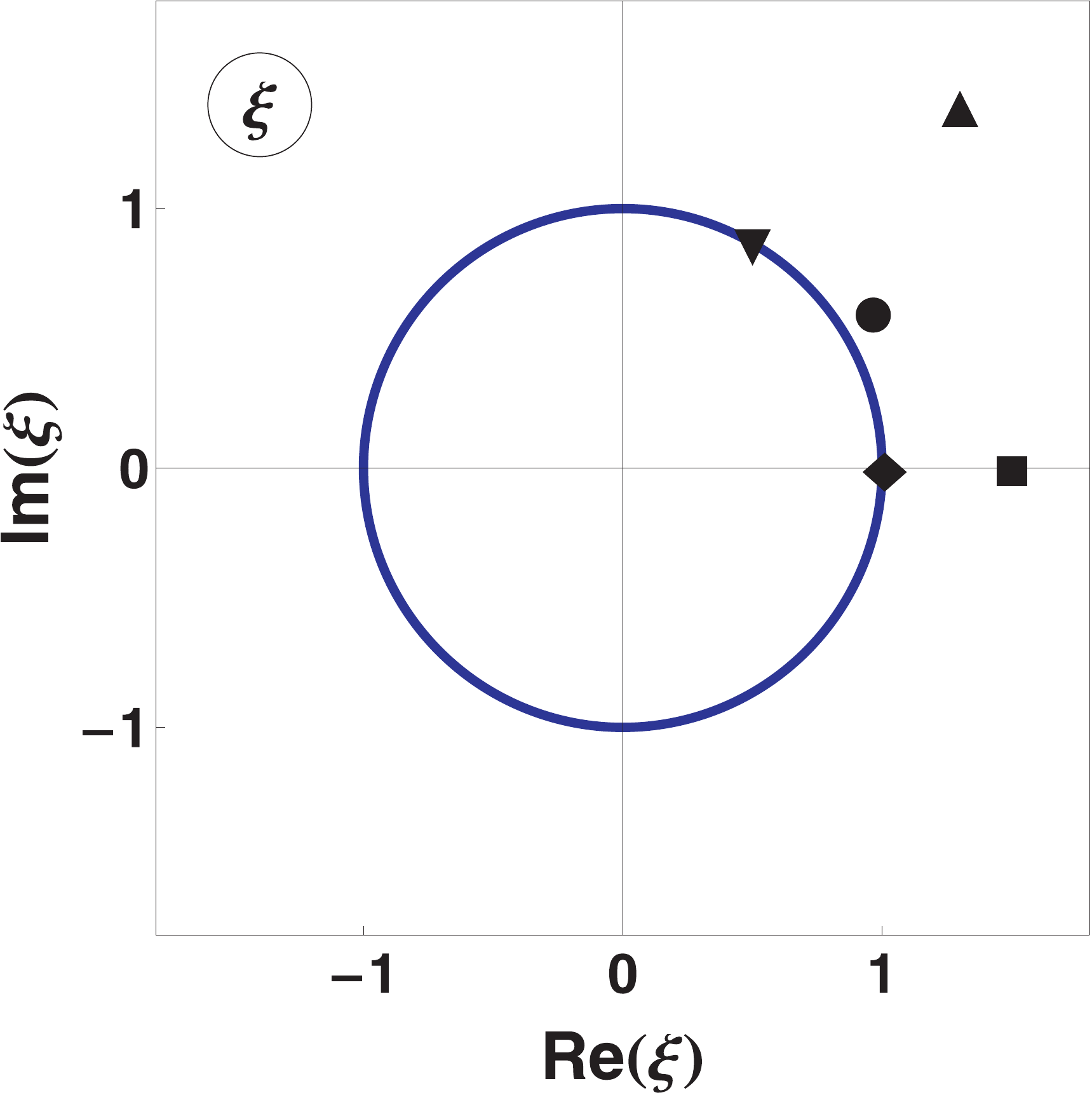}
\caption{\label{Poles family}
Species of one-solitonic solution. \fullsquare - Kuznetsov soliton, \fulltriangledown - Akhmediev breather \fulltriangle - General solution, \fullcircle - quasi-Akhmediev breather, \fulldiamond - Peregrine soliton.
}
\end{figure}

\begin{figure}[h]
\centering
\includegraphics[width=3in]{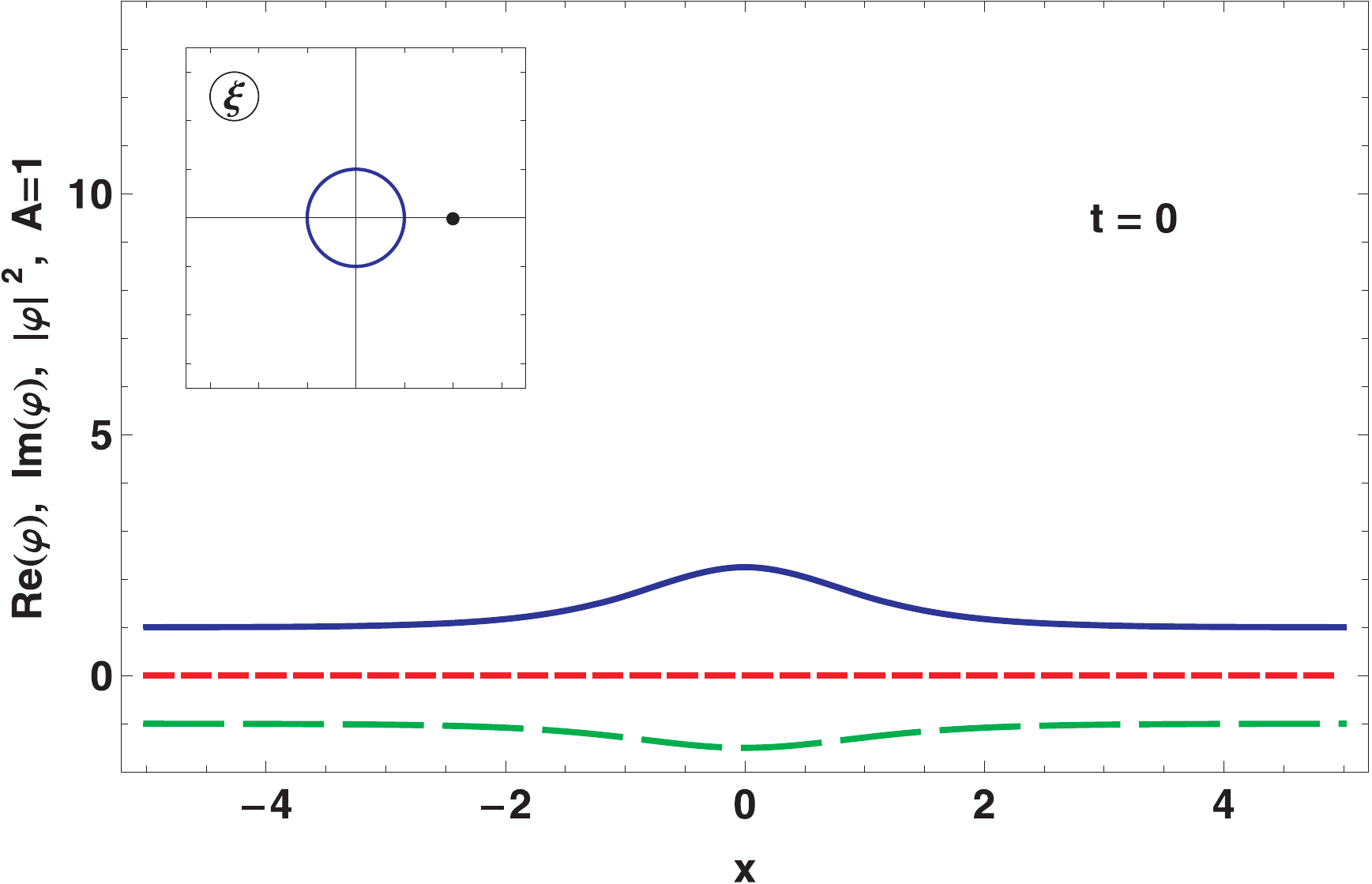}
\includegraphics[width=3in]{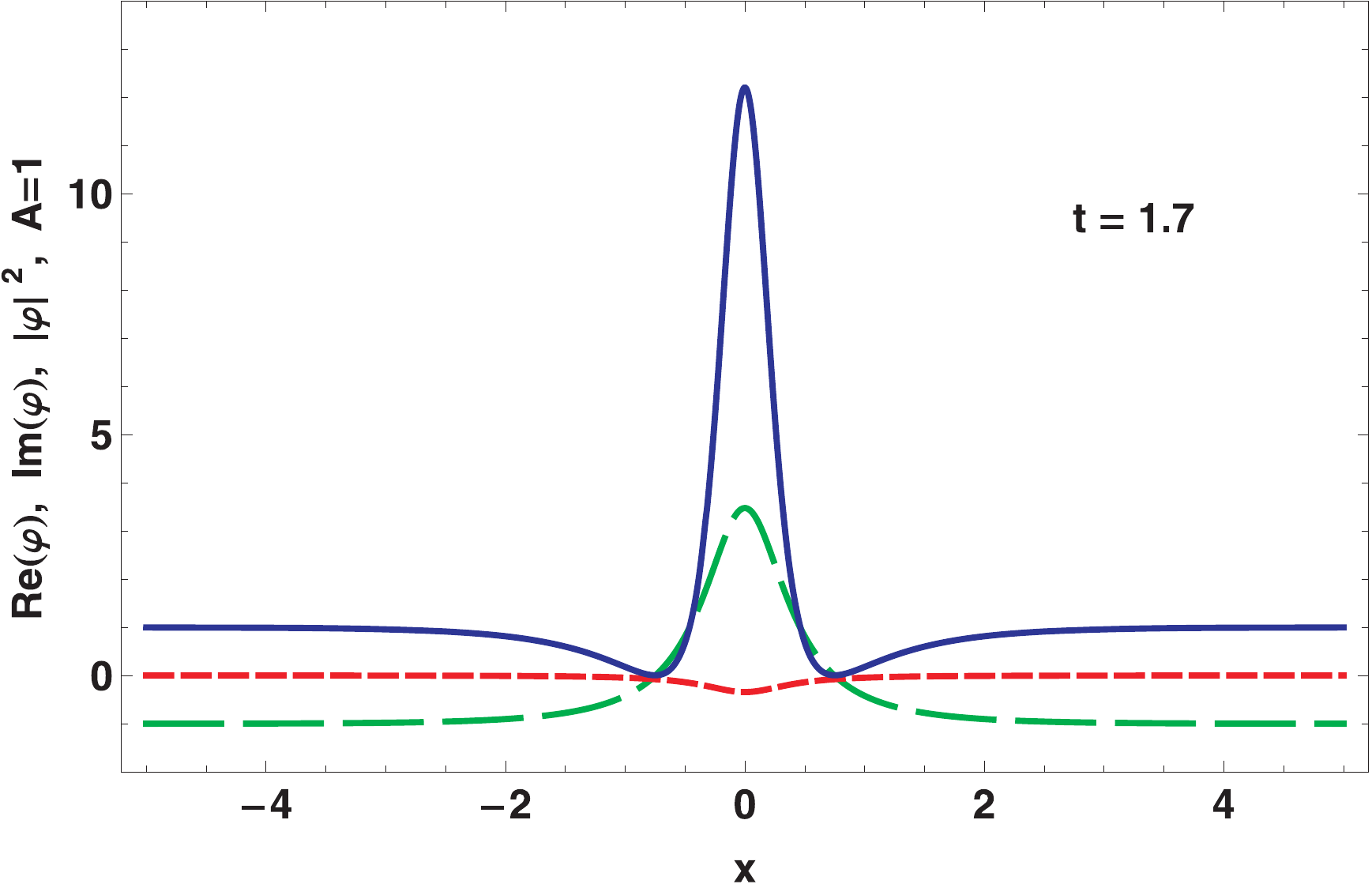}
\caption{\label{Kuznetsov}
Kuznetsov soliton $\varphi$ at the moment of minimum (left picture) and maximum (right picture) of its amplitude. $R=2,\; \mu=0,\; \theta=0$. Green dashed lines - real part of $\varphi$, red short dashed lines - imaginary part of $\varphi$ and blue solid lines - absolute squared value of $\varphi$.
}
\end{figure}

A one-solitonic solution on a condensate background can be obtained by implementing the results of the last part of \S 3. This solution is defined by only one complex eigenvalue $\lambda_1=\eta$ and one complex parameter $C_1=C$. Our standard parametrization is
\begin{equation}
\xi=R\exp(\rmi\alpha), $\qquad$ C=\exp(\rmi\theta+\mu), $\qquad$ R=\exp(z).
\nonumber
\end{equation}
Let us denote
\begin{eqnarray}
\phi=\phi(\eta), &\qquad& s=s(\eta).
\end{eqnarray}
Recall that by virtue of (\ref{k s phi properties})
\begin{eqnarray}
\phi(-\eta^*)=-\phi^*, &\qquad& s(-\eta^*)=-s^*.
\end{eqnarray}
We need only one complex vector $\bi{q}=(q_1,q_2)$:
\begin{eqnarray}
\fl q_1=\exp(-\phi)+\exp(-\rmi\alpha-z+\phi), &\qquad& q_2=\exp(-\rmi\alpha-z-\phi)+\exp(\phi).
\label{q for onesolitonic solution}
\end{eqnarray}
Here
\begin{eqnarray}
\phi=u+\rmi v, &\qquad&
\nonumber\\
u=\ae x-\gamma t+\mu/2, &\qquad&  v=kx -\omega t-\theta/2,
\nonumber\\
\ae=A\sinh z\cos\alpha, &\qquad& \gamma=-\frac{A^2}{2}\cosh 2z\sin 2\alpha, \nonumber\\
k=A\cosh z\sin\alpha, &\qquad& \omega=\frac{A^2}{2}\sinh 2z\cos 2\alpha.
\label{phi in uniformazing var.}
\end{eqnarray}
The general one-solitonic solution depends on four scalar parameters $R,\;\alpha,\;\theta,\;\mu$. Two of them $\theta,\;\mu$ are responsible for shifts in time and in space. If we put $\mu=0,\;\theta=0$, the one-solitonic solution can be written as follows
\begin{eqnarray}
\fl \varphi=-A(\cosh z\cosh 2u+\cos\alpha\cos 2v)^{-1}
[\cosh z\cos 2\alpha \cosh 2u+\cosh 2z\cos\alpha\cos 2v
\nonumber\\
\fl +\rmi(\cosh z\sin 2\alpha\sinh 2u+\sinh 2z\cos\alpha\sin 2v)].
\label{Onesolitonic solution.final form}
\end{eqnarray}
To obtain the general solution we replace $t \ra t-t_0$,  $x \ra x-x_0$ where $x_0=(\mu\omega+\theta\gamma)/2\Delta$, $t_0=(\mu k+\theta\ae)/2\Delta$. Here $\Delta=k\gamma-\ae\omega$. This solution is localized in space if $R \ne 1$. In this case the asymptotics of (\ref{Onesolitonic solution.final form}) are
\begin{eqnarray}
\varphi \ra -A\exp(-2\rmi\alpha), &\qquad& x \ra -\infty, \nonumber\\
\varphi \ra -A\exp(2\rmi\alpha), &\qquad& x \ra +\infty, \nonumber\\
|\varphi|^{2}=A^{2}, &\qquad& x \ra \pm\infty.
\end{eqnarray}
We see that solution (\ref{Onesolitonic solution.final form}) has identical asymptotics at $x \ra \pm \infty$ only in the case $\alpha=0$, when the pole is on the real axis. The position of pole defines the type of the soliton. Different possible positions are plotted in the figure \ref{Poles family}.
We describe below all these species of one-solitonic solutions. Let us put $\alpha=0,\;R>1,\;z>0$. This is exactly the Kuznetsov solution. If the solution is centered at $x=0$ and $\theta=0$ it is presented by following expression. Then
\begin{eqnarray}
\varphi=-A\frac{\cosh z \cosh 2u+\cosh 2z\cos 2v+\rmi\sinh 2z\sin 2v}{\cosh z\cosh 2u+\cos 2v}.
\label{Kuznetsov's Solution}
\end{eqnarray}
Here
\begin{eqnarray}
u=A\sinh(z)x, &\qquad& v=-\frac{A^2}{2}\sinh(2z)t.
\end{eqnarray}
This solution is periodic in time. Its oscillation period is
\begin{equation}
T=\frac{4\pi}{A^2\sinh 2z}.
\end{equation}
Note that
\begin{eqnarray}
T \ra \infty, &\qquad& z \ra 0,\; R \ra 1
\nonumber\\
T \ra 0, &\qquad& z \ra \infty,\; R \ra \infty.
\end{eqnarray}
It was reported that recently Kuznetsov soliton was observed experimentally in optical fibers \cite{Kibler2012}. The typical behavior of the Kuznetsov soliton is given in figure \ref{Kuznetsov}.
In the limit $R \ra 1$, $T \ra \infty$ the Kuznetsov solutions turns into the Peregrine homoclinic soliton given by expression
\begin{equation}
\varphi=-A+4A\frac{1-2\rmi A^2 t}{1+4 A^2 x^2+4 A^4 t^2}.
\end{equation}
This solution is actually "instanton" appearing from the condensate and disappearing. Indeed $\varphi \ra A$ at $t \ra \pm \infty$. The Peregrine soliton was observed experimentally in water wave tank ~\cite{Chabchoub2011} and optical fibers ~\cite{Kibler2010}. $N$-"instantonic" solutions can be obtained by analogous limit for $N$ poles. Multisolitonic solutions describe a special scenario of modulation instability development, when growing perturbation in the long run return back to condensate. We will show in this article that this is a very special scenario, badly unstable with respect to small deformations. In a more general case waves developed from small perturbation do exist infinitely long time.

\begin{figure}[h]
\centering
\includegraphics[width=2in]{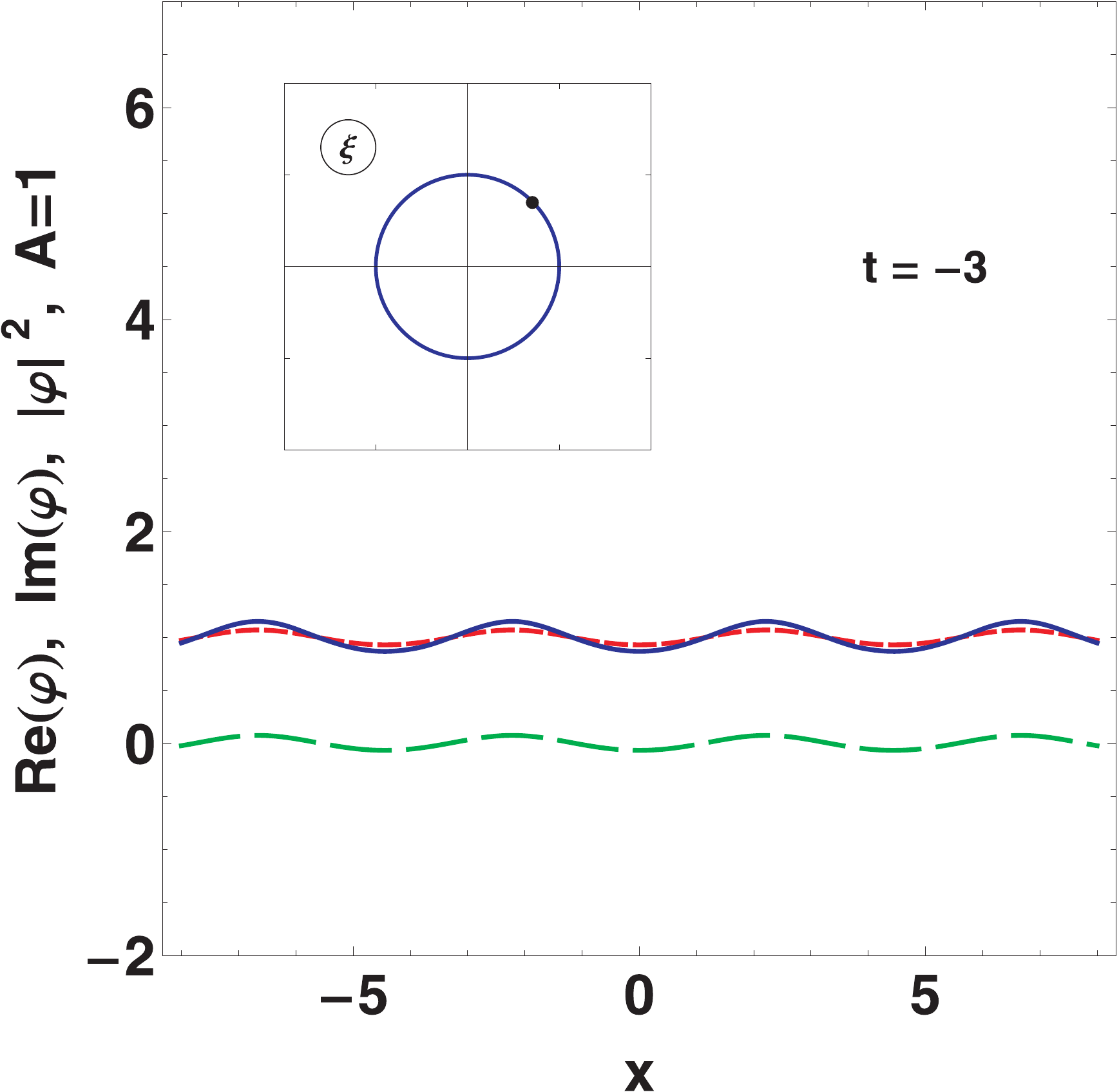}
\includegraphics[width=2in]{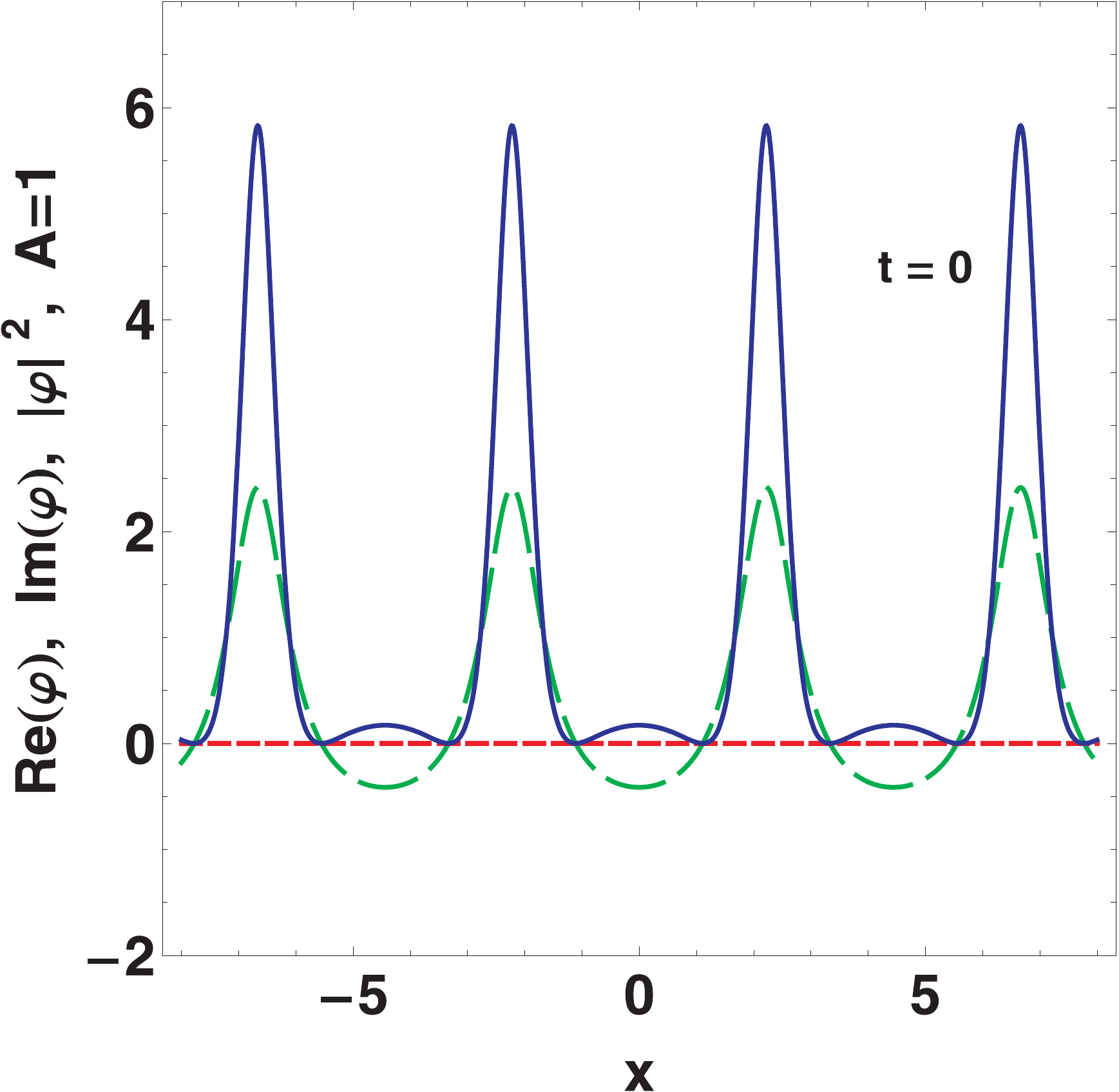}
\includegraphics[width=2in]{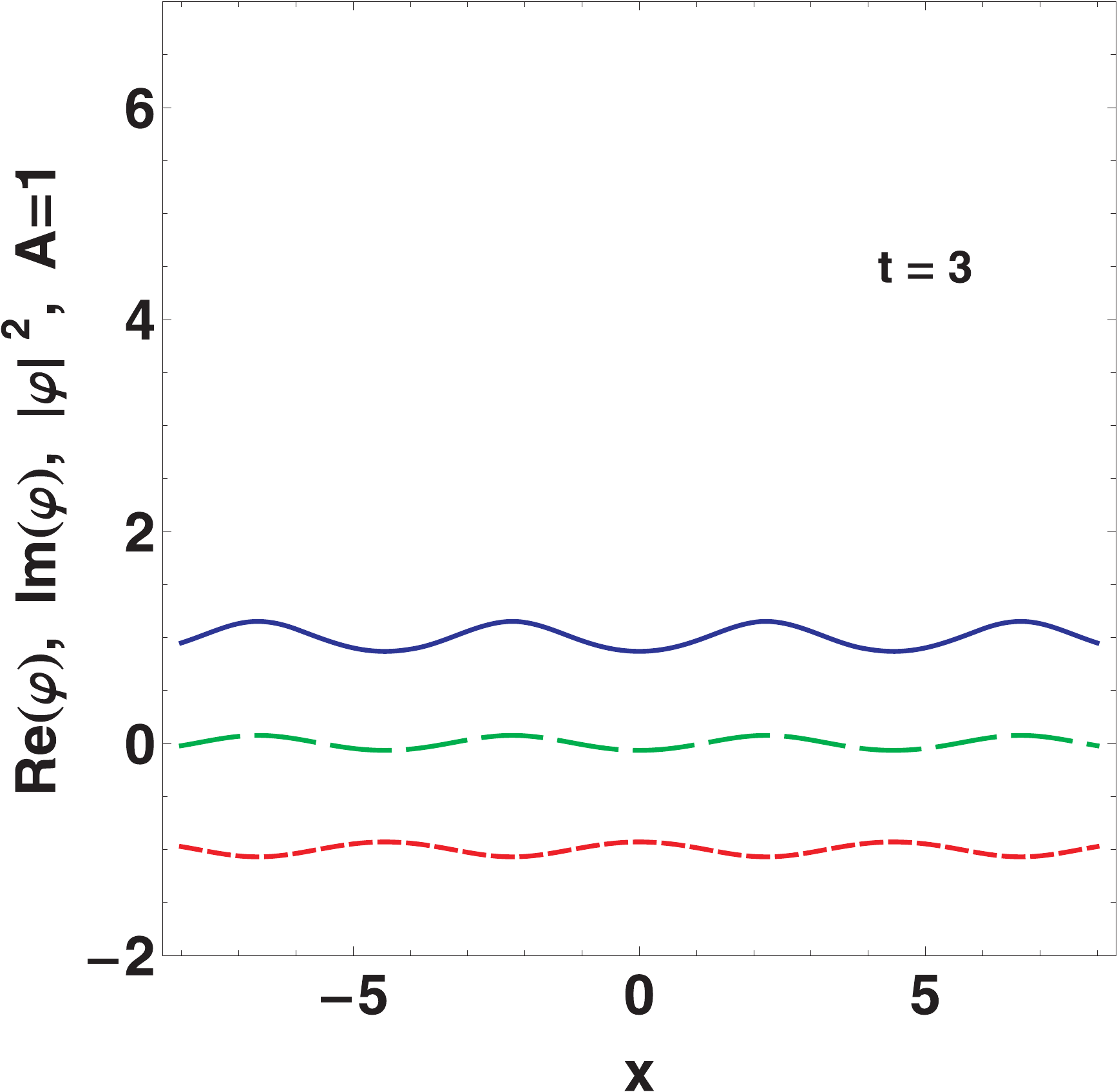}
\caption{\label{AB}
Akhmediev breather $\varphi$ at moments of time $t=-3$ (left picture) and $t=0$ (right picture). $\alpha=\pi/4,\;\mu=0,\;\theta=0$. Green dashed lines - real part of $\varphi$, red short dashed lines - imaginary part of $\varphi$ and blue solid lines - absolute squared value of $\varphi$.
}
\end{figure}

\begin{figure}[h]
\centering
\includegraphics[width=3in]{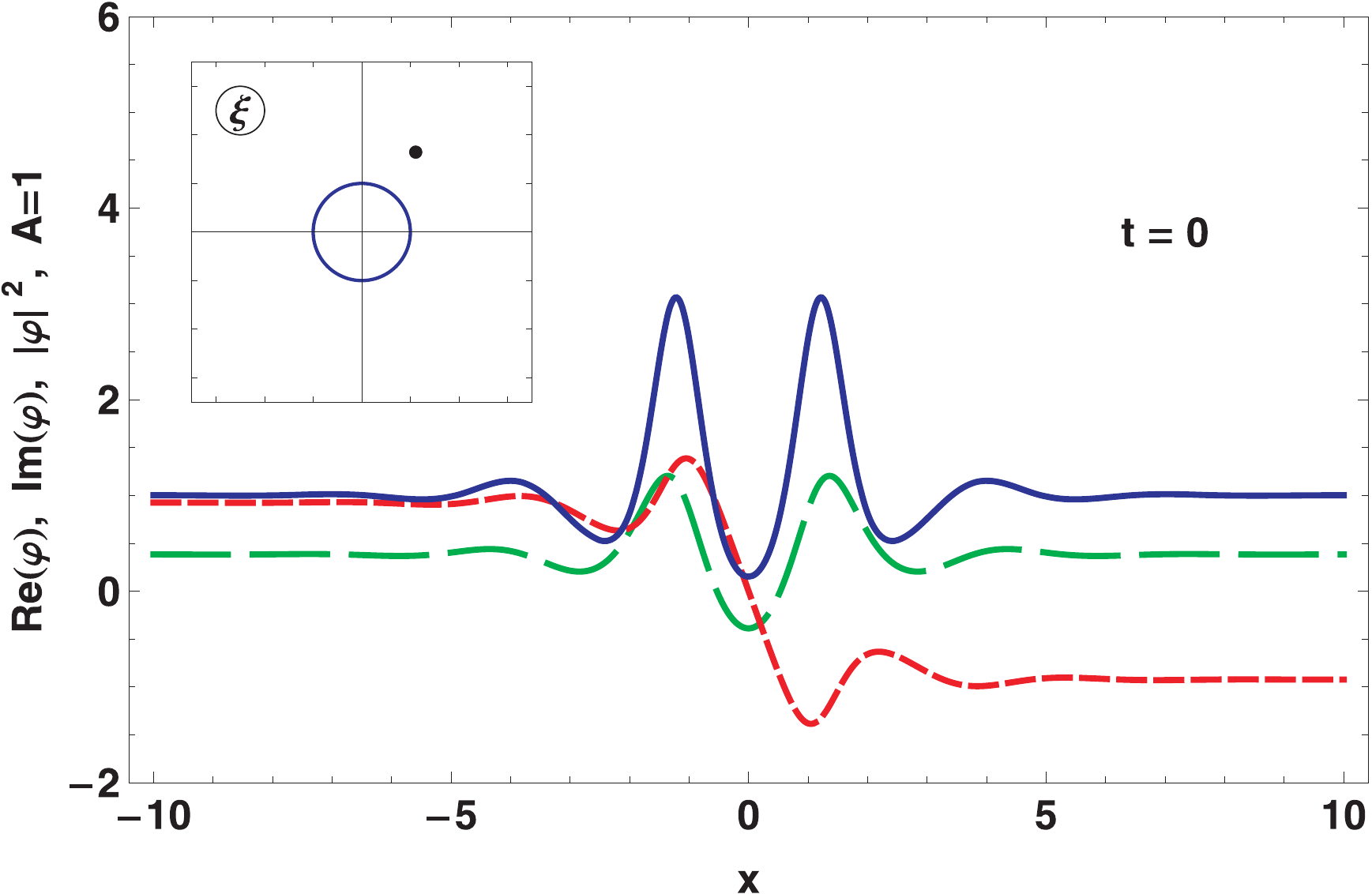}
\includegraphics[width=3in]{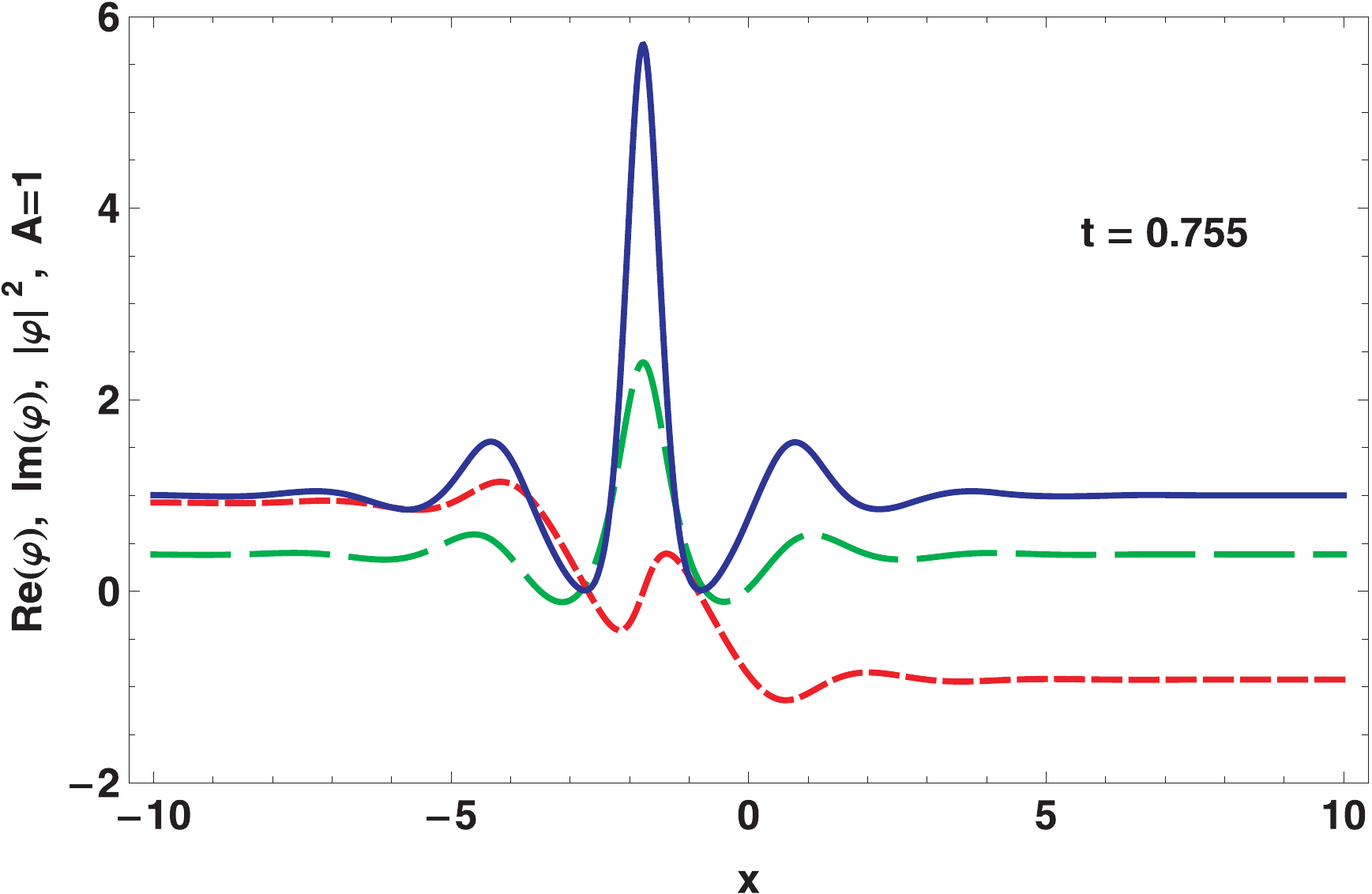}
\caption{\label{1S_General}
General one-solitonc solution $\varphi$  at the moment of minimum (left picture) and maximum (right picture) of its amplitude with parameters: $R=2,\;\alpha=5\pi/16,\;\mu=0,\;\theta=0$. Green dashed lines - real part of $\varphi$, red short dashed lines - imaginary part of $\varphi$ and blue solid lines - absolute squared value of $\varphi$.
}
\end{figure}

We now put $R=1,\;z=0$. Again we will put $\theta=0,\;\mu=0$. We obtain the famous Akhmediev breather, which is periodic in space and localized in time.
\begin{eqnarray}
\varphi=
-A
\frac
{\cos 2\alpha\cosh 2u+\cos\alpha\cos 2v+\rmi\sin 2\alpha\sinh 2u}
{\cosh 2u+\cos\alpha\cos 2v},
\label{Akhmediev's Breather}
\end{eqnarray}
here
\begin{eqnarray}
u=\frac{1}{2}A^2\sin(2\alpha)t, &\qquad&  v=A\sin(\alpha)x.
\nonumber
\end{eqnarray}
The Akhmediev breather is plotted in figure \ref{AB}. Note that the Akhmediev breather is identical at $\alpha$ and $-\alpha$. This is expected, because these two points are equal in the initial $\lambda$ plane. This leads to a modulus sign in asymptotic expressions at $t \ra \pm \infty$:
\begin{eqnarray}
\varphi \ra -A\exp(-2\rmi|\alpha|), &\qquad& t \ra -\infty \nonumber\\
\varphi \ra -A\exp(2\rmi|\alpha|), &\qquad& t \ra +\infty.
\nonumber
\end{eqnarray}
The Akhmediev breather is homoclinic in a weak sense, i.e. $|A^+|^2=|A^-|^2$, but $A^+ \ne A^-$. We can see it by comparison of the solution plotted in figure \ref{AB} at moments of time $t=-3$ and $t=3$. In the especially interesting case $\alpha=\pi/4$ we find that $A^+ = -A^-$. The spatial period of the Akhmediev breather
\begin{eqnarray}
L=\frac{2\pi}{A\sin\alpha},
\end{eqnarray}
tends to infinity if $\alpha \ra 0$. In this limit the solution tends to a periodic set of Peregrine solitons remote from one another.
Now we return to the general one-solitonic solution. This is a localized object of size $l \approx (A\sinh z\cos\alpha)^{-1}$ propagating along the condensate with group velocity
\begin{equation}
V_{gr}=\frac{\gamma}{\ae}=-\frac{A\cosh 2z\sin\alpha}{\sinh z}.
\end{equation}
The soliton's amplitude oscillates with angular frequency $\omega$. The soliton is filled with a carrying wave propagating with phase velocity.
\begin{equation}
V_{ph}=\frac{\omega}{k}=\frac{A\sinh z\cos 2\alpha}{\sin\alpha}.
\end{equation}
If $\alpha \ra 0$ this carrying wave vanishes. A typical general one-solitonic solution is plotted in figure \ref{1S_General}.

\begin{figure}[h]
\centering
\includegraphics[width=3in]{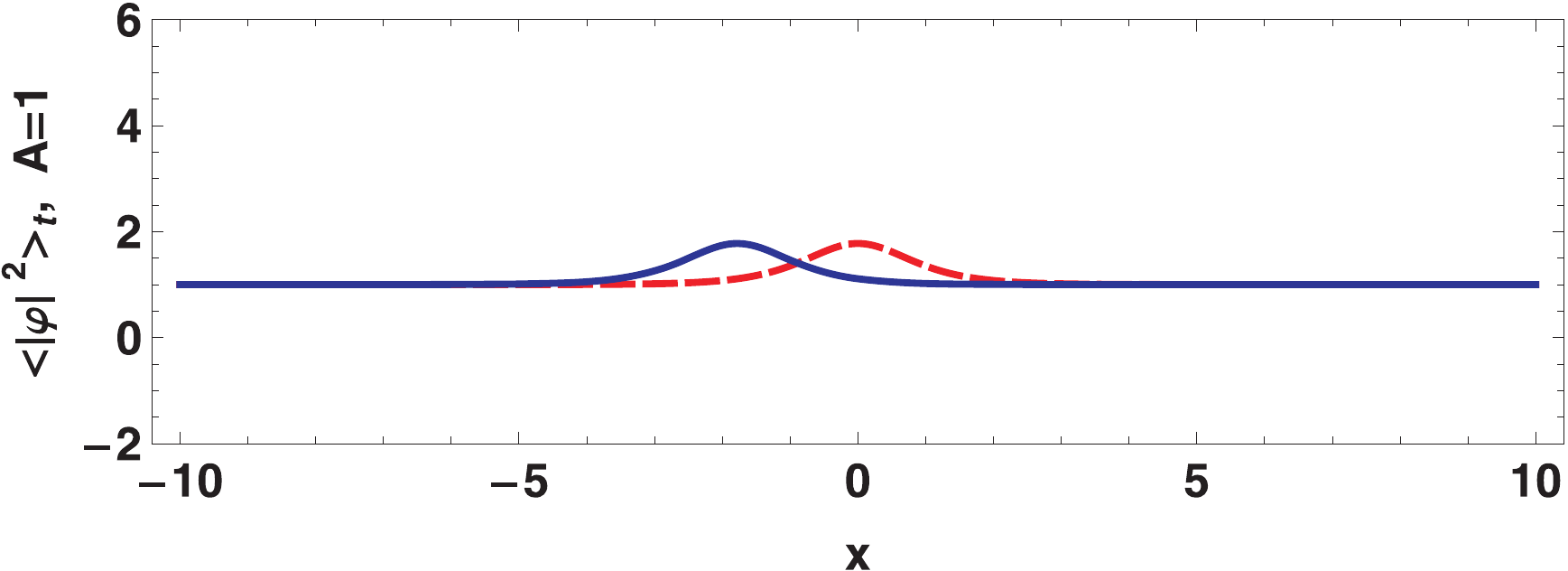}
\includegraphics[width=3in]{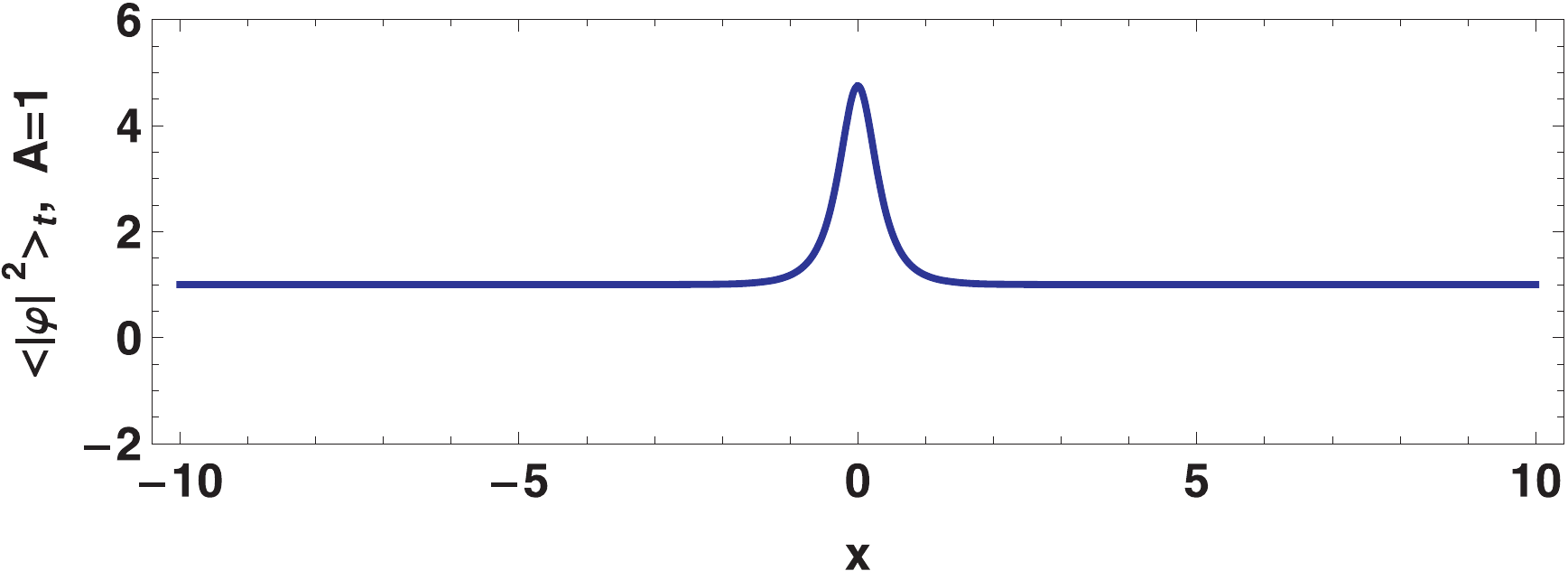}
\caption{\label{TAVG}
Time-averaged absolute squared value of one-solitonc solutions presented in figure \ref{1S_General} (left picture) and by formula \ref{averaged_Kuznetsov} (right picture). Dashed and solid lines in the left picture correspond to the moments of time $t=0$, $t=0.755$ from figure \ref{1S_General} respectively.
}
\end{figure}

It is important to average by integrating over an oscillation period $T=2\pi/\omega$ the squared absolute value of onesolitonic solution (\ref{Onesolitonic solution.final form}). This demonstrates the behavior of the average value of the number of particle integral. Remarkably, the answer can be obtained analytically in very simple form. We move to the coordinate system which travels with group velocity $V_{gr}$
\begin{eqnarray}
x \ra x+\frac{\gamma}{\ae}t.
\end{eqnarray}
Then the dependence on time in hyperbolic functions disappears and they can be regarded as constant when integrating over time. As a result
\begin{eqnarray}
<|\varphi|^2>_T=
\frac{A^2}{2\pi}
\int_0^{2\pi} d\tau
(\cosh z\cosh 2u+\cos\alpha\cos\tau)^{-2}
\nonumber\\
\bigl[ (\cosh z\cos 2\alpha \cosh 2u+\cosh 2z\cos\alpha\cos\tau)^2
\nonumber\\
+(\cosh z \sin 2\alpha \sinh 2u+\sinh 2z\cos\alpha\sin\tau)^2 \bigr].
\end{eqnarray}
The integral is evaluated using residues after the standard change of variables $w=\exp(\rmi\tau)$. The path of integration now $|w|=1$. The integrand has one pole of the first order at point $w_0=0$ and two poles of the second order in points
\begin{equation}
w_{1,2}=-\cosh z\cosh 2u\cos^{-1}\alpha \pm \sqrt{\cosh^2 z\cosh^2 2u\cos^{-2}\alpha-1}.
\end{equation}
However, only $w_0$ and $w_1$ lie inside the circle $|w| = 1$. After integration we need to change back to initial coordinate system. The final answer is
\begin{eqnarray}
\fl <|\varphi|^2>_{T}
=A^2+\frac{4A^2\cosh 2u}
{
(
\cosh^2 2u-\cos^2\alpha\cosh^{-2} z
)^{3/2}
}
\frac
{
\sinh^2 z\cos^2 \alpha(\sinh^2 z+\sin^2 \alpha)
}
{
\cosh^2 z
}.
\label{onesolitonic solution TAVG}
\end{eqnarray}
The case from figure \ref{1S_General} averaged by time is presented in figure \ref{TAVG}.

\begin{figure}[h]
\centering
\includegraphics[width=3in]{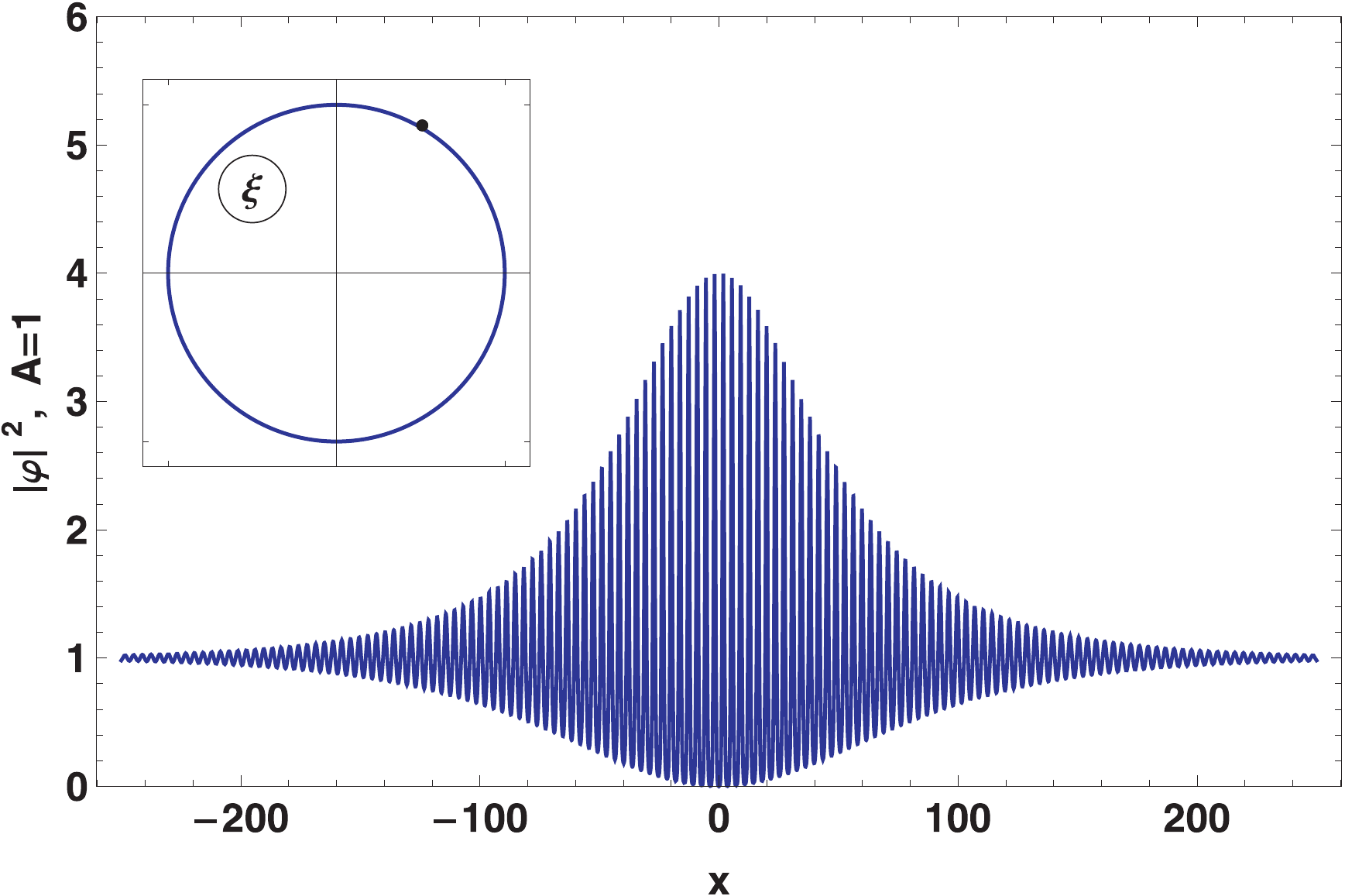}
\includegraphics[width=3in]{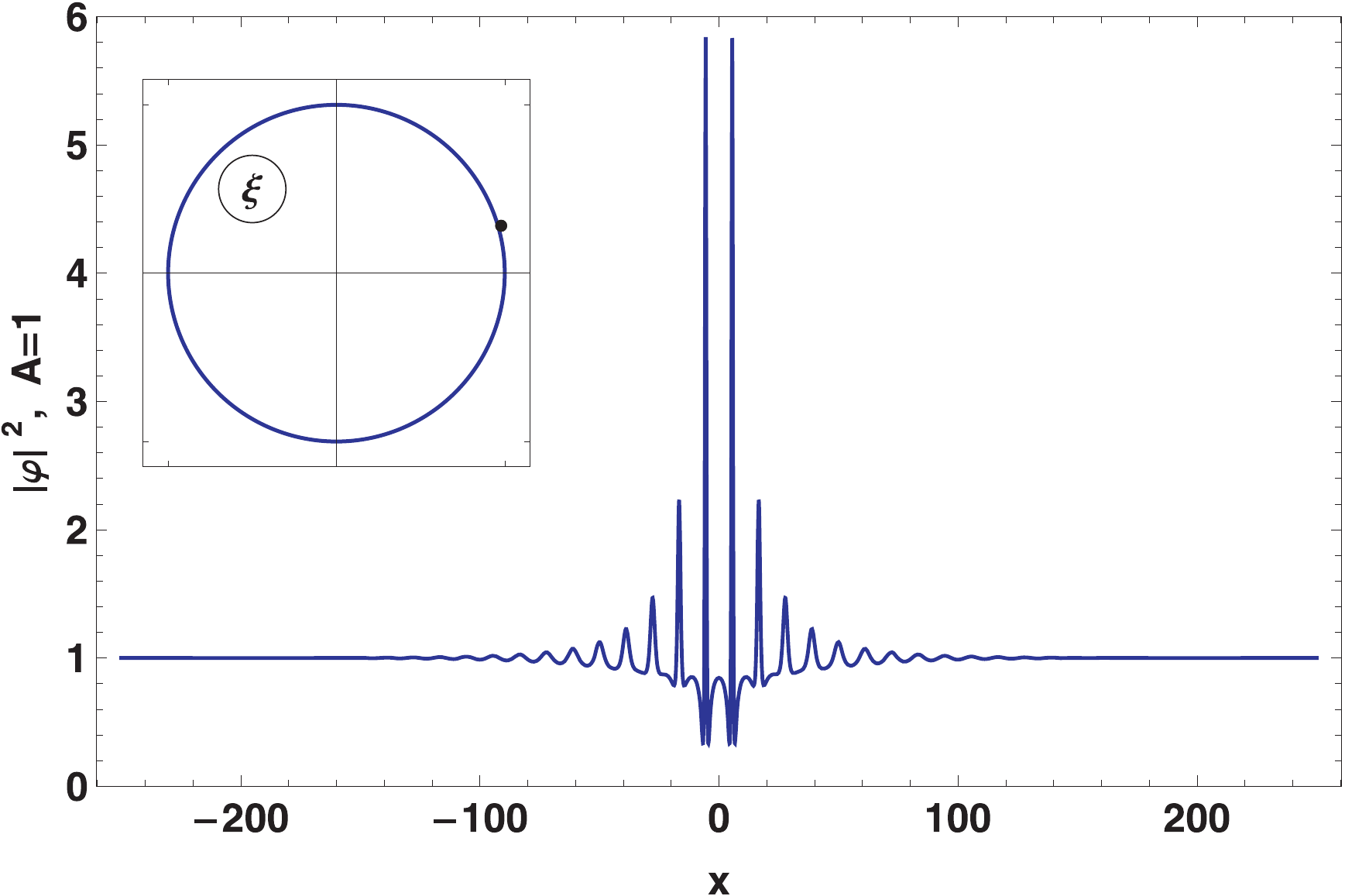}
\caption{\label{Near_Akhmediev}
Absolute squared value of "quasi-Akhmediev" solutions $\varphi$ with different $\alpha$. Left picture: $R=1.02,\;\alpha=\pi/3,\mu=0,\;\theta=0$, right picture: $R=1.02,\;\alpha=\pi/11,\mu=0,\;\theta=0$.
}
\end{figure}

It is interesting to perform an averaging of the Kuznetsov solution at the point $x=0$. In this case $\alpha=0$, $u=0$ and
\begin{equation}<|\varphi|^2>_{T}=A^2+A^2\frac{\sinh^4 z}{\cosh^2 z}.
\label{averaged_Kuznetsov}
\end{equation}
In particulary in the limiting case $R \ra 1,\;z \ra 0$ we obtain the expected result
\begin{equation}
<|\varphi|^2>_{T}=A^2.
\end{equation}
We now pay attention to the special case of general one-solitonic solution when $R \ra 1, z \ra 0$. This is the "quasi-Akhmediev" breather of very large size
\begin{equation}
L \approx \frac{1}{Az\cos\alpha}.
\end{equation}
moving with very high group velocity
\begin{equation}
V_{group} \approx -\frac{A\sin\alpha}{z},
\end{equation}
and very low phase velocity
\begin{equation}
V_{ph} \approx \frac{Az\cos 2\alpha}{\sin\alpha}.
\end{equation}
The soliton has inner quasiperiodic structure with characteristic scale $l \approx 2\pi (A\sin\alpha)^{-1}$. Observing this solution from a fixed point (for example $x=0$) the total passing time of this soliton is
\begin{equation}
T \approx \frac{L}{V_{group}} =\frac{1}{A^2\cos\alpha\sin\alpha}.
\end{equation}
$T \ra \infty$ if $\alpha \ra 0$. It is interesting that this time does not depend on $z$. It is important to stress that the "quasi-Akhmediev" breather remain after its passing slowly decaying "tails".
The "quasi-Akhmediev" breather is plotted in figure \ref{Near_Akhmediev}.
\section{Two-solitonic solution}
\begin{figure}[h]
\centering
\includegraphics[width=3in]{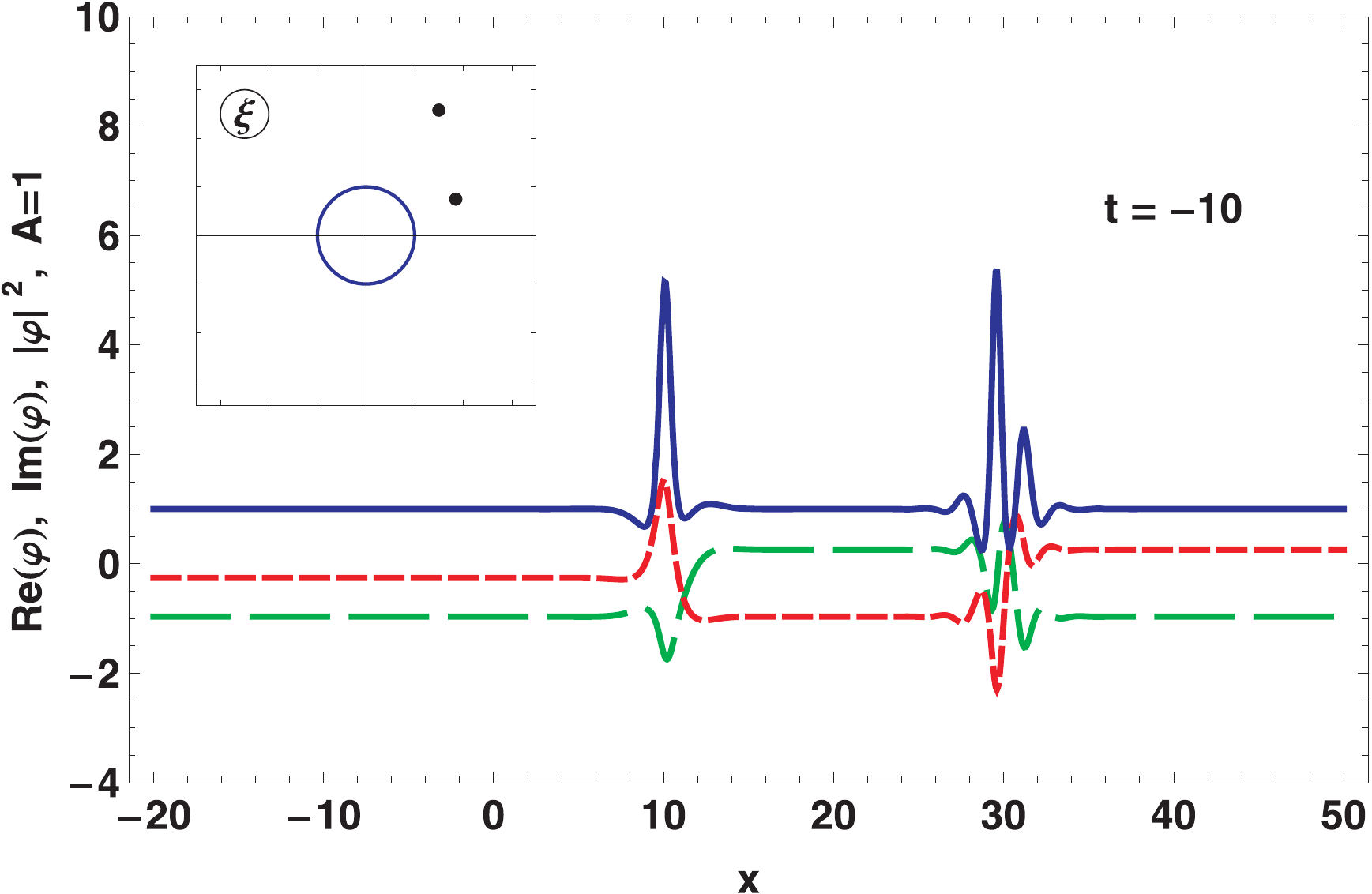}
\includegraphics[width=3in]{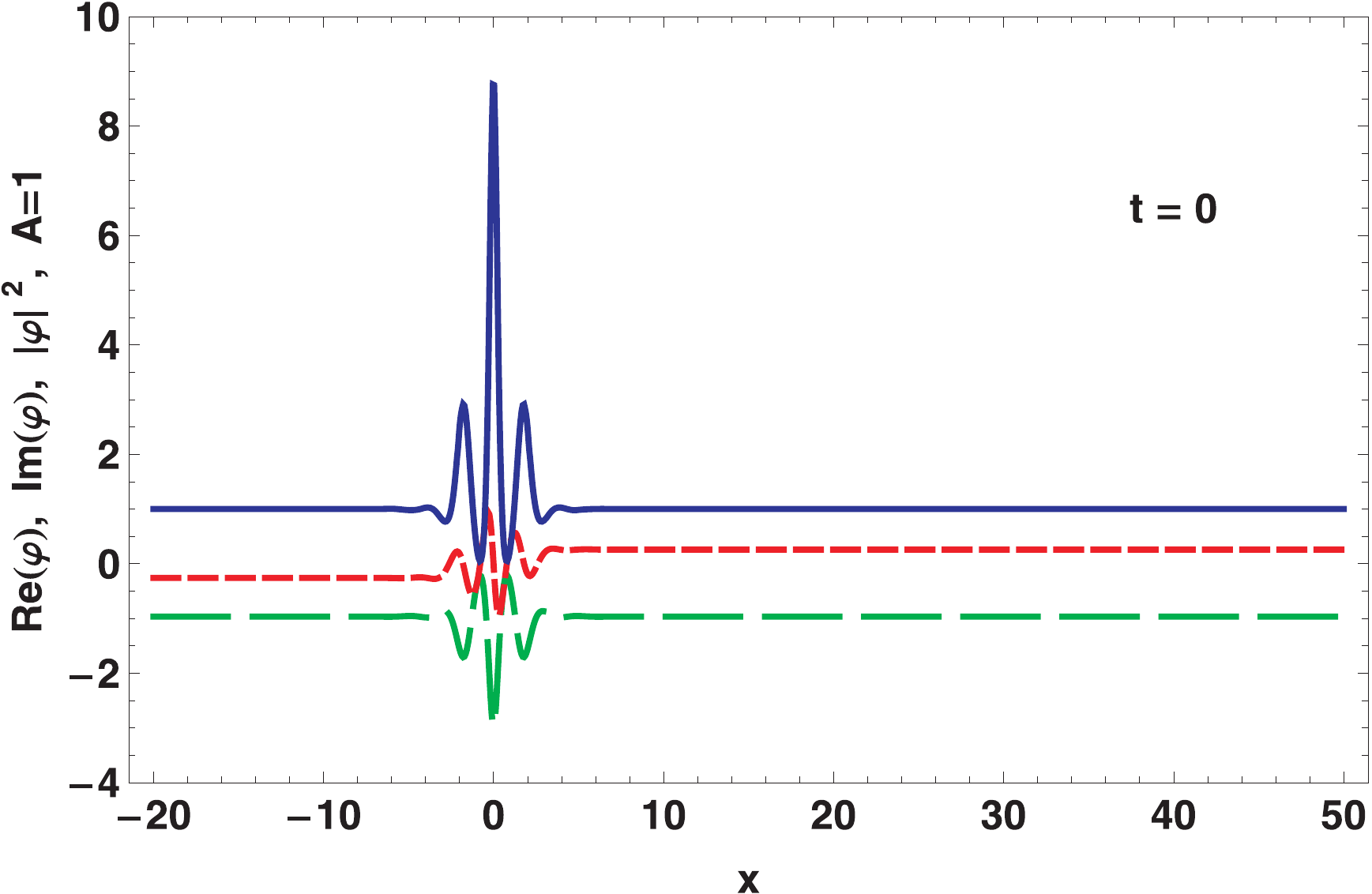}
\caption{\label{2S_G}
General two-solitonic solution $\varphi$ at moments of time $t=-10$ (left picture) and $t=0$ (right picture) with parameters: $R_1=2,\;\alpha_1=\pi/8,\;R_2=3,\;\alpha_2=\pi/3,\;\mu_1=\mu_2=0,\;\theta_1=\theta_2=0$. Green dashed lines - real part of $\varphi$, red short dashed lines - imaginary part of $\varphi$ and blue solid lines - absolute squared value of $\varphi$.
}
\end{figure}

\begin{figure}[h]
\centering
\includegraphics[width=3in]{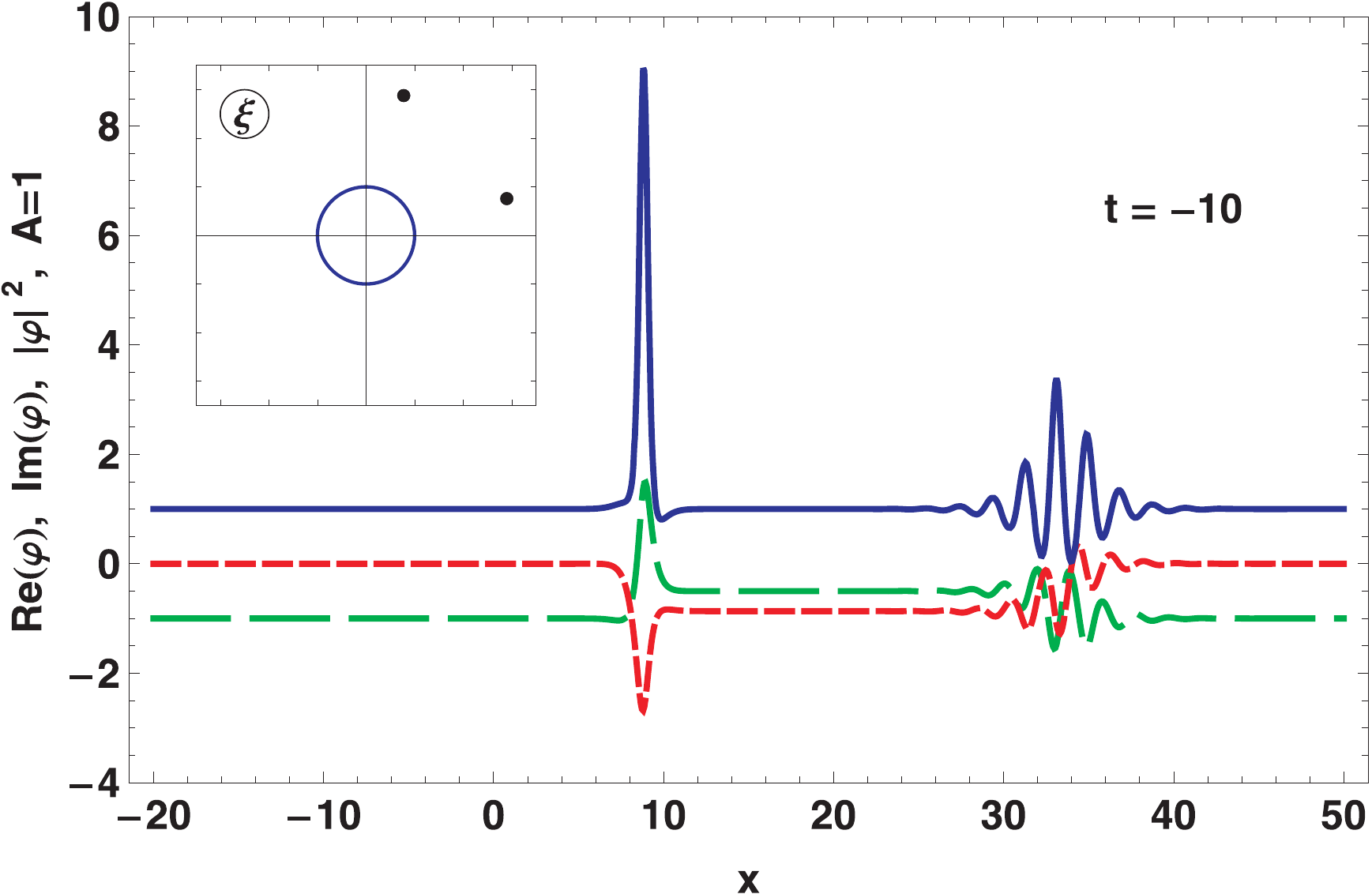}
\includegraphics[width=3in]{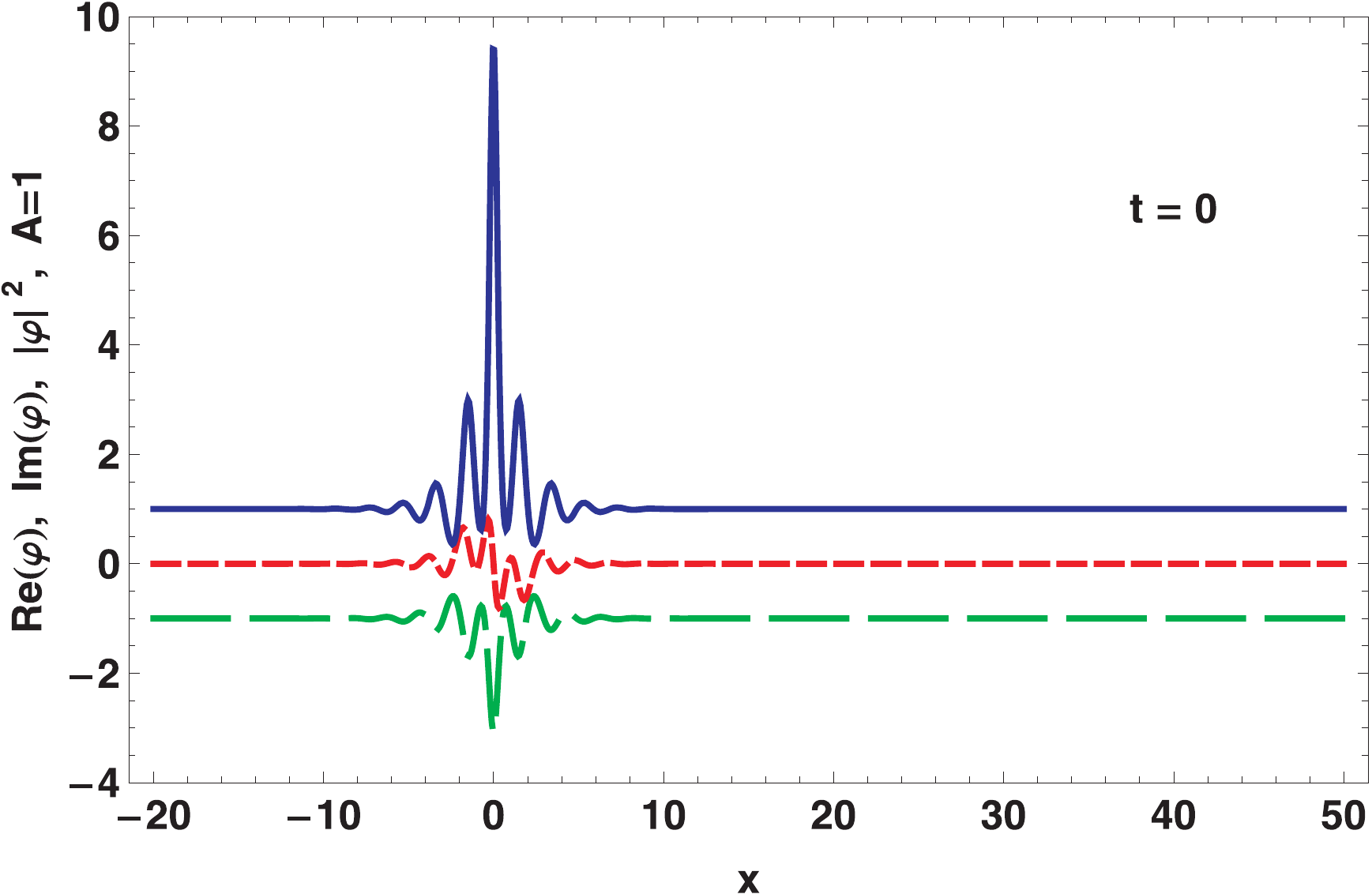}
\caption{\label{2S_R2Type}
Regular two-solitonic solution $\varphi$ of the second type at moments of time $t=-10$ (left picture) and $t=0$ (right picture) with parameters: $R_1=3,\;\alpha_1=\pi/12,\;R_2=3,\;\alpha_2=5\pi/12,\;\mu_1=\mu_2=0,\;\theta_1=\theta_2=0$. Green dashed lines - real part of $\varphi$, red short dashed lines - imaginary part of $\varphi$ and blue solid lines - absolute squared value of $\varphi$.
}
\end{figure}

A general two-solitonic solution on a condensate background can be obtained by applying the dressing procedure described in \S 2. As before we write solution in uniformazing variables. However, for intermediate calculations in the next paragraphs is more convenient to use two-solitonic solution in $\lambda$ variable:
\begin{eqnarray}
\varphi=
A-2\frac{N_{\lambda}}{\Delta_{\lambda}},
\nonumber\\
N_{\lambda}=
\frac{|\bi{q}_1|^2 q^*_{21}q_{22}}{\lambda_1+\lambda_1^*}-\frac{(\bi{q}^*_1 \bi{q}_2) q^*_{21}q_{12}}{\lambda_1^*+\lambda_2}
-\frac{(\bi{q}_1 \bi{q}^*_2) q^*_{11}q_{22}}{\lambda_2^*+\lambda_1}+\frac{|\bi{q}_2|^2 q^*_{11}q_{12}}{\lambda_2+\lambda_2^*},
\nonumber\\
\Delta_{\lambda}=
\frac{|\bi{q}_1|^2|\bi{q}_2|^2}
{(\lambda_1+\lambda_1^*)(\lambda_2+\lambda_2^*)}
-
\frac{
(\bi{q}_1\bi{q}^*_2)
(\bi{q}^*_1\bi{q}_2)
}
{(\lambda^*_1+\lambda_2)(\lambda^*_2+\lambda_1)}.
\label{2S_general_1form}
\end{eqnarray}
In uniformizing variables two-solitonic solution is given in the following form
\begin{equation}
\varphi=
A-2A\frac{N_{\xi}}{\Delta_{\xi}}.
\label{2S_general}
\end{equation}
Here
\begin{eqnarray}
\fl N_{\xi}=H_1 |\bi{q}_1|^2 q^*_{21}q_{22}+H_2 |\bi{q}_2|^2 q^*_{11}q_{12} +
iH_3 [(\bi{q}_1 \cdot \bi{q}^*_2) q^*_{11}q_{22}-(\bi{q}^*_1 \cdot \bi{q}_2) q^*_{21}q_{12}],
\nonumber\\
\fl \Delta_{\xi}=H_4 |\bi{q}_1|^2|\bi{q}_2|^2-H_5 |(\bi{q}_1 \cdot \bi{q}^*_2)|^2.
\label{2S_general_2form}
\end{eqnarray}
The vectors $\bi{q}_1$ and $\bi{q}_2$ are determined by the general expression (\ref{vectors q_n}) and
\begin{eqnarray}
\fl H_1=2\cos\alpha_2\cosh z_2
[(\sin\alpha_1\sinh z_1-\sin\alpha_2\sinh z_2)^2
-
\nonumber\\
\cos^2\alpha_1\cosh^2 z_1+\cos^2 \alpha_2\cosh^2 z_2],
\nonumber\\
\fl H_2=2\cos\alpha_1\cosh z_1
[(\sin\alpha_1\sinh z_1-\sin\alpha_2\sinh z_2)^2
+
\nonumber\\
\cos^2\alpha_1\cosh^2 z_1-\cos^2\alpha_2\cosh^2 z_2],
\nonumber\\
\fl H_3=4\cos\alpha_1\cos\alpha_2\cosh z_1\cosh z_2
\bigl(
\sin\alpha_1\sinh z_1-\sin\alpha_2\sinh z_2
\bigr),
\nonumber\\
\fl H_4=
\bigl(
\sin\alpha_1\sinh z_1-\sin\alpha_2\sinh z_2
\bigr)^2
+
\bigl(
\cos\alpha_1\cosh z_1+\cos\alpha_2\cosh z_2
\bigl)^2,
\nonumber\\
\fl H_5=4\cos\alpha_1\cos\alpha_2\cosh z_1\cosh z_2.
\label{H_defenition}
\end{eqnarray}
One can check that following identities are valid
\begin{eqnarray}
\fl |\bi{q}_1|^2 q^*_{21}q_{22}-(\bi{q}^*_1\cdot\bi{q}_2) q^*_{21}q_{12}-(\bi{q}_1\cdot\bi{q}^*_2) q^*_{11}q_{22}+|\bi{q}_2|^2 q^*_{11}q_{12} \equiv 0,
\nonumber\\
\fl (\bi{q}_1 \cdot \bi{q}^*_2) q^*_{11}q_{22}-(\bi{q}^*_1 \cdot \bi{q}_2) q^*_{21}q_{12}=(|q_{11}|^2-|q_{12}|^2)q^*_{21}q_{22}+(|q_{21}|^2-|q_{22}|^2)q^*_{11}q_{12},
\nonumber\\
\fl |\bi{q}_1|^2 |\bi{q}_2|^2-|(\bi{q}_1 \cdot \bi{q}^*_2)|^2=|q_{11}q_{22}-q_{12}q_{21}|^2.
\label{q_identities}
\end{eqnarray}
Let us note that when poles are not on the real axes ($\alpha_1 \ne 0,\; \alpha_2 \ne 0$) and group velocities are not equal ($V_{Gr_1} \ne V_{Gr_2}$) we can make $\mu_1=0$ and $\mu_2=0$ by shifts in space and time. Now the solitons collide at $(x=0,t=0)$. In what follows we put $\mu_1=0$ and $\mu_2=0$ in these cases. The signs of $\alpha_{1}$ and $\alpha_{2}$ determine the signs of the group velocities. In general two-solitonic solution changes the phase of the condensate. The example of two solitons which move in one direction and collide is presented in figure \ref{2S_G}.
We can write two-solitonic solution in explicit form by using expressions (\ref{f(q)_a}) and
\begin{eqnarray}
(\bi{q}_1 \cdot \bi{q}^*_2)=4\exp\biggl(\frac{z_2-z_1}{2}+\rmi\frac{\alpha_2-\alpha_1}{2} \biggr)\times
\nonumber\\
\biggl[
(\cos(v_1-v_2)\cosh(u_1+u_2) +\rmi \sin(v_1-v_2)\sinh(u_1+u_2) )\times
\nonumber\\
\biggl(\cos\frac{\alpha_1-\alpha_2}{2}\cosh\frac{z_1+z_2}{2}+\rmi \sin\frac{\alpha_1-\alpha_2}{2}\sinh\frac{z_1+z_2}{2} \biggr)+
\nonumber\\
(\cos(v_1+v_2)\cosh(u_1-u_2) +\rmi \sin(v_1+v_2)\sinh(u_1-u_2) )\times
\nonumber\\
\biggl(\cos\frac{\alpha_1+\alpha_2}{2}\cosh\frac{z_1-z_2}{2}+\rmi \sin\frac{\alpha_1+\alpha_2}{2}\sinh\frac{z_1-z_2}{2}\biggr)
\biggr].
\label{f(q)_b}
\end{eqnarray}
Let us write another one useful expression
\begin{eqnarray}
q_{11}q_{22}-q_{12}q_{21} = -4\exp\biggl(-\frac{z_1+z_2}{2}-\rmi\frac{\alpha_1+\alpha_2}{2}\biggr)\times
\nonumber\\
\biggl[
(\cos(v_1+v_2)\cosh(u_1+u_2) +\rmi \sin(v_1+v_2)\sinh(u_1+u_2) )\times
\nonumber\\
\biggl(\cos\frac{\alpha_1-\alpha_2}{2}\cosh\frac{z_1-z_2}{2}+\rmi \sin\frac{\alpha_1-\alpha_2}{2}\sinh\frac{z_1-z_2}{2} \biggr)+
\nonumber\\
(\cos(v_1-v_2)\cosh(u_1-u_2) + \rmi \sin(v_1-v_2)\sinh(u_1-u_2) )\times
\nonumber\\
\biggl(\cos\frac{\alpha_1+\alpha_2}{2}\cosh\frac{z_1+z_2}{2}+\rmi \sin\frac{\alpha_1+\alpha_2}{2}\sinh\frac{z_1+z_2}{2}\biggr)
\biggr].
\label{f(q)_c}
\end{eqnarray}
In \S 4 we obtained that the solution is regular when the sum of angular parameters is equal to zero or $\pm \pi/2$. For the two-solitonic case this leads to the existence of two types of regular solutions. When $\alpha_1=\alpha,\;\alpha_2=-\alpha$ the solution is a regular two-solitonic solution of the first type. It consists of two solitons which move in opposite directions and collide. The group velocities are
\begin{eqnarray}
V_{Gr1}=-\frac{\cosh 2z_1}{\sinh z_1}\sin\alpha,
&\qquad&
V_{Gr2}=\frac{\cosh 2z_2}{\sinh z_2}\sin\alpha.
\end{eqnarray}
We set $\mu_1=0$ and $\mu_2=0$, and finally a regular two-solitonic solution of the first type depends on five real parameters $R_1,R_2,\alpha,\theta_1,\theta_2$.
The regular two-solitonic solution of the first type is plotted in figure \ref{2S_R1Type}.
\begin{figure}[h]
\centering
\includegraphics[width=3in]{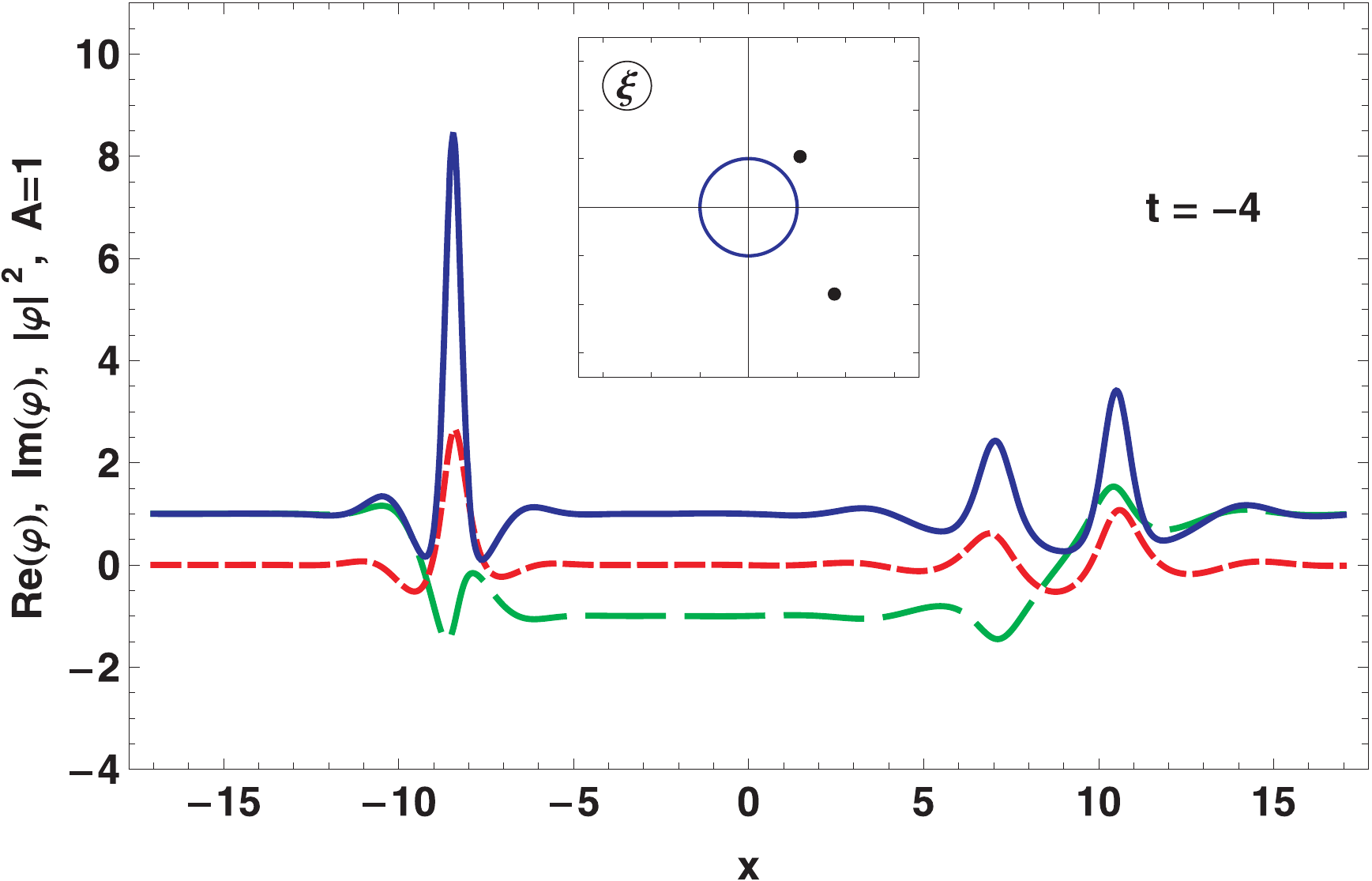}
\includegraphics[width=3in]{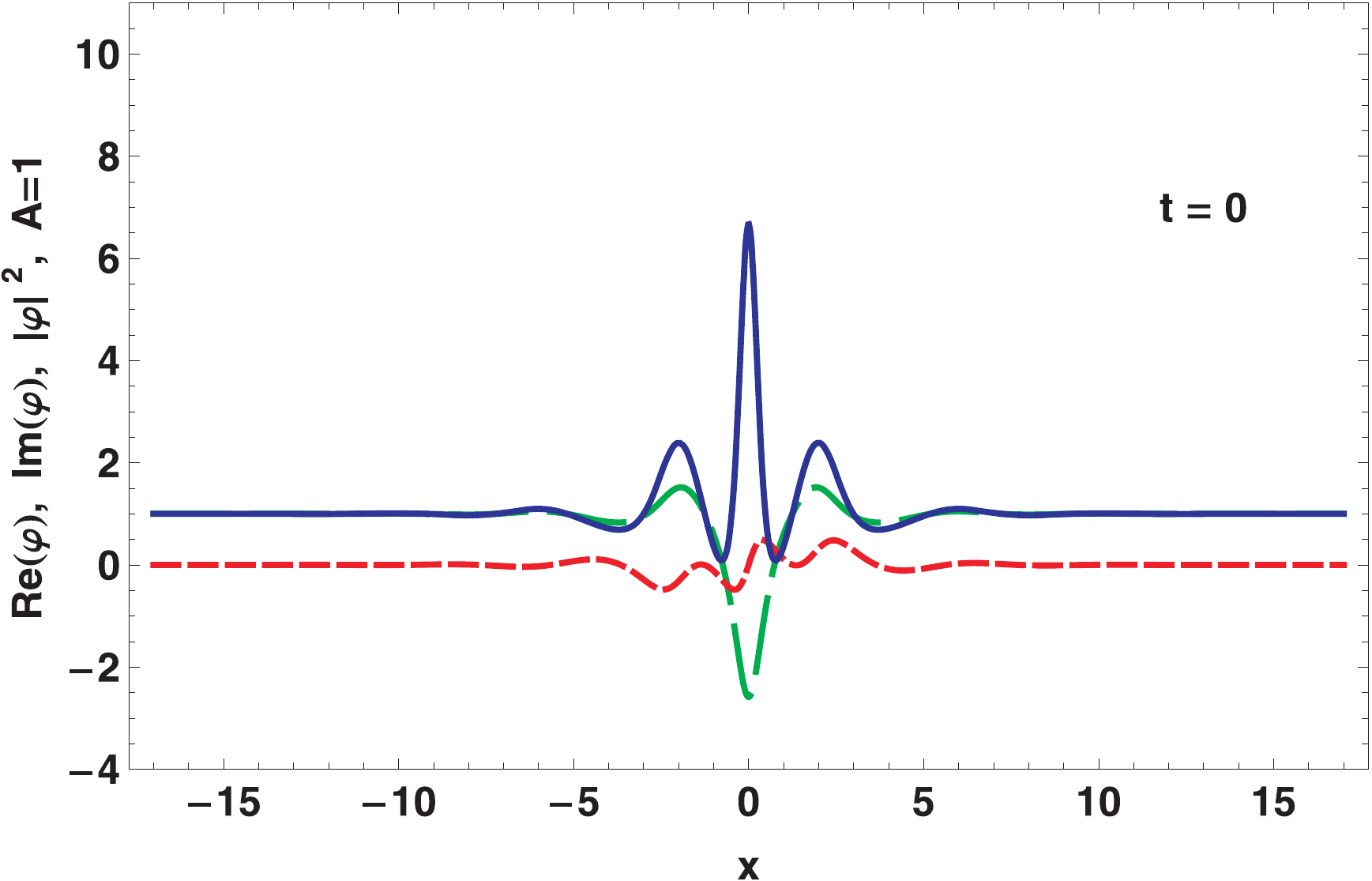}
\caption{\label{2S_R1Type}
Regular two-solitonic solution $\varphi$ of the first type at moments of time $t=-4$ (left picture) and $t=0$ (right picture) with parameters: $R_1=1.5,\;\alpha_1=\pi/4,\;R_2=2.5,\;\alpha_2=-\pi/4,\;\mu_1=\mu_2=0,\;\theta_1=\theta_2=0$. Green dashed lines - real part of $\varphi$, red short dashed lines - imaginary part of $\varphi$ and blue solid lines - absolute squared value of $\varphi$.
}
\end{figure}
In the special symmetric case $z_1=z_2=z$ the eigenvalues are complex conjugates. Then $\phi_1$ and $\phi_2$ are given by its definition (\ref{phi_n}) with
\begin{eqnarray}
\ae_2=\ae_1=\ae=A\sinh z\cos\alpha,
\nonumber\\
k_1=-k_2=k=A\cosh z\sin\alpha,
\nonumber\\
\gamma_1=-\gamma_2=\gamma=-\frac{A^2}{2}\cosh 2z\sin 2\alpha,
\nonumber\\
\omega_1=\omega_2=\omega=\frac{A^2}{2}\sinh 2z\cos 2\alpha.
\end{eqnarray}
$N_{\xi}$ and $\Delta_{\xi}$ for two-solitonic solution (\ref{2S_general}) in symmetric case are given by
\begin{eqnarray}
\fl N_{\xi}=\sinh 2z\sinh z\sin 2\alpha\sin\alpha(|\bi{q}_1|^2 q^*_{21}q_{22}+|\bi{q}_2|^2 q^*_{11}q_{12}),
\nonumber\\
\fl +\rmi\cosh z\cos\alpha((\bi{q}_1 \cdot \bi{q}^*_2) q^*_{11}q_{22}-(\bi{q}^*_1 \cdot \bi{q}_2) q^*_{21}q_{12}),
\nonumber\\
\fl \Delta_{\xi}=2\cosh^2 z\cos^2\alpha|q_{11}q_{22}-q_{12}q_{21}|^2+2\sinh^2 z\sin^2\alpha|\bi{q}_1|^2|\bi{q}_2|^2,
\label{2S_symmetric}
\end{eqnarray}
and
\begin{eqnarray}
\fl q_{11}=\exp(-\phi_1)+\exp(-\rmi\alpha-z+\phi_1),
&\qquad&
q_{21}=\exp(-\phi_2)+\exp(\rmi\alpha-z+\phi_2),
\nonumber\\
\fl q_{12}=\exp(-\rmi\alpha-z-\phi_1)+\exp(\phi_1),
&\qquad&
q_{22}=\exp(\rmi\alpha-z-\phi_2)+\exp(\phi_2).
\end{eqnarray}
\begin{figure}[h]
\centering
\includegraphics[width=3in]{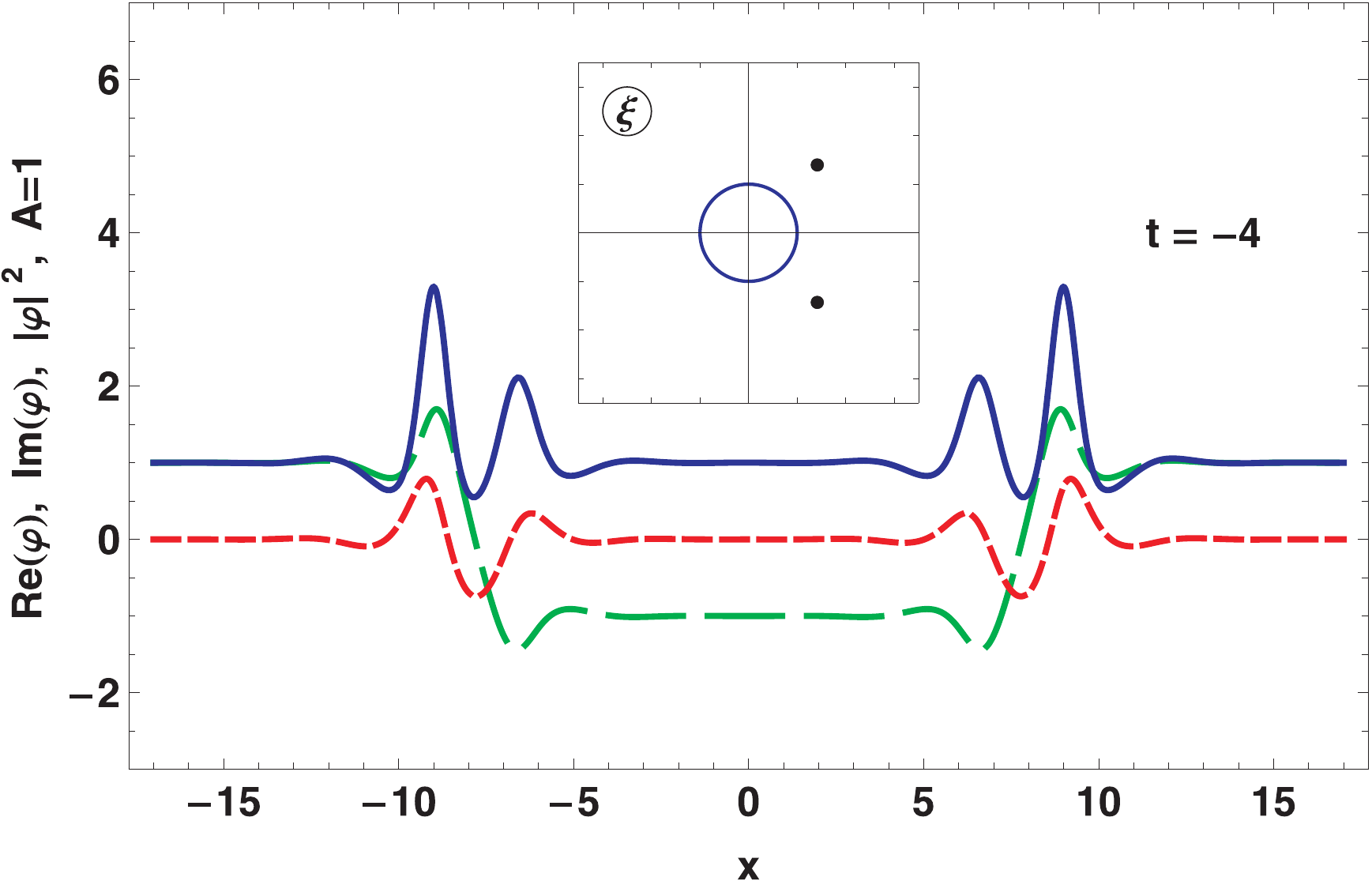}
\includegraphics[width=3in]{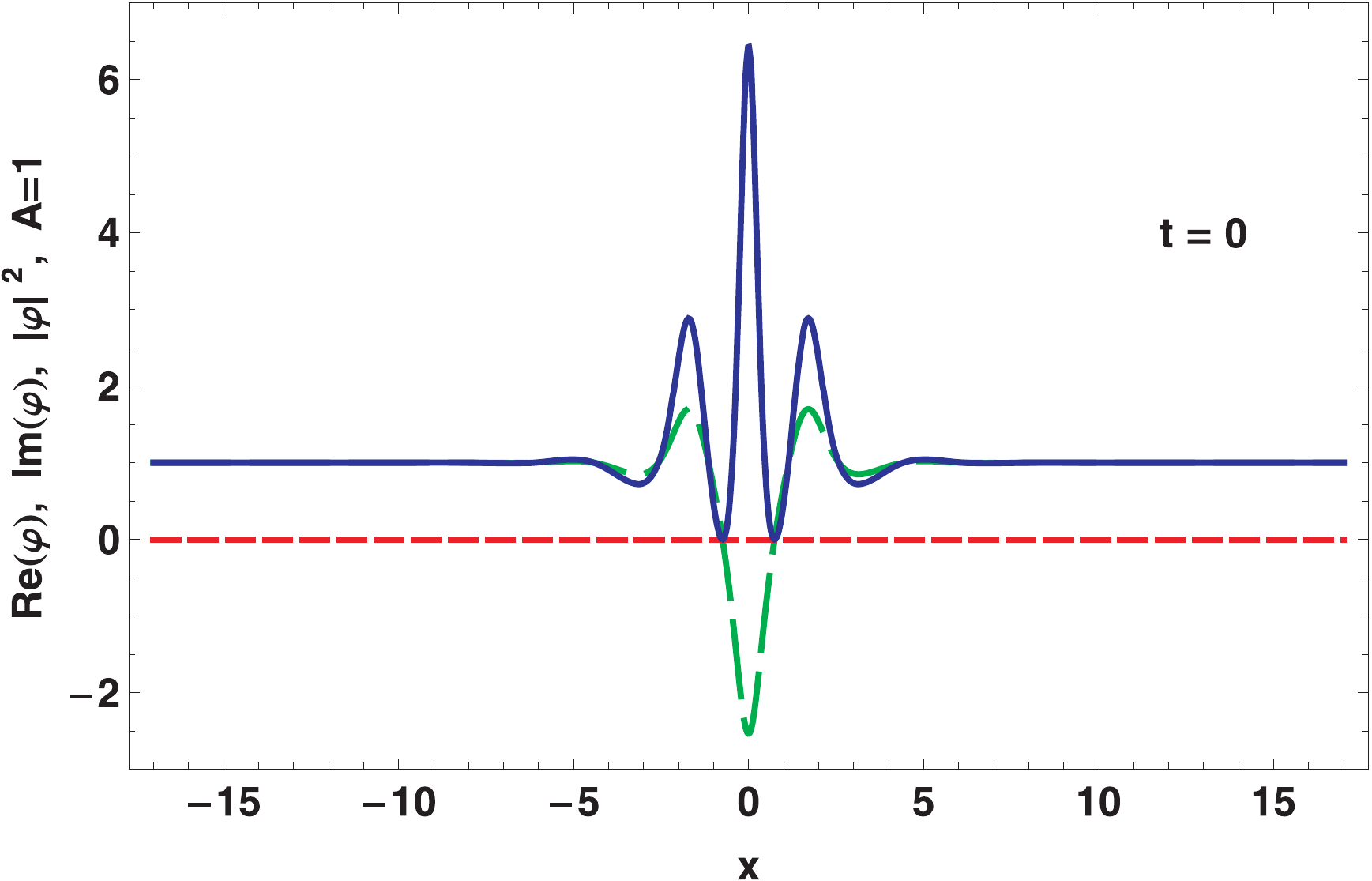}
\caption{\label{2S_R1Type}
Regular two-solitonic solution $\varphi$ in symmetric case at moments of time $t=-4$ (left picture) and $t=0$ (right picture) with parameters: $R_1=2,\;\alpha_1=\pi/4,\;R_2=2,\;\alpha_2=-\pi/4,\;\mu_1=\mu_2=0,\;\theta_1=\theta_2=0$. Green dashed lines - real part of $\varphi$, red short dashed lines - imaginary part of $\varphi$ and blue solid lines - absolute squared value of $\varphi$.
}
\end{figure}
When $\alpha_1=\alpha,\;\alpha_2=\pi/2-\alpha,\;\alpha>0$ or $\alpha_1=\alpha,\;\alpha_2=-\pi/2-\alpha,\;\alpha<0$ the solution is a regular two-solitonic solution of the second type. It consists of two solitons moving in one direction and in the general case colliding (unless the group velocities coincide). Let us consider only the case $\alpha>0$ (the case $\alpha<0$ is different only in the direction of movement). Note that now we cannot put $\mu_1=0$ and $\mu_2=0$ simultaneously because the group velocities can coincide. A regular two-solitonic solution of the second type is plotted in figure \ref{2S_R2Type}.
For regular solutions of the second type the condition $z_1=z_2=z$ does not lead to serious simplifications as for the first type. Therefore we will omit expressions for this case.

\begin{figure}[h]
\centering
\includegraphics[width=2in]{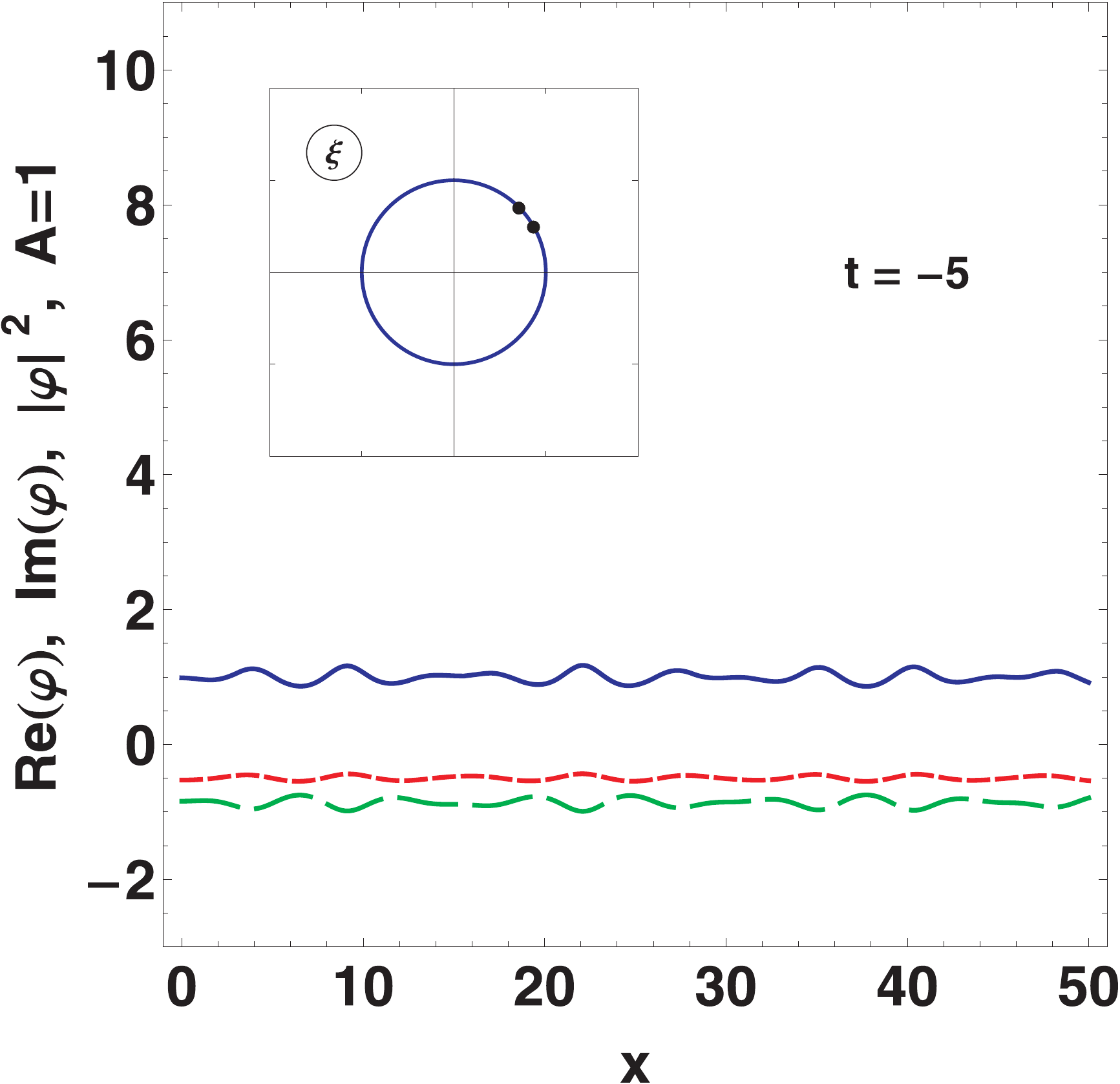}
\includegraphics[width=2in]{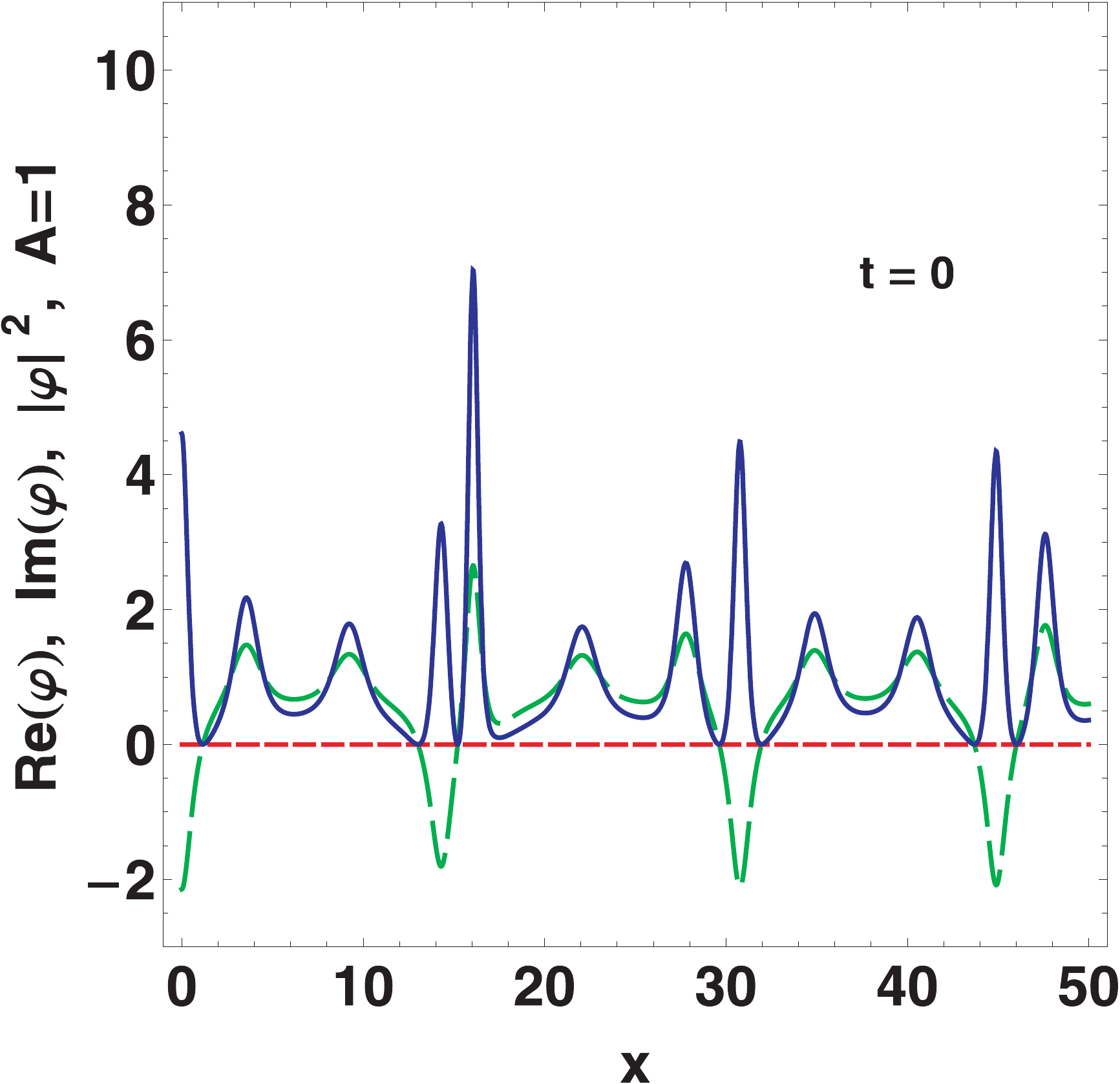}
\includegraphics[width=2in]{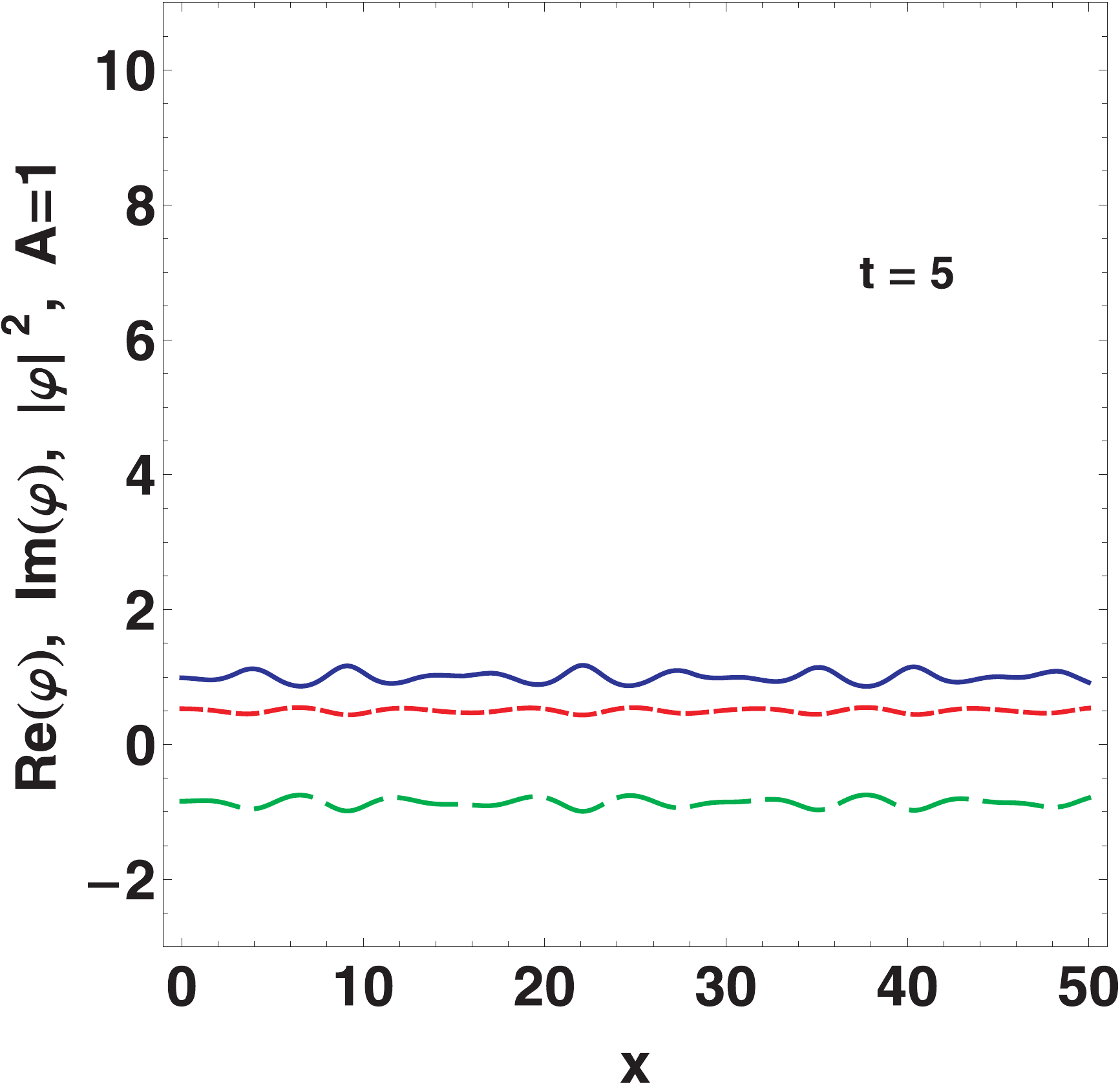}
\includegraphics[width=2in]{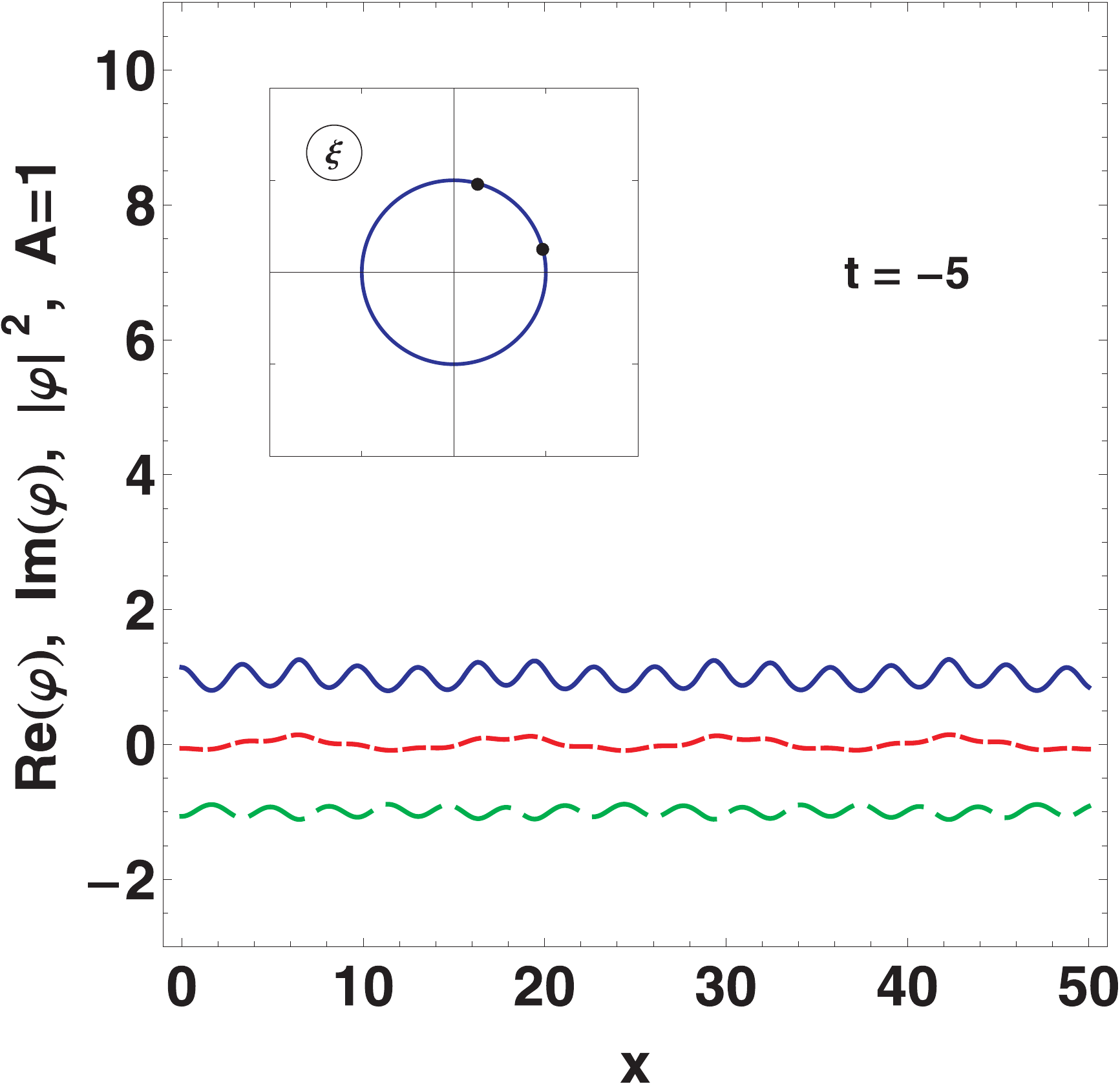}
\includegraphics[width=2in]{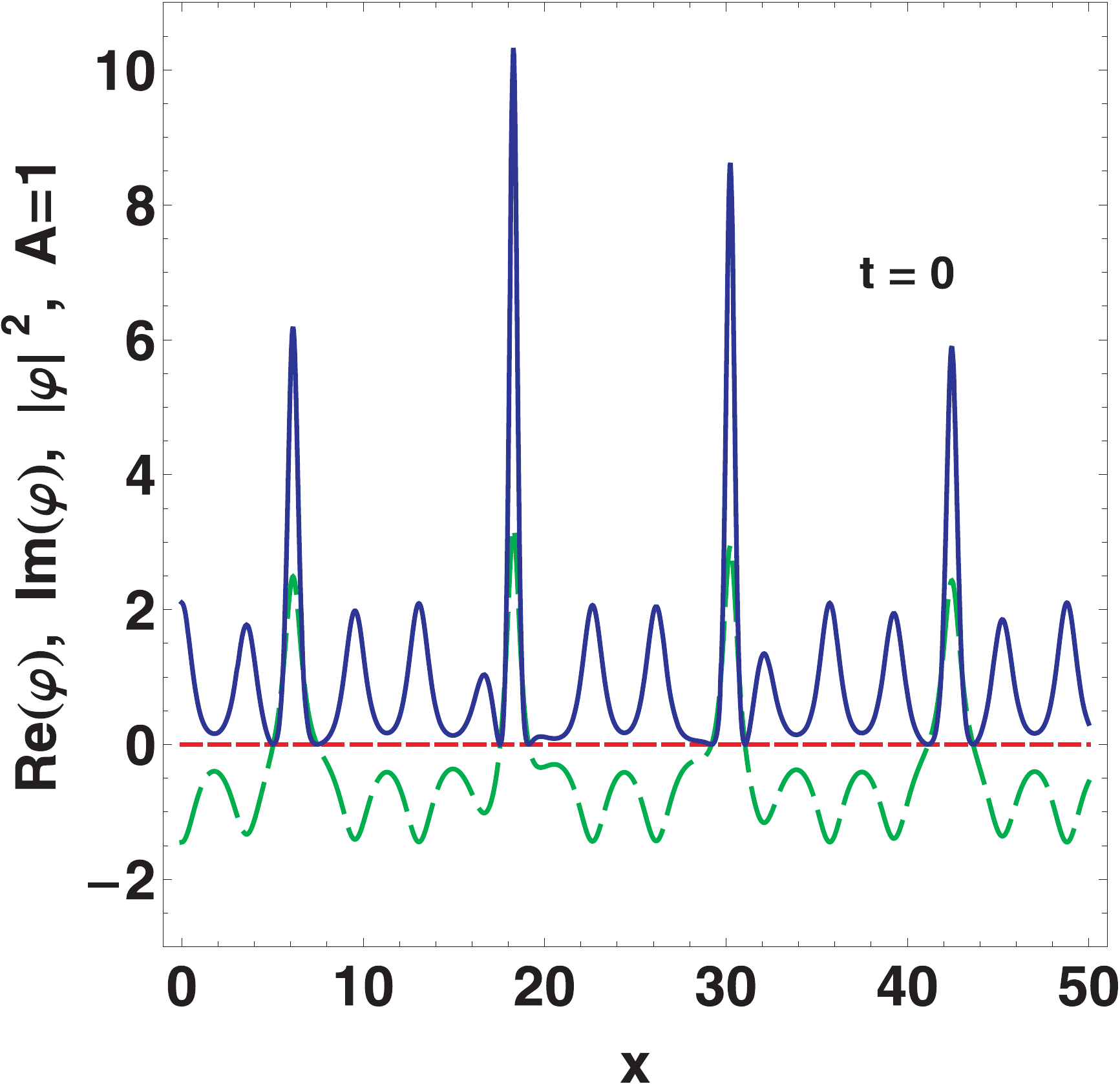}
\includegraphics[width=2in]{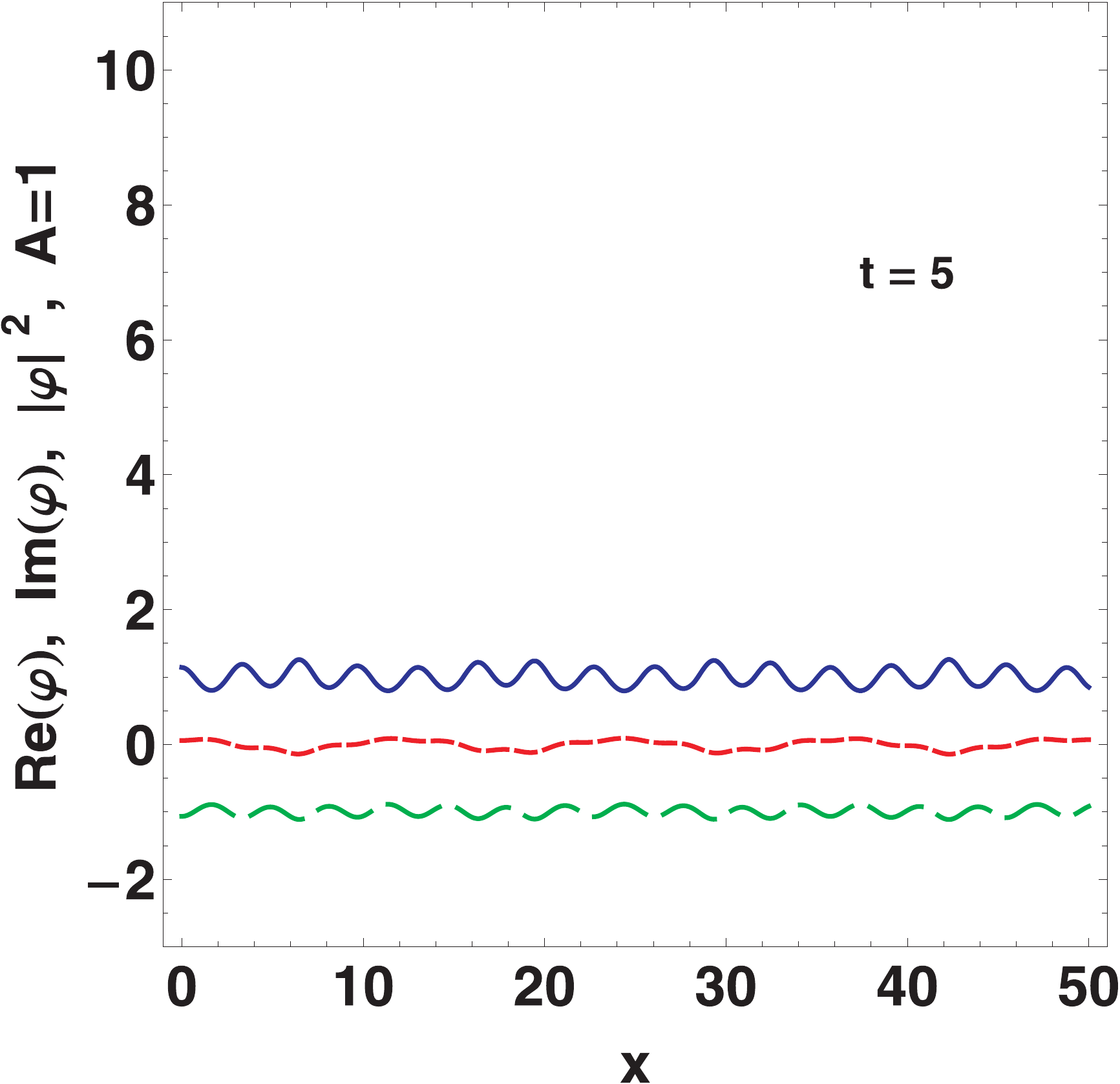}
\caption{\label{2S_Akhmediev}
Double Akhmediev breathers $\varphi$ at moments of time $t=-5$ (left pictures), $t=0$ (middle pictures) and  with $t=5$ (right pictures). Top pictures  - double Akhmediev breather with parameters $\alpha_1=\pi/6,\;\alpha_2=\pi/4,\;\mu=0,\;\theta=0$. Bottom pictures - homoclinic double Akhmediev breather with parameters $\alpha_1=\pi/12,\;\alpha_2=5\pi/12,\;\mu=0,\;\theta=0$.
}
\end{figure}

When $z_1=z_2=0$ the solution is a double Akhmediev breather. This is a solution periodic in space and localized in time. In general a double Akhmediev breather is almost homoclinic because the phases of condensate at $t \ra \pm \infty$ are not equal: $\varphi \ra \exp(\pm \rmi (|\alpha_1|+|\alpha_2|))$. Now expression (\ref{2S_general}) can be written as
\begin{eqnarray}
\fl \varphi=A+4A
\frac
{(\cos^2\alpha_1-\cos^2\alpha_2)
(\cos\alpha_2 |\bi{q}_1|^2 q^*_{21}q_{22}-\cos\alpha_1 |\bi{q}_2|^2 q^*_{11}q_{12})}
{(\cos\alpha_1+\cos\alpha_2)^2 |\bi{q}_1|^2|\bi{q}_2|^2-4\cos\alpha_1\cos\alpha_2 |(\bi{q}_1 \cdot \bi{q}^*_2)|^2}.
\end{eqnarray}
Double Akhmediev breather is shown in figure \ref{2S_Akhmediev}. For the double Akhmediev breather we can put $\mu_1=0$ and $\theta_1=0$. Let us denote $\mu_2=\mu$ and $\theta_2=\theta$. Now the first component of the Akhmediev breather is centered by $x$ and $t$ and $\mu,\theta$ correspond to phase shifts with the second component. Now
\begin{eqnarray}
\phi_1=\frac{A^2}{2}\sin(2\alpha_1) t + \rmi A\sin(\alpha_1) x,
\nonumber\\
\phi_2=\frac{A^2}{2}\sin(2\alpha_2) t +\mu/2+ \rmi ( A\sin(\alpha_2) x - \theta/2),
\end{eqnarray}
and
\begin{eqnarray}
\fl q_{11}=\exp(-\phi_1)+\exp(-\rmi \alpha_1+\phi_1),
&\qquad&
q_{21}=\exp(-\phi_2)+\exp(-\rmi \alpha_2+\phi_2),
\nonumber\\
\fl q_{12}=\exp(-\rmi \alpha_1-\phi_1)+\exp(\phi_1),
&\qquad&
q_{22}=\exp(-\rmi \alpha_2-\phi_2)+\exp(\phi_2).
\end{eqnarray}
If $\mu >> A^2\sin 2\alpha_2$ the solution is two Akhmediev breathers which appear at different moments of time. In other case the solution is complicated nonlinear superposition of two Akhmediev breathers.

When $|\alpha_2|=\pi/2-|\alpha_1|$ double Akhmediev breather is homoclinic. The solution is
\begin{eqnarray}
\fl \varphi=A+4A
\frac
{
(\cos^2\alpha-\sin^2\alpha)(\sin\alpha |\bi{q}_1|^2 q^*_{21}q_{22}-\cos\alpha |\bi{q}_2|^2 q^*_{11}q_{12})
}
{
(\cos\alpha+\sin\alpha)^2 |\bi{q}_1|^2|\bi{q}_2|^2-2\sin 2\alpha |(\bi{q}_1 \cdot \bi{q}^*_2)|^2
}.
\end{eqnarray}
Now
\begin{eqnarray}
\phi_1=\frac{A^2}{2}\sin(2\alpha)t + \rmi A\sin(\alpha) x,
\nonumber\\
\phi_2=\frac{A^2}{2}\sin(2\alpha)t +\mu/2+ \rmi (A\cos(\alpha)x - \theta/2),
\end{eqnarray}
and
\begin{eqnarray}
\fl q_{11}=\exp(-\phi_1)+\exp(-\rmi\alpha+\phi_1),
&\qquad&
q_{21}=\exp(-\phi_2)-\rmi \exp(\rmi\alpha+\phi_2),
\nonumber\\
\fl q_{12}=\exp(-\rmi\alpha-\phi_1)+\exp(\phi_1),
&\qquad&
q_{22}=-\rmi \exp(\rmi\alpha-\phi_2)+\exp(\phi_2).
\label{RTSS vectors q}
\end{eqnarray}
This solution has equal phases at $t \ra \pm \infty$. An example is shown in figure \ref{2S_Akhmediev}.

\begin{figure}[h]
\centering
\includegraphics[width=3in]{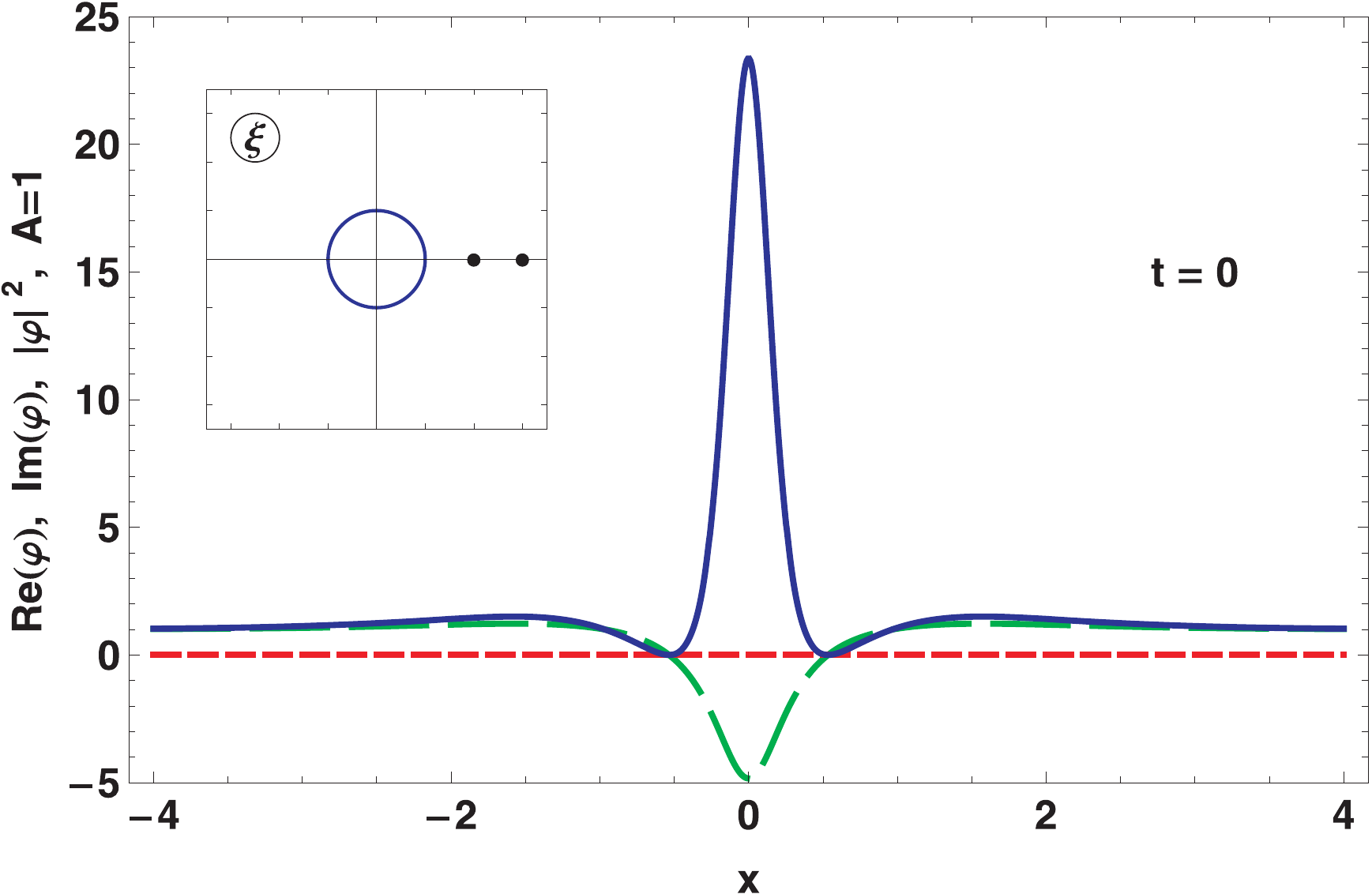}
\includegraphics[width=3in]{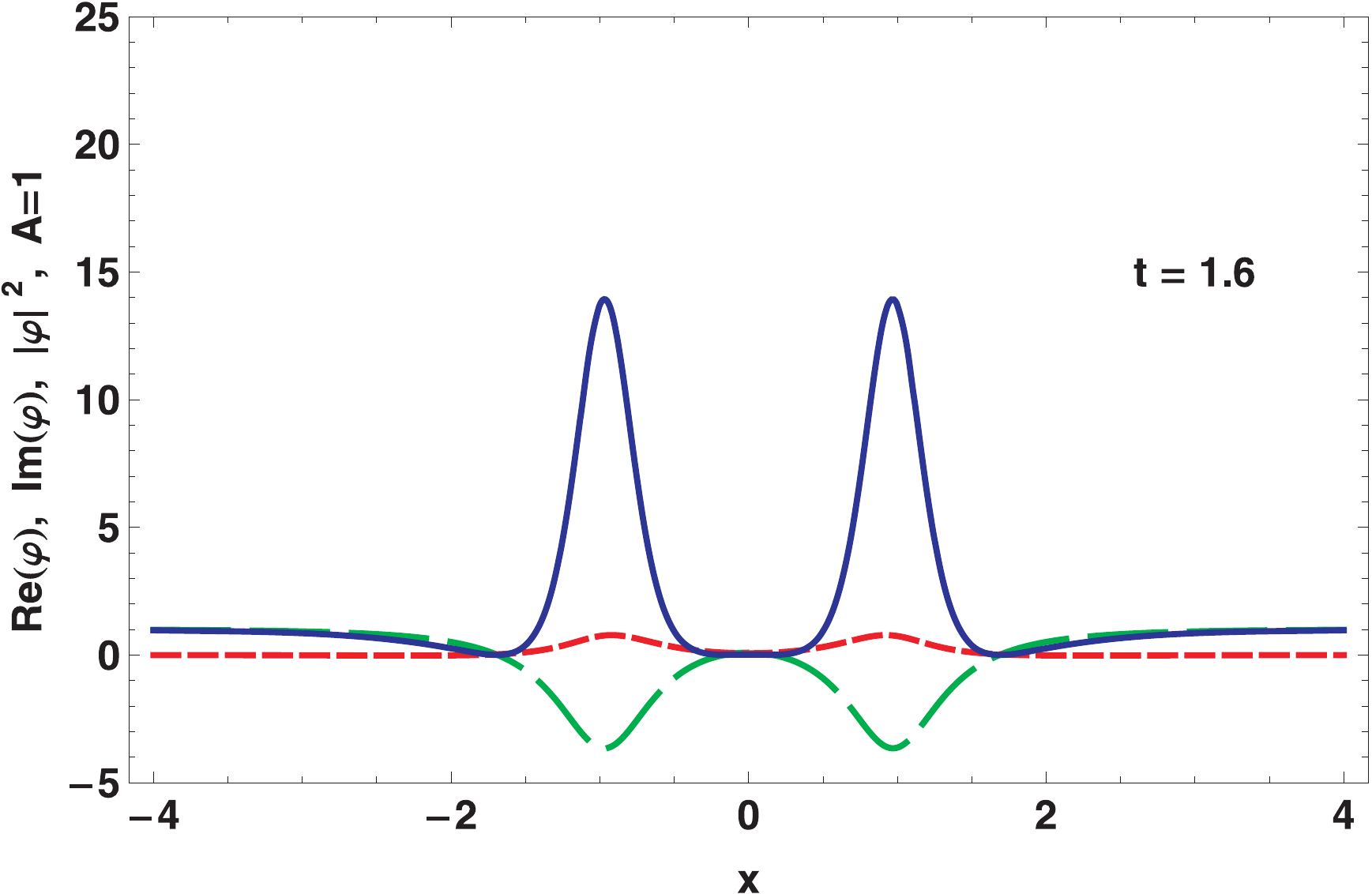}
\caption{\label{2S_bounded}
Bounded two-solitonic solution $\varphi$ at moments of collision (left picture) and maximum distance (right picture) with parameters: $R_1=2,\;\alpha_1=0,\;R_2=3,\;\alpha_2=0,\;\mu_1=\mu_2=0,\;\theta_1=\theta_2=0$. Green dashed lines - real part of $\varphi$, red short dashed lines - imaginary part of $\varphi$ and blue solid lines - absolute squared value of $\varphi$.
}
\end{figure}

\begin{figure}[h]
\centering
\includegraphics[width=3in]{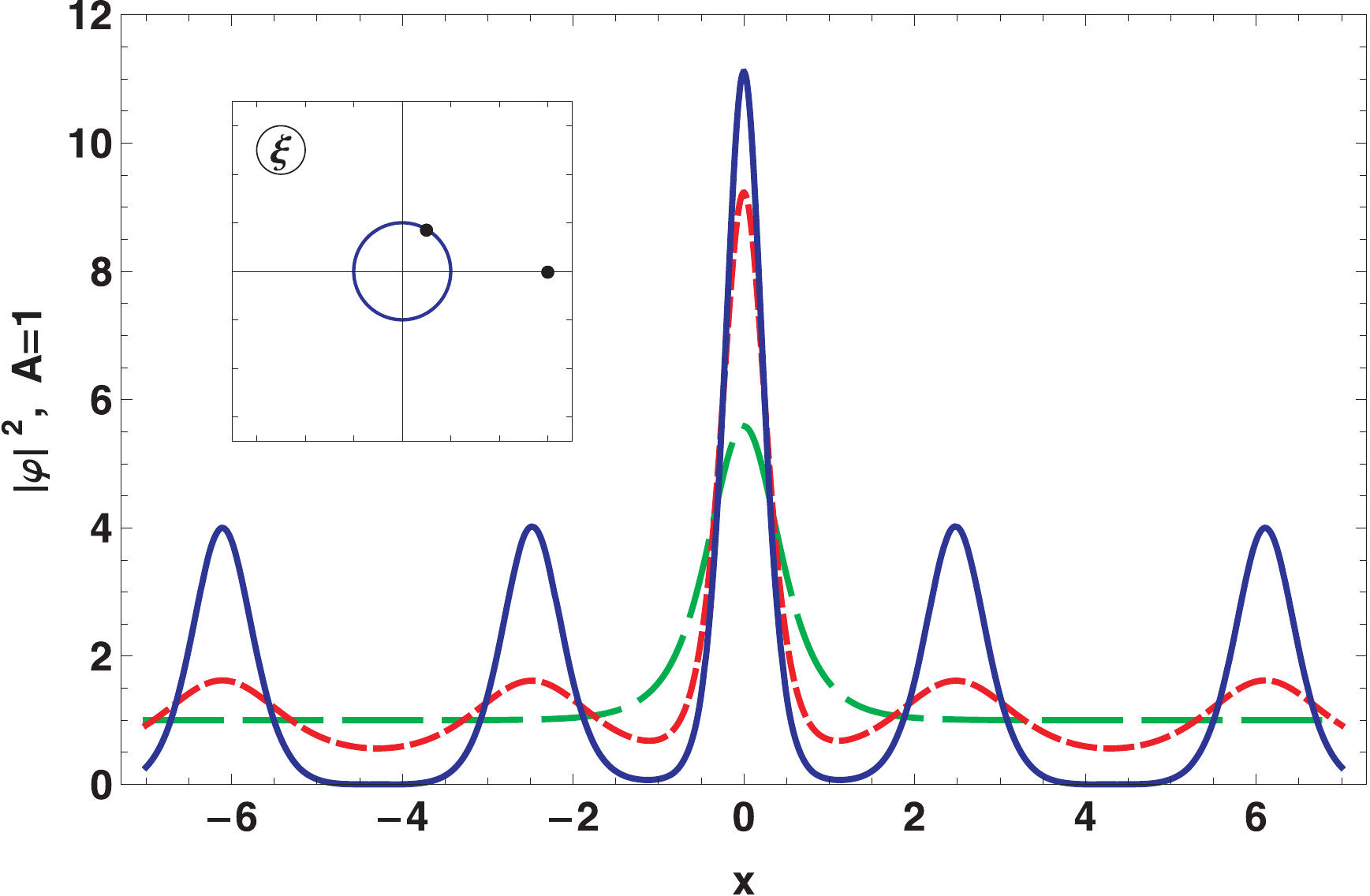}
\caption{\label{2S_AB_plus_KM}
Combination of Akhmediev breather and Kuznetsov soliton. Absolute squared value of $\varphi$ at the moment of time at moments of time $t=-10$ (green dashed line), $t=-2$ (red short dashed line) and $t=0$ (blue solid line) with parameters: $R_1=3,\;\alpha_1=0,\;R_2=1,\;\alpha_2=\frac{\pi}{3},\;\mu_1=\mu_2=0,\;\theta_1=\theta_2=0$.
}
\end{figure}

When both poles are on the real axis ($\alpha_1=0,\alpha_2=0$) the solution is a bounded state. Now $N_{\xi}$ and $\Delta_{\xi}$ for (\ref{2S_general}) are given by
\begin{eqnarray}
\fl N_{\xi}=2\cosh z_2(\cosh^2 z_2-\cosh^2 z_1)|\bi{q}_1|^2 q^*_{21}q_{22}+2\cosh z_1(\cosh^2 z_1-\cosh^2 z_2)|\bi{q}_2|^2 q^*_{11}q_{12},
\nonumber\\
\fl \Delta_{\xi} = (\cosh z_1+\cosh z_2)^2 |\bi{q}_1|^2|\bi{q}_2|^2-4\cosh z_1\cosh z_2|(\bi{q}_1\cdot \bi{q}^*_2)|^2
\end{eqnarray}
One can put as in previous case $\mu_1=0,\;\theta_1=0,\;\mu_2=\mu,\;\theta_2=\theta$. Then
\begin{eqnarray}
\phi_1=A\sinh(z_1)x-\rmi \frac{A^2}{2}\sinh(2 z_1)t,
\nonumber\\
\phi_2=A \sinh(z_2)x+\mu/2-\rmi\biggl(\frac{A^2}{2}\sinh(2 z_1)t-\theta/2 \biggr).
\end{eqnarray}
The example of bounded solution is presented in figure \ref{2S_bounded}.
Many two-soliton combination types are possible. The example of a combination of an Akhmediev breather and a Kuznetsov soliton is plotted in figure \ref{2S_AB_plus_KM}.

\section{Solitonic atoms}

\begin{figure}[h]
\centering
\includegraphics[width=3in]{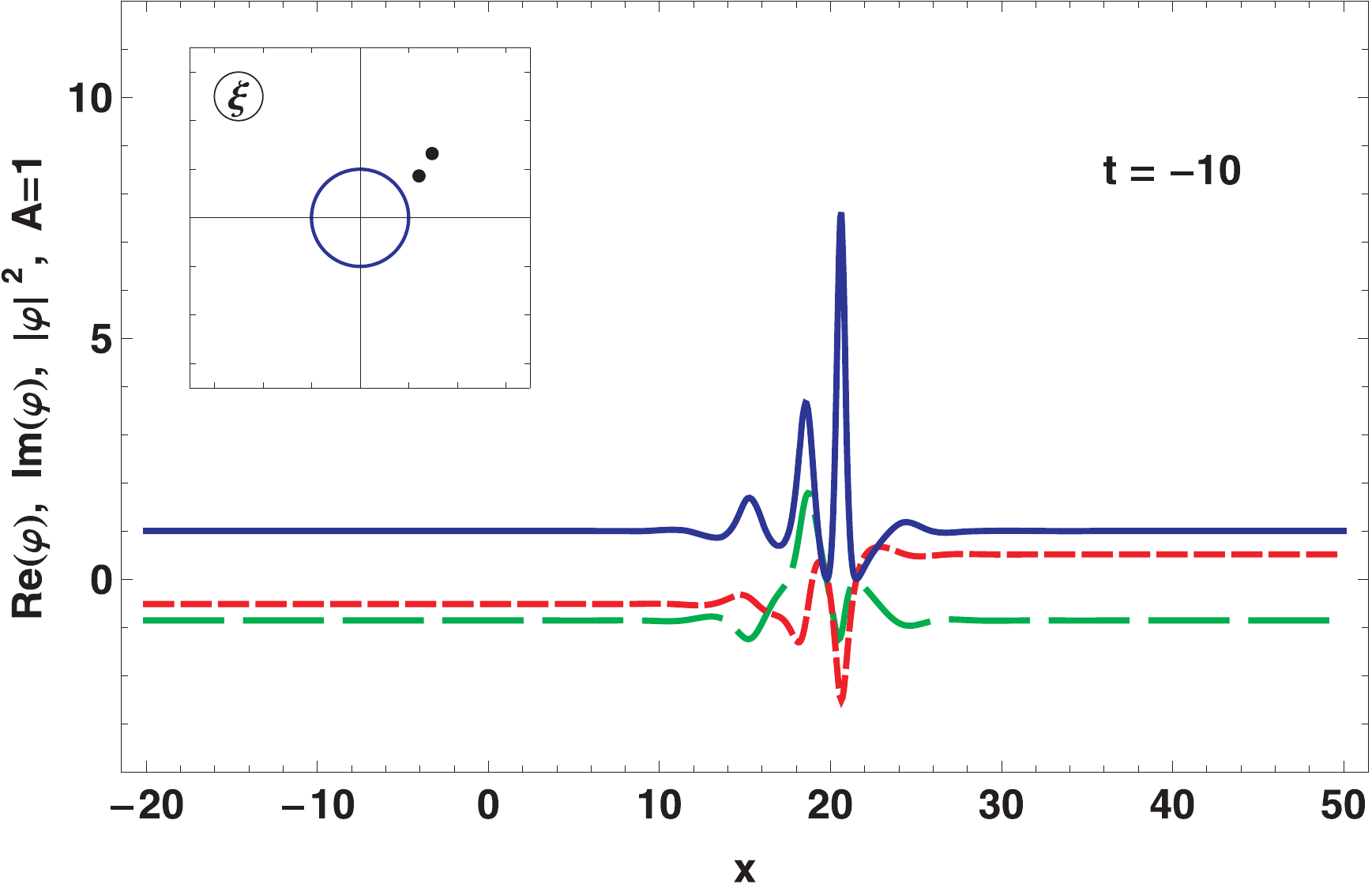}
\includegraphics[width=3in]{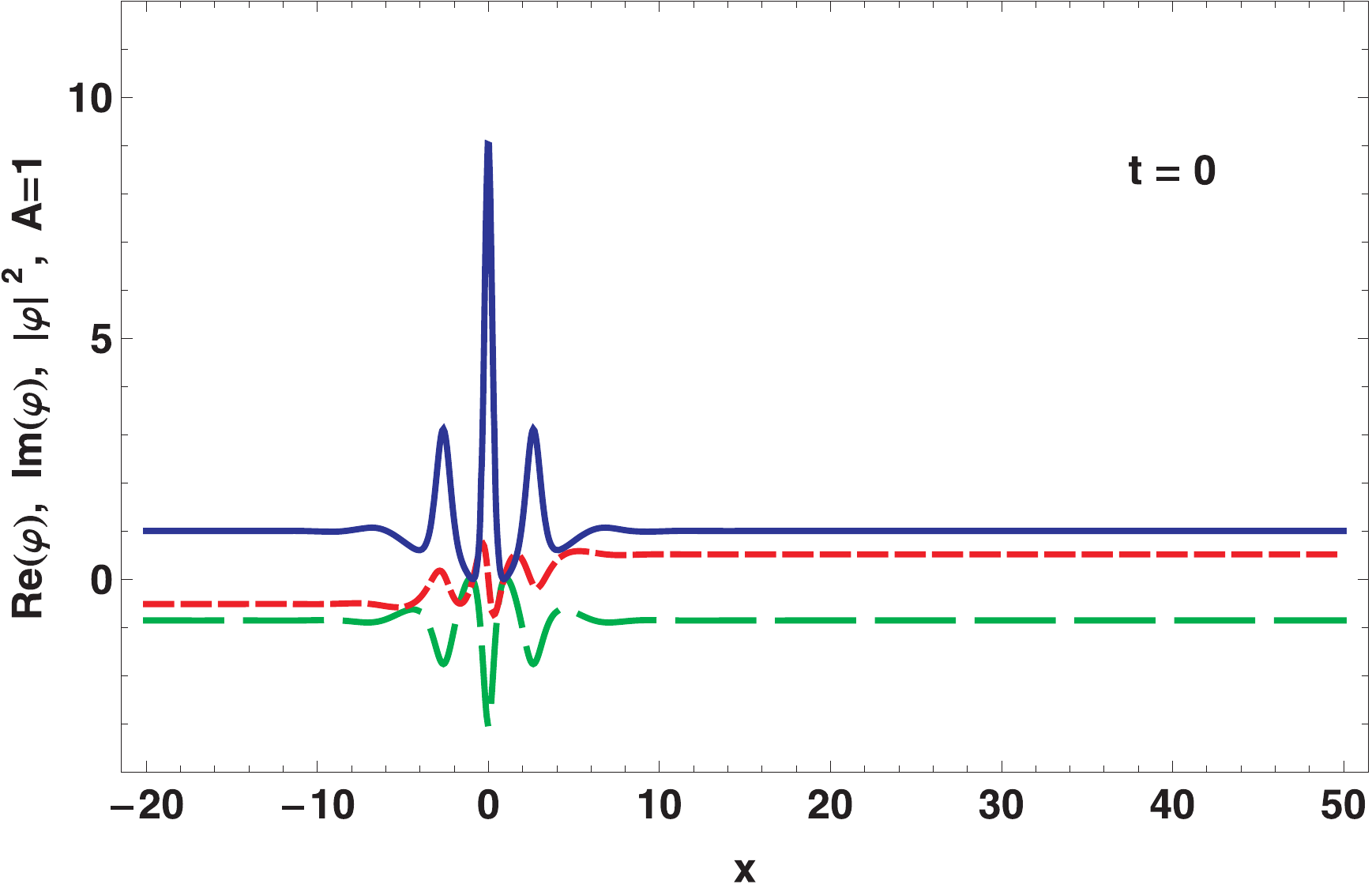}
\caption{\label{2S_Atom}
Two-solitonic atom $\varphi$ at moments of time $t=-10$ (left picture) and $t=0$ (right picture) with parameters: $R_1=1.5,\;R_2=2,\;\alpha_1=\pi/5,\;\alpha_2=0.735242,\;\mu_{1,2}=0,\;\theta_{1,2}=0$.
Green dashed lines - real part of $\varphi$, red short dashed lines - imaginary part of $\varphi$ and blue solid lines - absolute squared value of $\varphi$.
}
\end{figure}

\begin{figure}[h]
\centering
\includegraphics[width=3in]{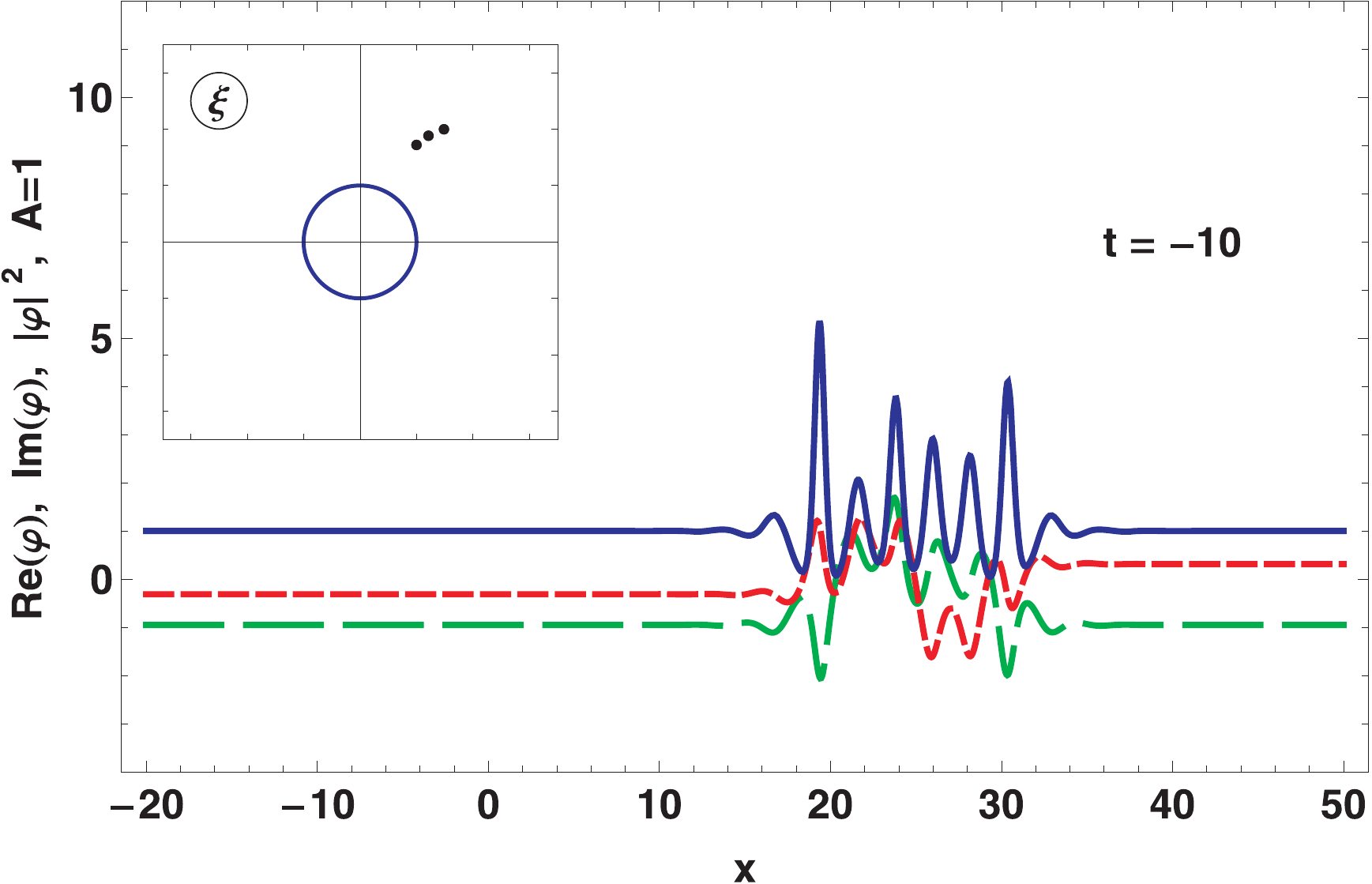}
\includegraphics[width=3in]{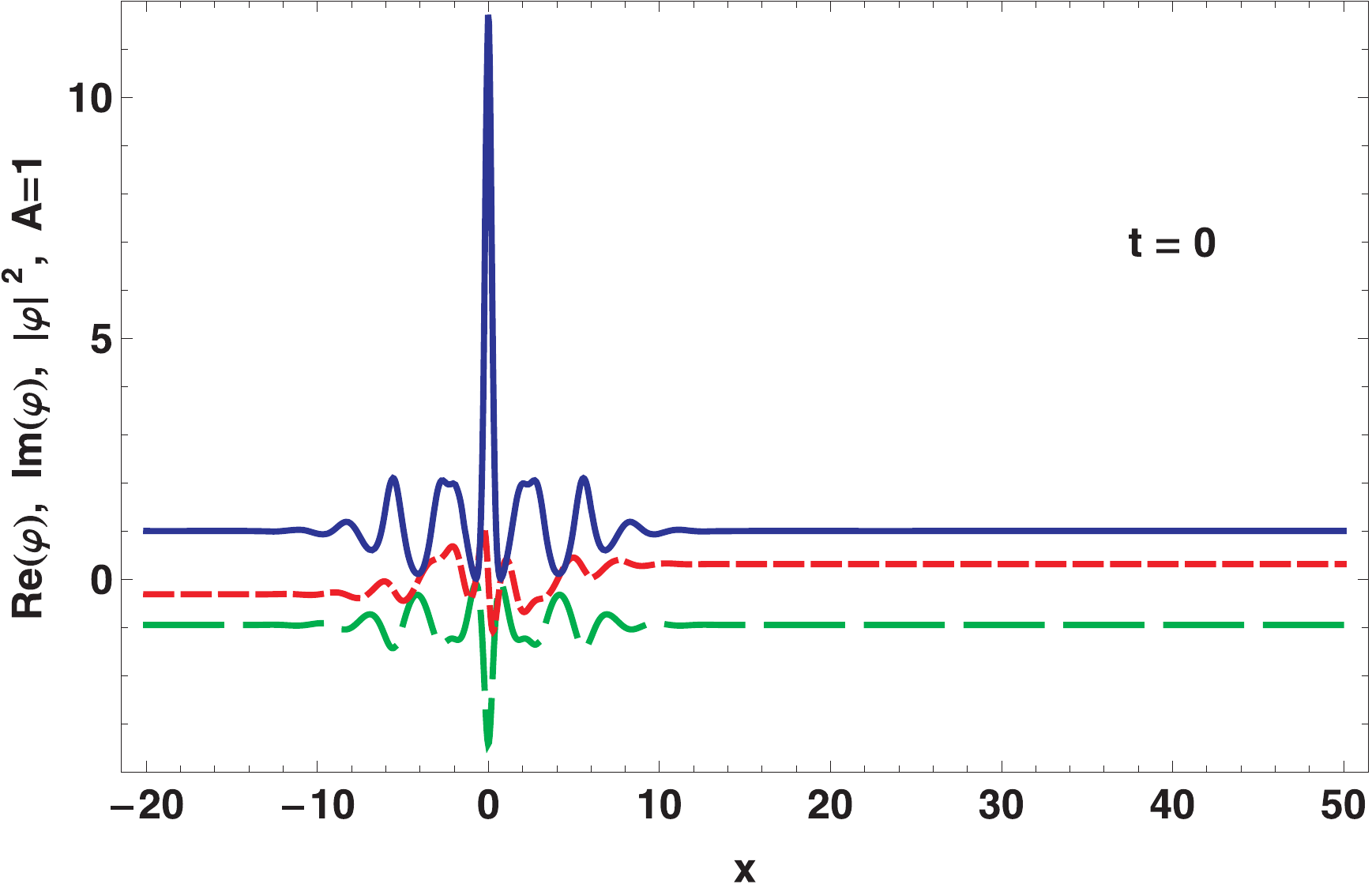}
\caption{\label{3S_Atom}
Three-solitonic atom $\varphi$ at moments of time $t=-10$ (left picture) and $t=0$ (right picture) with parameters: $R_1=2,\;R_2=2.25,\;R_3=2.5,\;\alpha_1=\pi/3,\;\alpha_2=1.00148,\;\alpha_3=0.933785,\;\mu_{1,2,3}=0,\;\theta_{1,2,3}=0$. Green dashed lines - real part of $\varphi$, red short dashed lines - imaginary part of $\varphi$ and blue solid lines - absolute squared value of $\varphi$.
}
\end{figure}

In the case when all group velocities coincide the solution represents a solitonic atom (first mentioned in ~\cite{Zakharov-Shabat1979}) - a complicated configuration of solitons moving together. Complicated solitonic atoms containing a large number of solitons should be described by methods of statistical mechanics. In the two-solitonic case
\begin{eqnarray}
\frac{\cosh 2z_1}{\sinh z_1}\sin\alpha_1=\frac{\cosh 2z_2}{\sinh z_2}\sin\alpha_2.
\end{eqnarray}
The shapes of typical two-solitonic and three-solitonic atom are presented in figure \ref{2S_Atom} and figure \ref{3S_Atom}.

\begin{figure}[h]
\centering
\includegraphics[width=3in]{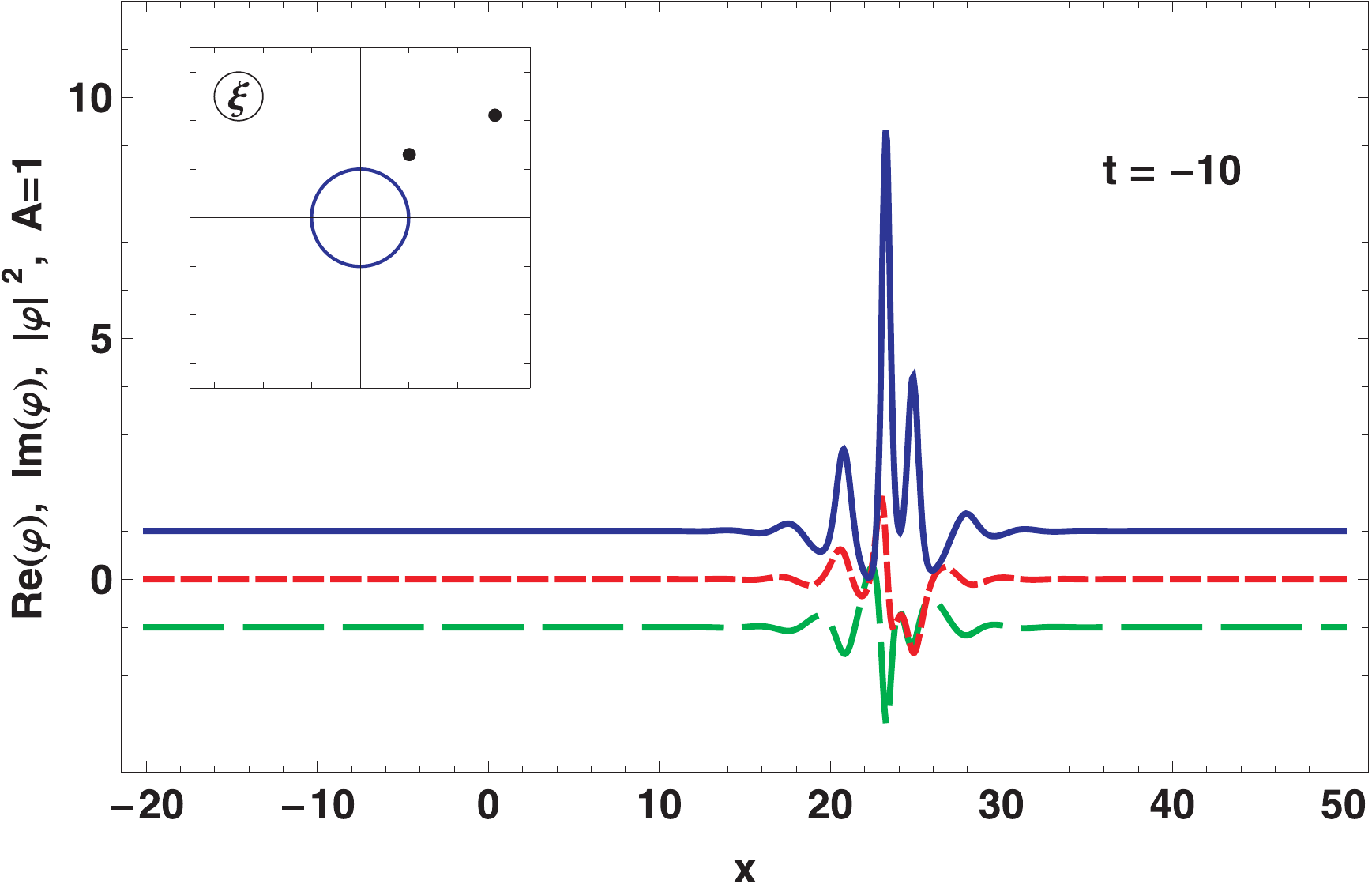}
\includegraphics[width=3in]{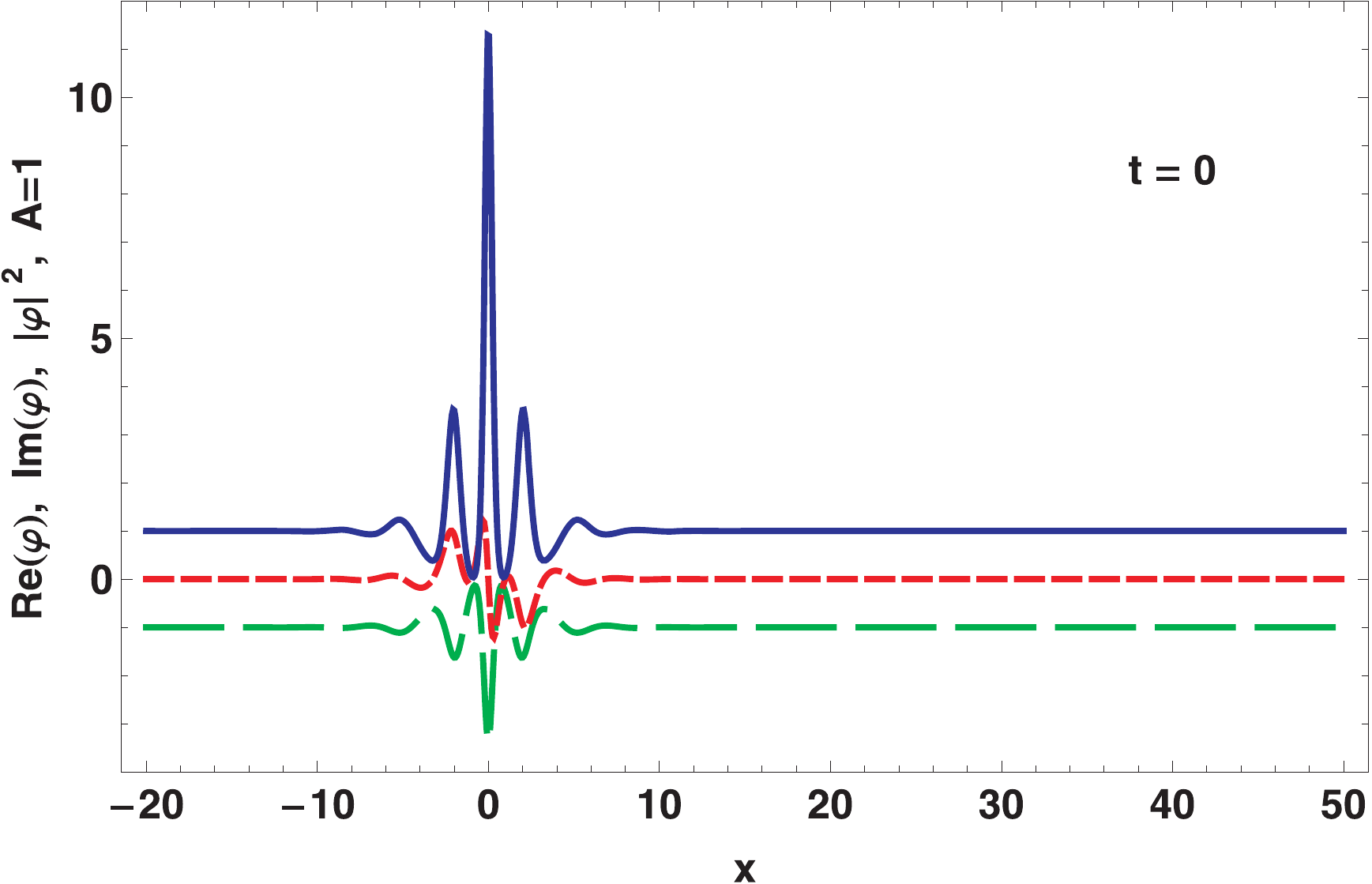}
\caption{\label{2S_Atom_Regular}
Regular two-solitonic atom $\varphi$ with parameters $R_1=3.5,\; R_2=1.66456,\;\alpha_1=5\pi/24,\;\alpha_2=7\pi/24,\;\mu_{1,2}=0,\;\theta_{1,2}=0$. Green dashed lines - real part of $\varphi$, red short dashed lines - imaginary part of $\varphi$ and blue solid lines - absolute squared value of $\varphi$.
}
\end{figure}

We can construct a "regular $N$-solitonic atom". This is possible only when
\begin{eqnarray}
|\alpha_1|+ \cdots +|\alpha_n|=\pm \frac{\pi}{2}.
\end{eqnarray}
An example of a regular two-solitonic atom is given in figure \ref{2S_Atom_Regular}.

\section{Annulation of solitons}
Let us consider the limiting case
\begin{eqnarray}
R_1=R_2=1, &\qquad& \alpha_1=-\alpha_2=-\alpha.
\end{eqnarray}
In this paragraph we use $\lambda$ - variables. Now
\begin{equation}
\lambda_1=\lambda_1^*=\lambda_2=\lambda_2^*=A\cos\alpha.
\end{equation}
Then (see (\ref{q_identities}))
\begin{eqnarray}
\fl N_{\lambda}=\frac{|\bi{q}_1|^2 q^*_{21}q_{22}-(\bi{q}^*_1\cdot\bi{q}_2) q^*_{21}q_{12}-(\bi{q}_1\cdot\bi{q}^*_2) q^*_{11}q_{22}+|\bi{q}_2|^2 q^*_{11}q_{12}}{2A\cos\alpha} \equiv 0,
\nonumber\\
\fl \Delta_{\lambda}=\frac{|q_{11}q_{22}-q_{12}q_{21}|^2}{4A^2\cos^2 \alpha},
\end{eqnarray}
as well
\begin{eqnarray}
\ae_1=\ae_2=0, &\qquad& \omega_1=\omega_2=0,
\nonumber\\
k_1=-k_2=k=A\sin\alpha, &\qquad& \gamma_1=-\gamma_2=\gamma=\frac{A^2}{2}\sin 2\alpha.
\end{eqnarray}
\begin{eqnarray}
u_1=-\gamma t+\mu_1/2, &\qquad& u_2=\gamma t+\mu_2/2,
\nonumber\\
v_1=k x-\theta_1/2, &\qquad& v_2=-k x-\theta_2/2.
\end{eqnarray}
Then vectors $\bi{q}_1$, $\bi{q}_2$ are periodic functions of $x$ and exponential functions of time. We denote
\begin{eqnarray}
\mu_1 \pm \mu_2 = \mu^{\pm}, &\qquad& \theta_1 \pm \theta_2 = \theta^{\pm}.
\end{eqnarray}
From (\ref{f(q)_c}) we obtain that
\begin{eqnarray}
\fl q_{11}q_{22}-q_{12}q_{21}=
-4\sin\alpha\biggl(\cos\frac{\theta^+}{2}\sinh\frac{\mu^+}{2}-
\rmi \sin\frac{\theta^+}{2}\cosh\frac{\mu^+}{2}\biggr).
\end{eqnarray}
Finally
\begin{eqnarray}
\fl \Delta_{\lambda}=\frac{4\sin^2\alpha}{A^2\cos^2\alpha}
\biggl(\cos^2\frac{\theta^+}{2}\sinh^2\frac{\mu^+}{2}+
\sin^2\frac{\theta^+}{2}\cosh^2\frac{\mu^+}{2}\biggr).
\label{denominator leading order}
\end{eqnarray}
The denominator $\Delta$ in this case does not depend on $x$ and $t$. It is just a number. $\Delta \ne 0$ if $\theta^+ \ne 0$ or $\mu^+ \ne 0$ (we do not consider the special Peregrine case $\alpha=0$). Since $N=0$ this means that in a general case $\theta^+ \ne 0$ two pure Akhmediev breathers with opposite values of $\alpha$ completely annihilate each other. In this case the dressing function $\bchi$ is given by the scalar matrix
\begin{equation}
\bchi=
\biggl(
1+\frac{2A\cos\alpha}{\lambda-A\cos\alpha}
\biggr)
\left(
  \begin{array}{cc}
    1 & 0 \\
    0 & 1 \\
  \end{array}
\right).
\label{bchi 2 solitonic}
\end{equation}
When $\theta_1=-\theta_2 = \theta$ and $\mu_1=-\mu_2 = \mu$, both the numerator $N$ and the denominator $\Delta$ in (\ref{denominator leading order}) are zero. We consider this case separately.

The annihilation of solitons takes place for the general case of $2N$ pairs of poles. This can easily be shown mathematically. Now
\begin{eqnarray}
R_{N+k}=R_k=1, &\qquad& \alpha_{N+k}=-\alpha_k.
\end{eqnarray}
Here $k=1,...N$. Let us discuss the system (\ref{p and q* system})
\begin{equation}
\sum_{m} \frac{(\bi{q}_{n}\cdot \bi{q}^*_{m})}{\lambda_{n}+\lambda^*_{m}}\bi{p}^*_m=\bi{q}_n.
\label{p and q* system.2}
\end{equation}
This system can be solved explicitly if we assume that all $\lambda_n$ are real. Their total number is $2N$ and
\begin{eqnarray}
\lambda_{N+k}=\lambda_k, &\qquad& k=1,...N.
\end{eqnarray}
We introduce following new variables
\begin{eqnarray}
X_n=p_{n,1}q_{n,1}+p_{N+n,1}q_{N+n,1}, &\qquad& Y_n=p_{n,1}q_{n,2}+p_{N+n,1}q_{N+n,2}.
\end{eqnarray}
We check that system (\ref{p and q* system.2}) is satisfied if $Y_n=0$ while $X_k$ satisfies following system of equations.
\begin{equation}
\sum_{k=1}^N \frac{X_k}{\lambda_n+\lambda_k}=1.
\end{equation}
This system has a unique solution if $\lambda_i \ne \lambda_j$ as long as $X_k$ are known. If we assume $S_n=q_{n,1}q_{N+n,2}-q_{n,2}q_{N+n,1} \ne 0$, then
\begin{eqnarray}
\fl p_{n,1}=\frac{1}{q_{n,1}q_{N+n,2}-q_{n,2}q_{N+n,1}}
\left|
  \begin{array}{cc}
    X_n & q_{N+n,1} \\
    0 & q_{N+n,2} \\
  \end{array}
\right|
=
\frac{X_n q_{N+2}}{q_{n,1}q_{N+n,2}-q_{n,2}q_{N+n,1}},
\nonumber\\
\fl p_{N+2,1}=\frac{1}{q_{n,1}q_{N+n,2}-q_{n,2}q_{N+n,1}}
\left|
  \begin{array}{cc}
    q_{n,1} & X_n \\
    q_{n,2} & 0 \\
  \end{array}
\right|
=
-\frac{X_n}{S_n}.
\end{eqnarray}
The dressing function $\bchi$ is diagonal
\begin{equation}
\bchi=
E\biggl(
1+\sum_k \frac{X_k}{\lambda-\lambda_k}
\biggr).
\end{equation}
But the condition $Y_n=0$ is equivalent to the pairwise annihilation of the solitons, which was to be proved. It is important that here we do not specify the form of vectors $\bi{q}_n$. This means that the annulation of solitons takes place for wide class of background solutions (background solution must correspond to a cut in the plane of spectral parameter. It is a subject for discussion in another article), particularly on $N$-solitonic solution, describing in this paper.
\section{Superregular two-solitonic solutions}
This chapter is central to our paper. The remarkable fact of the exact cancelation of two Akhmediev breathers with opposite values of angular parameter $\alpha$ makes it possible to construct a special and very important class of $2N$-solitonic solutions describing the evolution of an infinitesimally small perturbation of condensate. These solutions form a subset of regular solutions. We call them superregular solitonic solutions. Early we shortly reported superregular solitonic solutions in \cite{Zakharov-Gelash2012}.

\begin{figure}[h]
\centering
\includegraphics[width=2in]{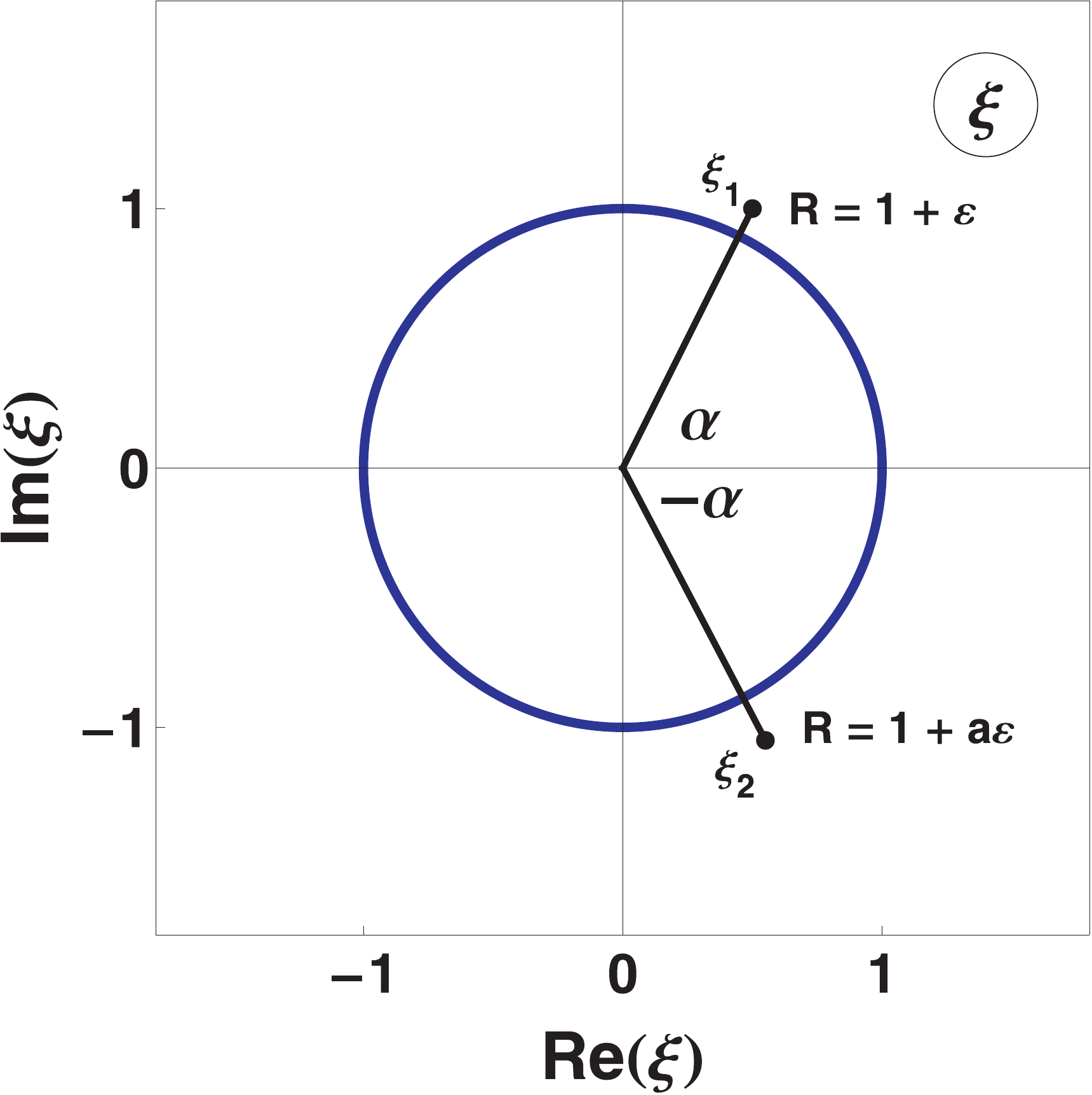}
\caption{\label{SP_poles}
Superregular pair of poles corresponds to a small perturbation of the condensate.
}
\end{figure}

\begin{figure}[h]
\centering
\includegraphics[width=3in]{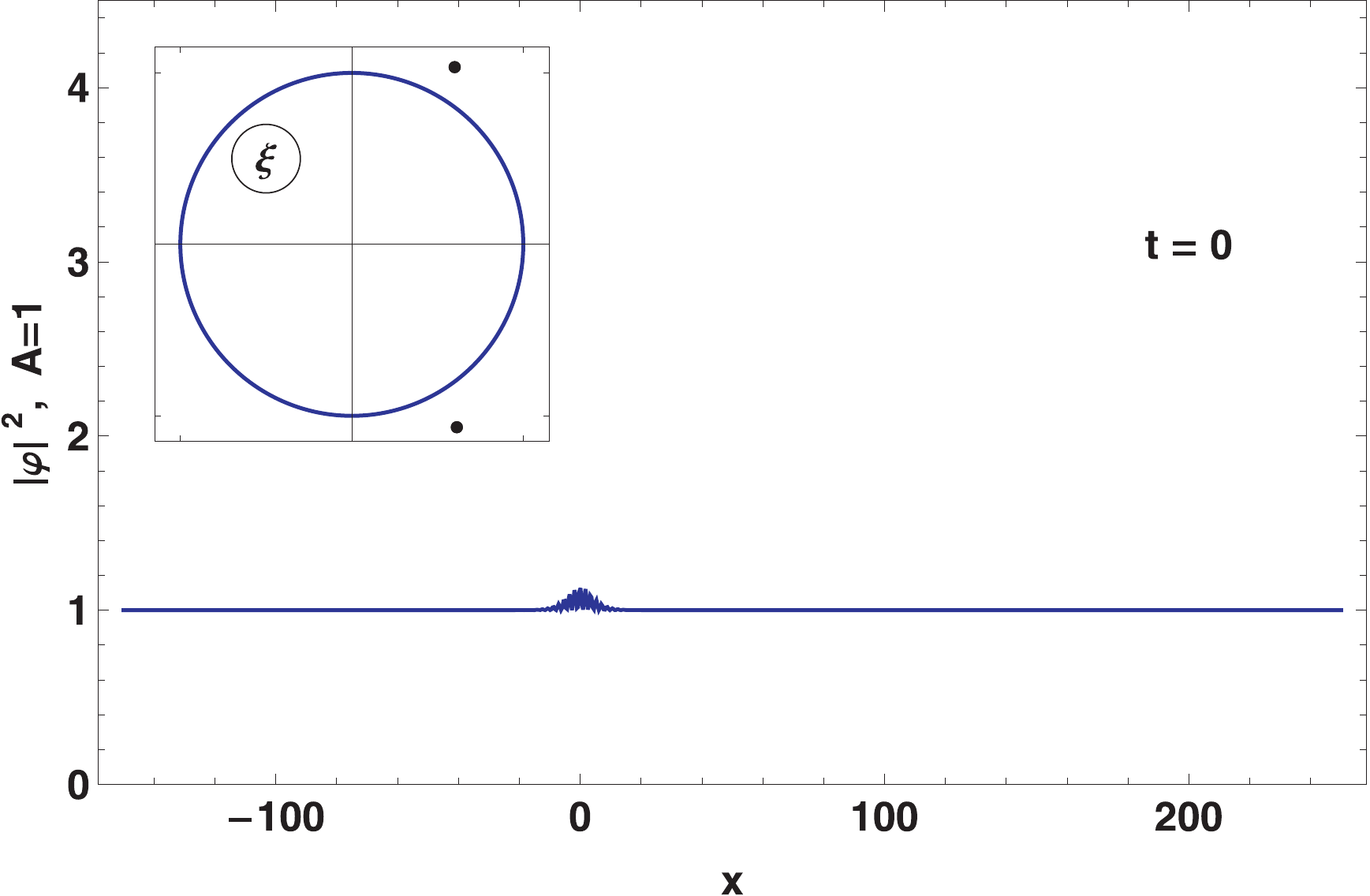}
\includegraphics[width=3in]{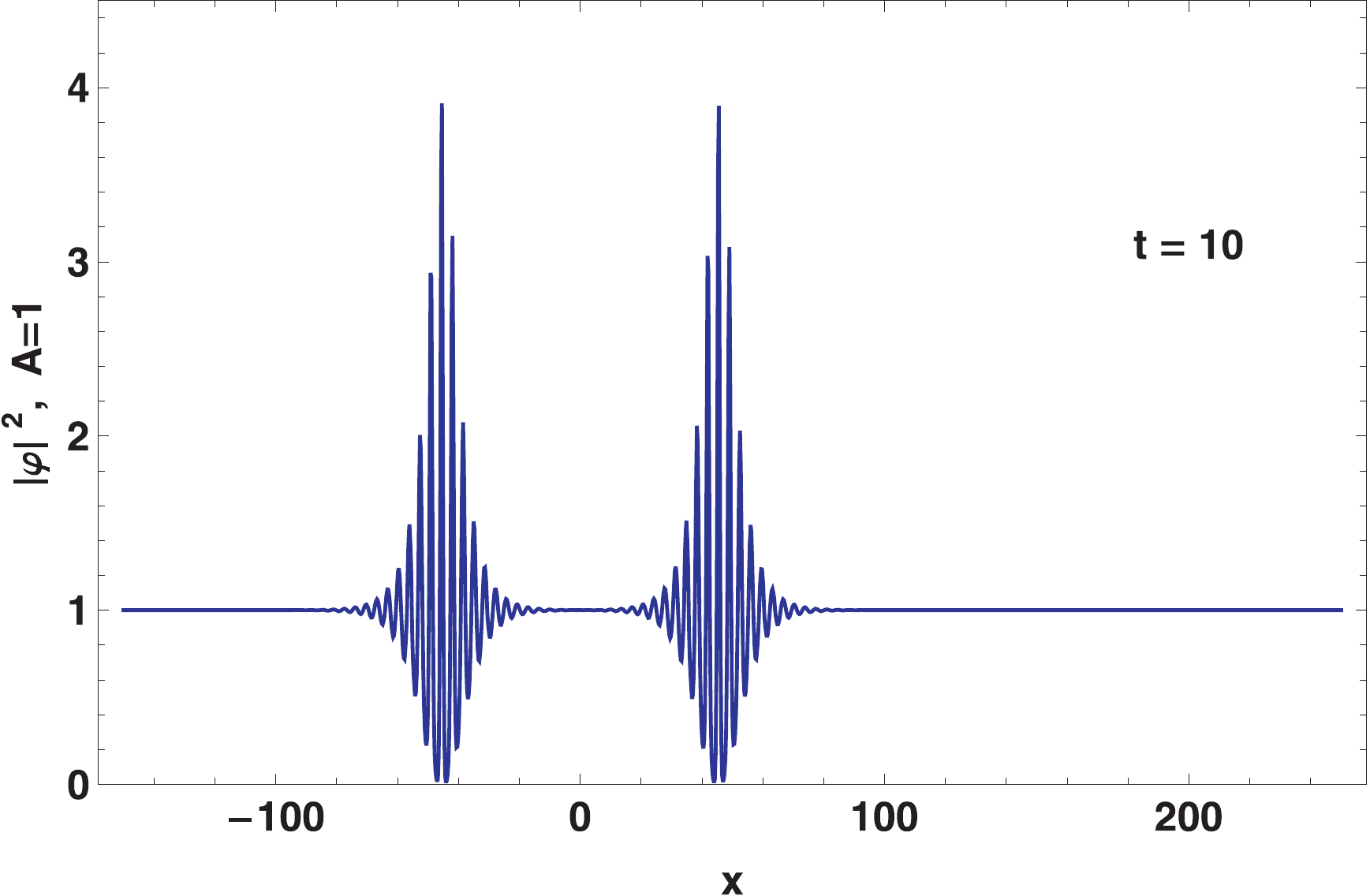}
\includegraphics[width=3in]{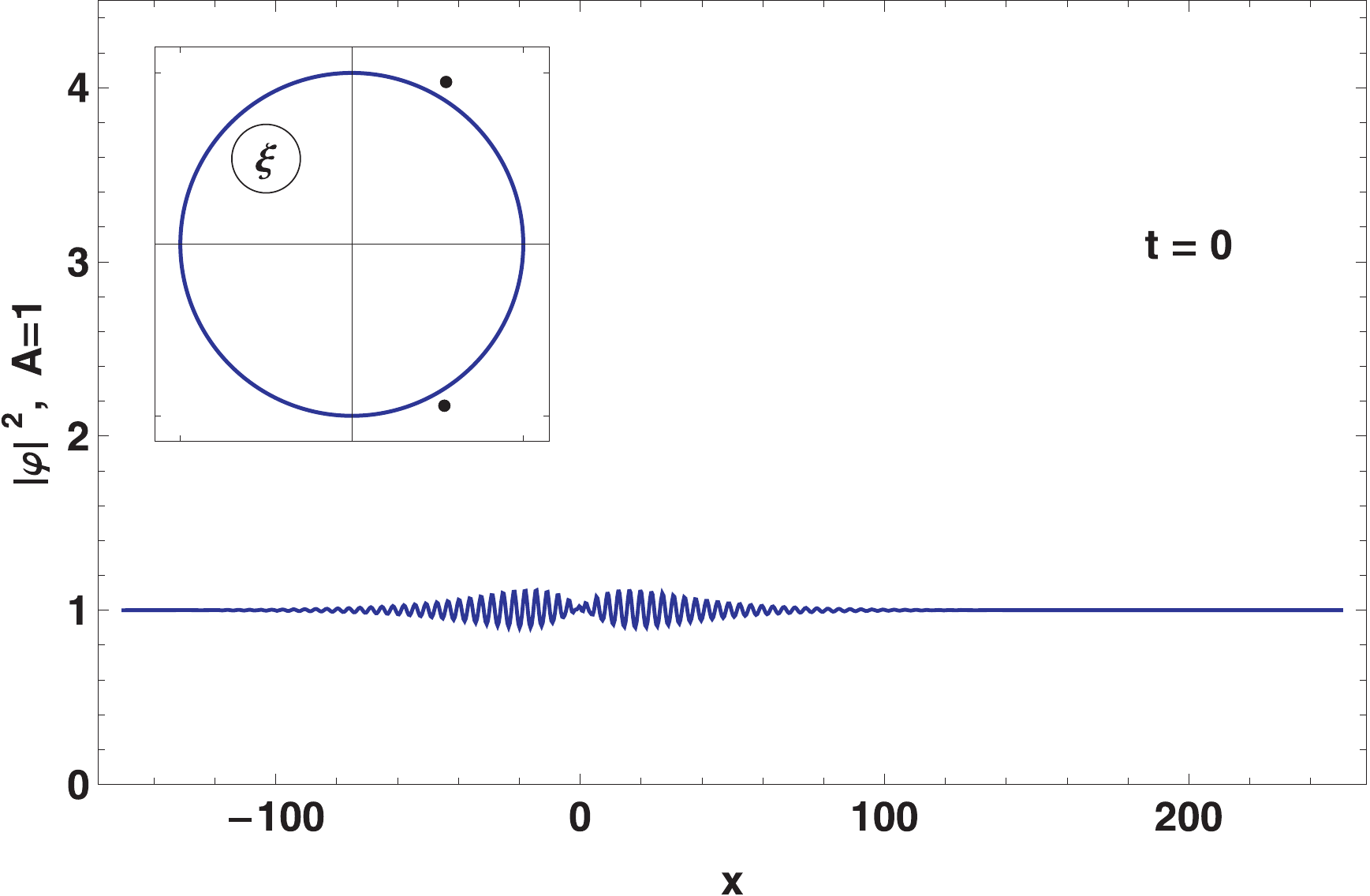}
\includegraphics[width=3in]{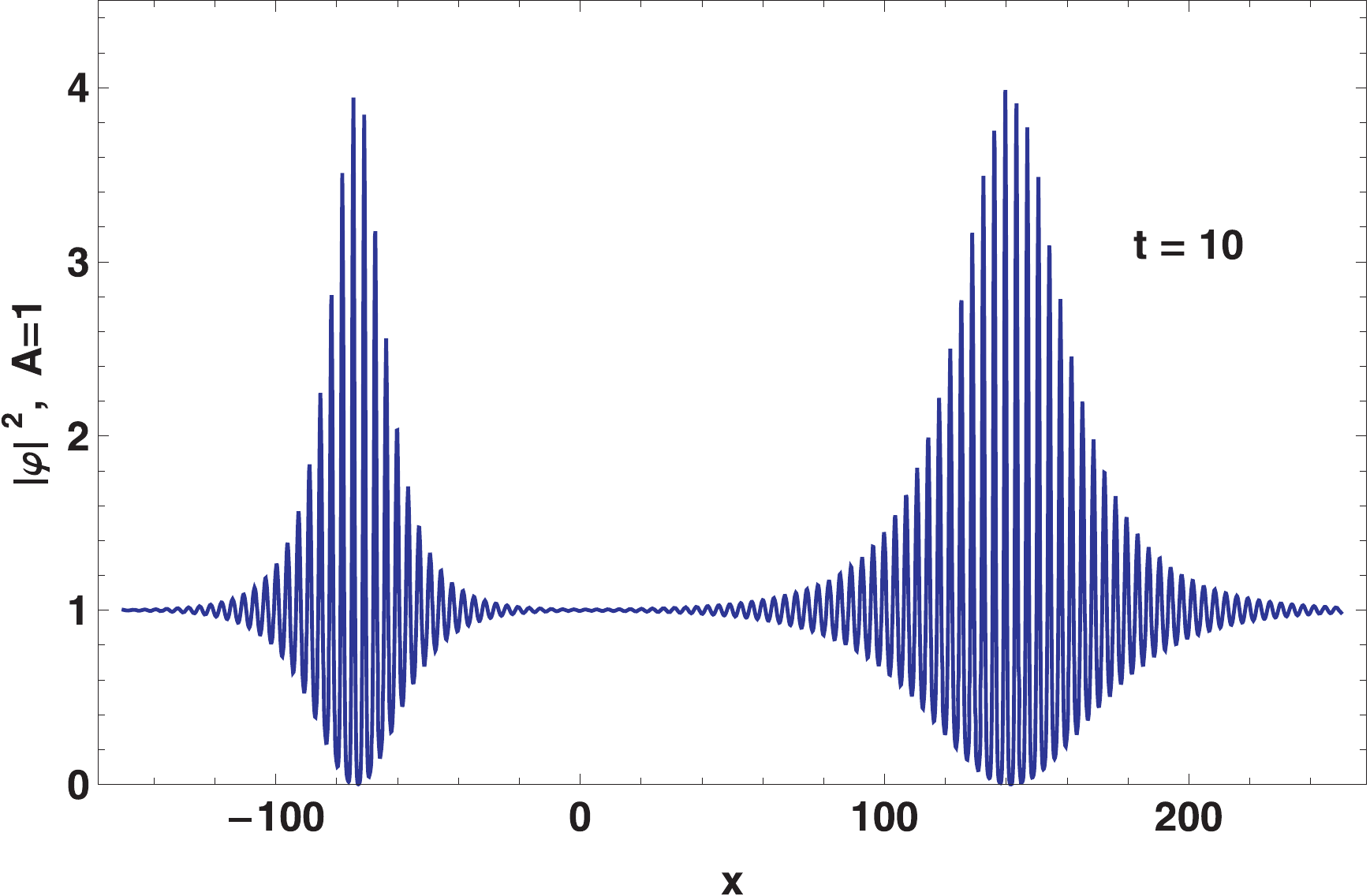}
\caption{\label{2S_SmallPert}
The development of superregular two-solitonic solutions. Absolute squared value of solution $\varphi$ at moments of time $t=0$ (left pictures) and $t=10$ (right pictures) with parameters: $R_1=R_2=1.2,\;\alpha_1=\pi/3,\;\alpha_2=-\pi/3,\;\mu_1=\mu_2=0,\;\theta_1=\theta_2=\pi/2$ (top pictures) and $R_1=1.1,\;R_2=1.05,\;\alpha_1=\pi/3,\;\alpha_2=-\pi/3,\;\mu_1=\mu_2=0,\;\theta_1=\theta_2=\pi/2$ (bottom pictures).
}
\end{figure}

We start with the case $N=1$ and assume that $R_1,R_2$ are close to one:
\begin{eqnarray}
R_1 \simeq 1+\varepsilon,
&\qquad&
R_2 \simeq 1+a\varepsilon,
\end{eqnarray}
$\varepsilon > 0, a > 0, |\varepsilon| \ll 1$.  We call such poles as superregular pair (see figure \ref{SP_poles}).
Denote the deviation of a solution from the condensate at the moment of solitons collision by $\delta \varphi$:
\begin{equation}
\varphi=A+\delta\varphi.
\end{equation}
Again we make intermediate calculations in $\lambda$ variable. Deviation $\delta \varphi$ is given by expression
\begin{equation}
\delta \varphi=-2\frac{\delta N_{\lambda}}{\widetilde{\Delta}_{\lambda}}.
\label{SP_variation}
\end{equation}
Here $\delta N_{\lambda}$ is a variation of the numerator while $\widetilde{\Delta}_{\lambda}$ is a modified  version of the denominator calculated with higher accuracy. In what follows to the calculate modified function we have to neglect the parameter $\varepsilon$ except for the case when $\varepsilon$ is multiplied by $x$. We do so because the product $\varepsilon x$ is not small at $x>\varepsilon^{-1}$. Actually such products appear in functions $\phi_{n}$. First of all we notice that
\begin{equation}
\lambda_1\approx A\cos\alpha+i\varepsilon A\sin\alpha,
\end{equation}
\begin{equation}
\lambda_2\approx A\cos\alpha-\rmi a\varepsilon A\sin\alpha,
\end{equation}
and
\begin{eqnarray}
\lambda_1+\lambda^*_2=2A\cos\alpha+\rmi(1+a)\varepsilon A\sin\alpha,
\nonumber\\
\lambda^*_1+\lambda_2=2A\cos\alpha-\rmi(1+a)\varepsilon A\sin\alpha.
\end{eqnarray}
Then
\begin{equation}
\widetilde{\Delta}_{\lambda}=
\frac{|\widetilde{q}_{11}\widetilde{q}_{22}-\widetilde{q}_{12}\widetilde{q}_{21}|^2}{4A^2\cos^2\alpha}.
\end{equation}
By analogy
\begin{equation}
\delta N_{\lambda}=
\frac{\rmi \varepsilon(1+a)\sin\alpha}{4A\cos^2\alpha}
((\bi{\widetilde{q}}_1 \cdot \bi{\widetilde{q}}^*_2) \widetilde{q}^*_{11}\widetilde{q}_{22}-(\bi{\widetilde{q}}^*_1 \cdot \bi{\widetilde{q}}_2) \widetilde{q}^*_{21}\widetilde{q}_{12}).
\end{equation}
The vectors $\bi{\widetilde{q}}_1,\bi{\widetilde{q}}_2$ are given by formulae (\ref{RTSS vectors q}) but now
\begin{eqnarray}
k_1=-k_2=k=A\sin\alpha,
\nonumber\\
\ae_1=\varepsilon A\cos\alpha = \varepsilon \xi,
\nonumber\\
\ae_2=a \varepsilon A\cos\alpha = a\varepsilon \xi.
\end{eqnarray}
Here we denote $\xi = A\cos\alpha$. In what follows in this paragraph we put $\mu_1=\mu_2=0$, which corresponds to solitons colliding at $x=0,\;t=0$. Now
\begin{eqnarray}
\phi_1=\rmi k x+\varepsilon \xi x-\rmi\theta_1/2,
&\qquad&
\phi_2=-\rmi k x+a\varepsilon \xi x-\rmi\theta_2/2.
\end{eqnarray}
As a result the vectors $\bi{\widetilde{q}}_1$ and $\bi{\widetilde{q}}_2$ have following components
\begin{eqnarray}
\fl \widetilde{q}_{11}=\exp(-\rmi k x-\varepsilon \xi x+\rmi\theta_1/2)+\exp(-\rmi\alpha+\rmi k x+\varepsilon \xi x-\rmi\theta_1/2),
\nonumber\\
\fl \widetilde{q}_{21}=\exp(\rmi k x-a\varepsilon \xi x+\rmi\theta_2/2)+\exp(\rmi\alpha-\rmi k x+a\varepsilon \xi x-\rmi\theta_2/2),
\nonumber\\
\fl \widetilde{q}_{12}=\exp(-\rmi\alpha-\rmi k x-\varepsilon \xi x+\rmi\theta_1/2)+\exp(\rmi k x+\varepsilon \xi x-\rmi\theta_1/2),
\nonumber\\
\fl \widetilde{q}_{22}=\exp(\rmi\alpha+\rmi k x-a\varepsilon \xi x+\rmi\theta_2/2)+\exp(-\rmi k x+a\varepsilon \xi x-\rmi\theta_2/2).
\end{eqnarray}
We calculate $\widetilde{\Delta}$ assuming $\theta^+ \ne 0$
\begin{eqnarray}
\fl \widetilde{\Delta_{\lambda}}=
\frac{4\sin^2\alpha}{A^2\cos^2\alpha}
\biggl(
\cos^2\frac{\theta^+}{2}\sinh^2(\varepsilon(1+a)\xi x)+\sin^2\frac{\theta^+}{2}\cosh^2(\varepsilon(1+a)\xi x)
\biggr).
\end{eqnarray}
The denominator is a function slowly oscillating in space. The products $\widetilde{q}_{12}\widetilde{q}_{22}$ and $\widetilde{q}^*_{21}\widetilde{q}^*_{11}$ in the numerator consist of both fast and slowly changing components. Hence $\varphi$ can be given as a sum of two terms
\begin{eqnarray}
\delta\varphi=\delta\varphi_{slow}+\delta\varphi_{fast}.
\label{delta_varphi}
\end{eqnarray}
Here
\begin{eqnarray}
\fl \delta\varphi_{slow} \approx
\frac
{\varepsilon A\rmi(1+a)\sin\theta^+}
{\cos^2\frac{\theta^+}{2}\sinh^2(\varepsilon(1+a)\xi x)+\sin^2\frac{\theta^+}{2}\cosh^2(\varepsilon(1+a)\xi x)},
\end{eqnarray}
is a function slowly varying in space, while
\begin{eqnarray}
\fl \delta\varphi_{fast} \approx
\nonumber\\
\fl \biggl[-\varepsilon A(1+a)\sin\alpha
(\sinh(2\epsilon a \xi x)\sin(2k x-\theta_1)+\sinh(2\epsilon\xi x)\sin(2k x+\theta_2))
\nonumber\\
\fl +\rmi\cos\alpha
(\cosh(2\epsilon a \xi x)\sin(2k x-\theta_1)-\cosh(2\epsilon\xi x)\sin(2k x+\theta_2))
-\rmi \sin\theta^+ \biggr] \div
\nonumber\\
\fl \biggl[ \cos^2\frac{\theta^+}{2}\sinh^2(\varepsilon(1+a)\xi x)+\sin^2\frac{\theta^+}{2}\cosh^2(\varepsilon(1+a)\xi x) \biggr],
\end{eqnarray}
is rapidly oscillating. In the symmetric case $a=1$ and $\delta N_{\lambda}$ and $\widetilde{\Delta}_{\lambda}$ become
\begin{eqnarray}
\fl \delta N_{\lambda} = 2\varepsilon \sin\alpha \sinh(2\varepsilon\xi x) \sin\biggl(2k x-\frac{\theta^-}{2}\biggr)\cos\frac{\theta^+}{2}
\nonumber\\
\fl -\rmi\cos\alpha \cosh(2\varepsilon\xi x) \cos\biggr(2k x-\frac{\theta^-}{2}\biggl)\sin\frac{\theta^+}{2}
-\rmi \sin\theta^+
\nonumber\\
\fl \widetilde{\Delta}_{\lambda}=\cos^2\frac{\theta^+}{2}\sinh^2(2\varepsilon\xi x)+\sin^2\frac{\theta^+}{2}\cosh^2(2\varepsilon\xi x).
\end{eqnarray}
The simplest case appears when $\theta^+=\pi$ and $a=1$
\begin{eqnarray}
\delta\varphi \approx
4\rmi\varepsilon A
\frac{\cos\alpha\cos\biggl(2k x-\frac{\theta^-}{2}\biggr)}{\cosh(2\varepsilon \xi x)}.
\label{small_pert_simpliest}
\end{eqnarray}
It is important that this perturbation grows exponentially at the first time. It is described by well-known equations for the linear stage of modulation instability. This can be seen by looking at the expansion of $\varphi$ at $t \ne 0,\;\exp(\gamma t) < \varepsilon^{-1}$  (remember that $\gamma =-\frac{A^2}{2} \sin2\alpha$). Let us write only the simplest expression for the latter case (\ref{small_pert_simpliest}):
\begin{eqnarray}
\delta\varphi \approx
2\rmi\varepsilon A
\frac{(\exp(\rmi\alpha-2\gamma t)+\exp(-\rmi\alpha+2\gamma t))\cos\biggl(2k x-\frac{\theta^-}{2}\biggr)}{\cosh(2\varepsilon \xi x)}.
\end{eqnarray}
The derived two-solitonic solution describes the following physical process. An initially small localized perturbation of the condensate generates a pair of quasi-Akhmediev breathers propagating in opposite directions, as a rule having fast group velocity. In the symmetric case $a=1$ these solitons are symmetric. The typical development of these superregular solitonic solution is given in figure \ref{2S_SmallPert}. Two-solitonic superregular solution is described by five parameters $\alpha,\varepsilon,a,\theta^+,\theta^-$ and can be shifted in space and in time, thus the total number of free parameters is seven.
\begin{figure}[h]
\centering
\includegraphics[width=3in]{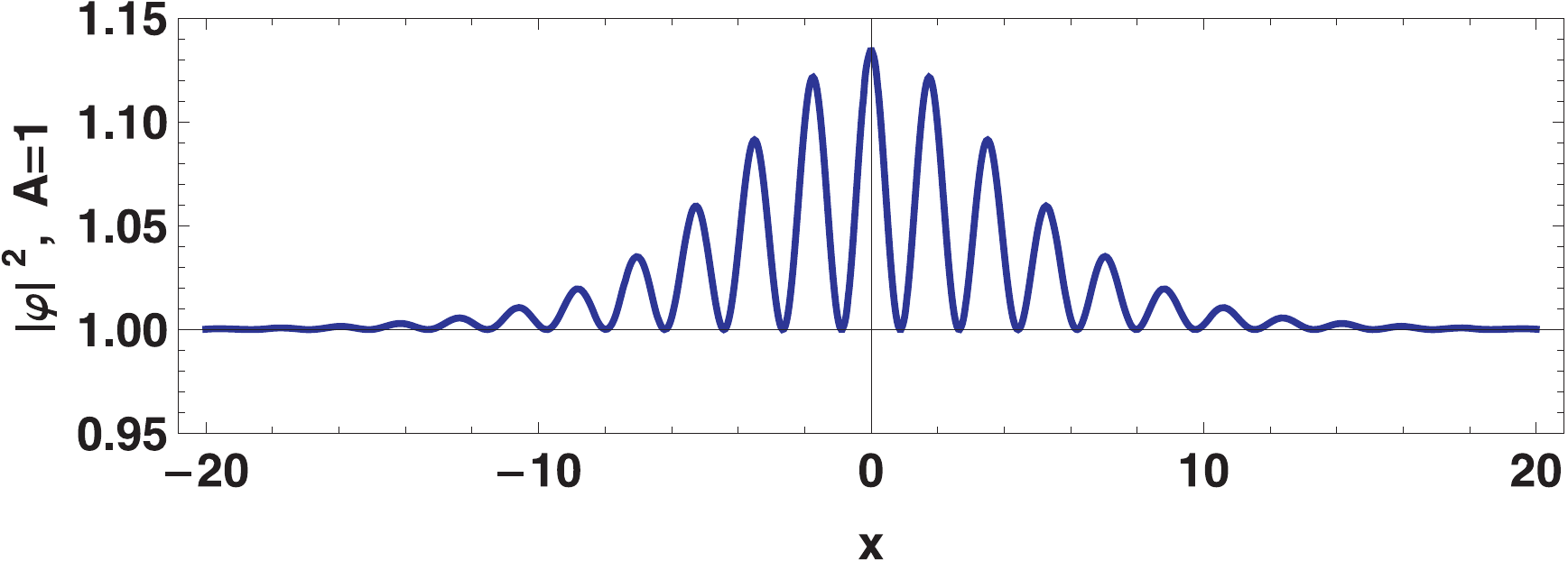}
\includegraphics[width=3in]{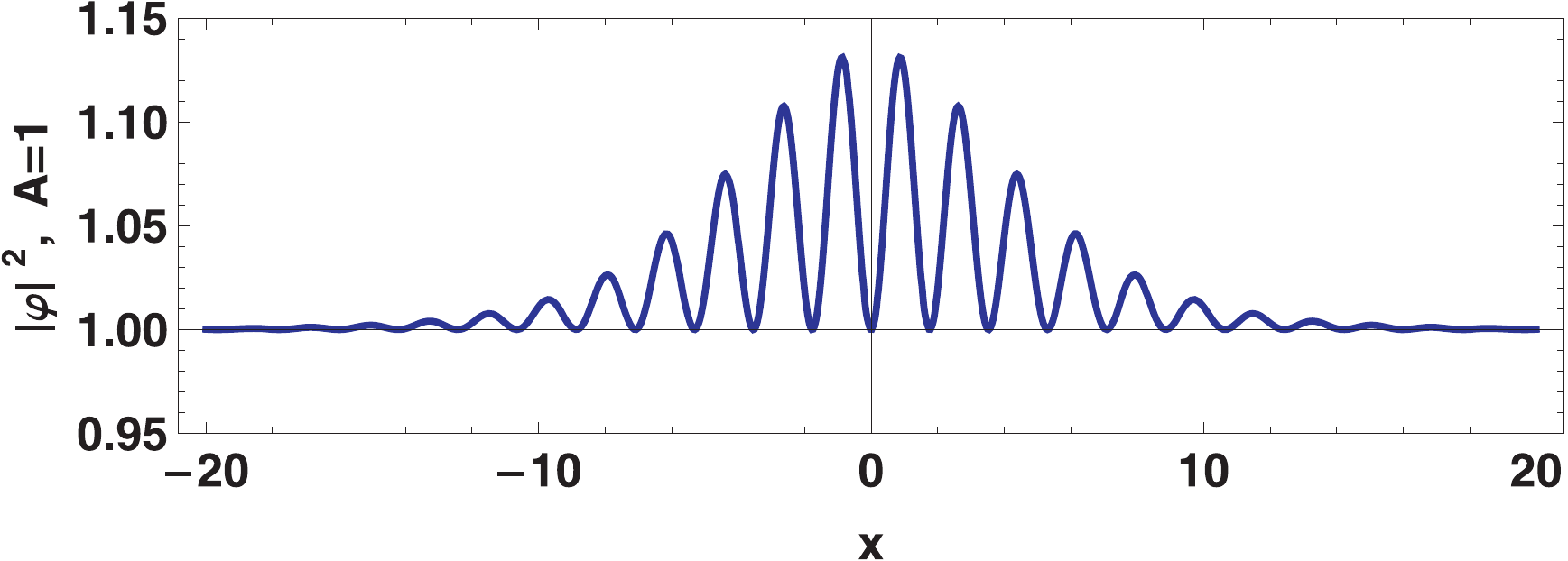}
\caption{\label{2S_SmallPert3}
The enlarged small perturbations at the moment of time $t=0$ are given on the top of figure \ref{2S_SmallPert} (left picture) and the same perturbation with different constants $C_1,C_2$: $\mu_1=0,\;\theta_1=0,\;\mu_2=0,\;\theta_1=\pi$ (right picture).
}
\end{figure}
Note that the generated solitons have a small number of oscillations if we choose the parameter $\alpha$ near zero. Figure \ref{2S_SmallPert3} demonstrates such a situation.
\begin{figure}[h]
\centering
\includegraphics[width=3in]{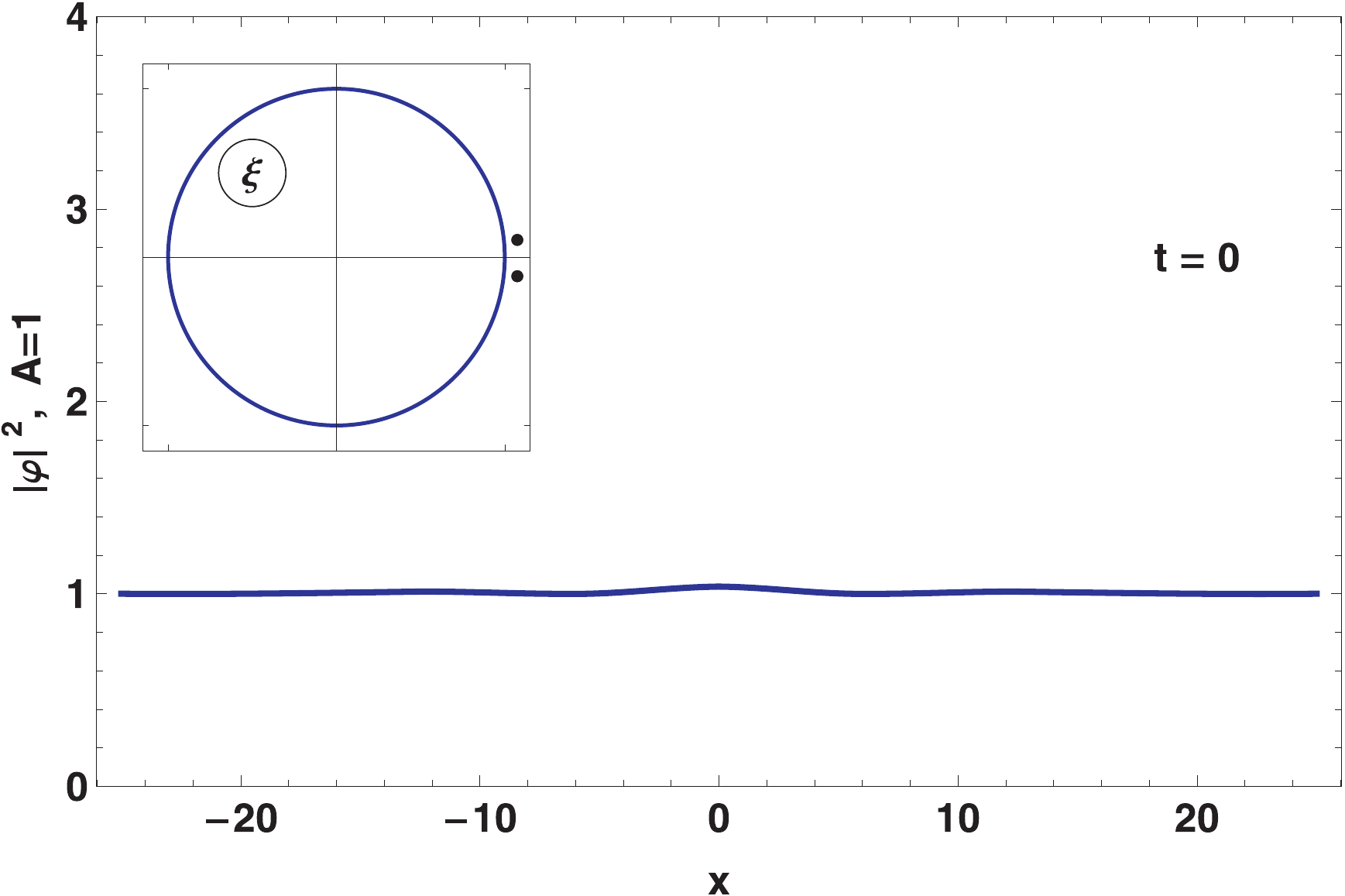}
\includegraphics[width=3in]{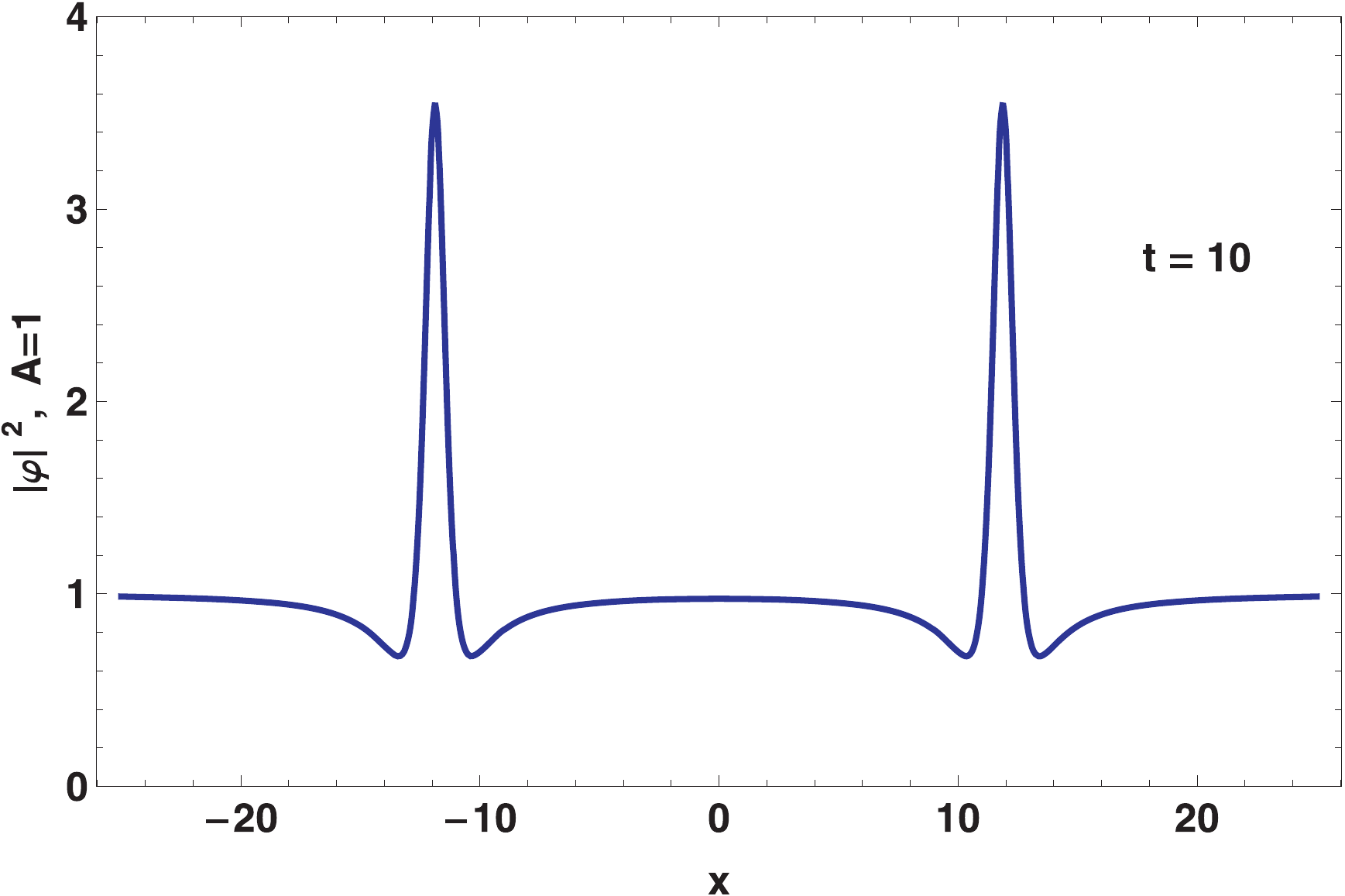}
\caption{\label{2S_SmallPert3}
The development of a superregular two-solitonic solution with small angle. Absolute squared value of solution $\varphi$ at moments of time $t=0$ (left picture) and $t=16$ (right picture) with parameters: $R_1=R_2=1.05,\;\alpha_1=0.1,\;\alpha_2=-0.1,\;\mu_1=\mu_2=0,\;\theta_1=\theta_2=\pi/2$.
}
\end{figure}
At the end of our paper we show that these results can be essentially generalized. We construct $2N$-solitonic superregular solutions representing small perturbation of the condensate at $t=0$.
\section{Annulation of solitons and degenerate solutions}
\begin{figure}[h]
\centering
\includegraphics[width=3in]{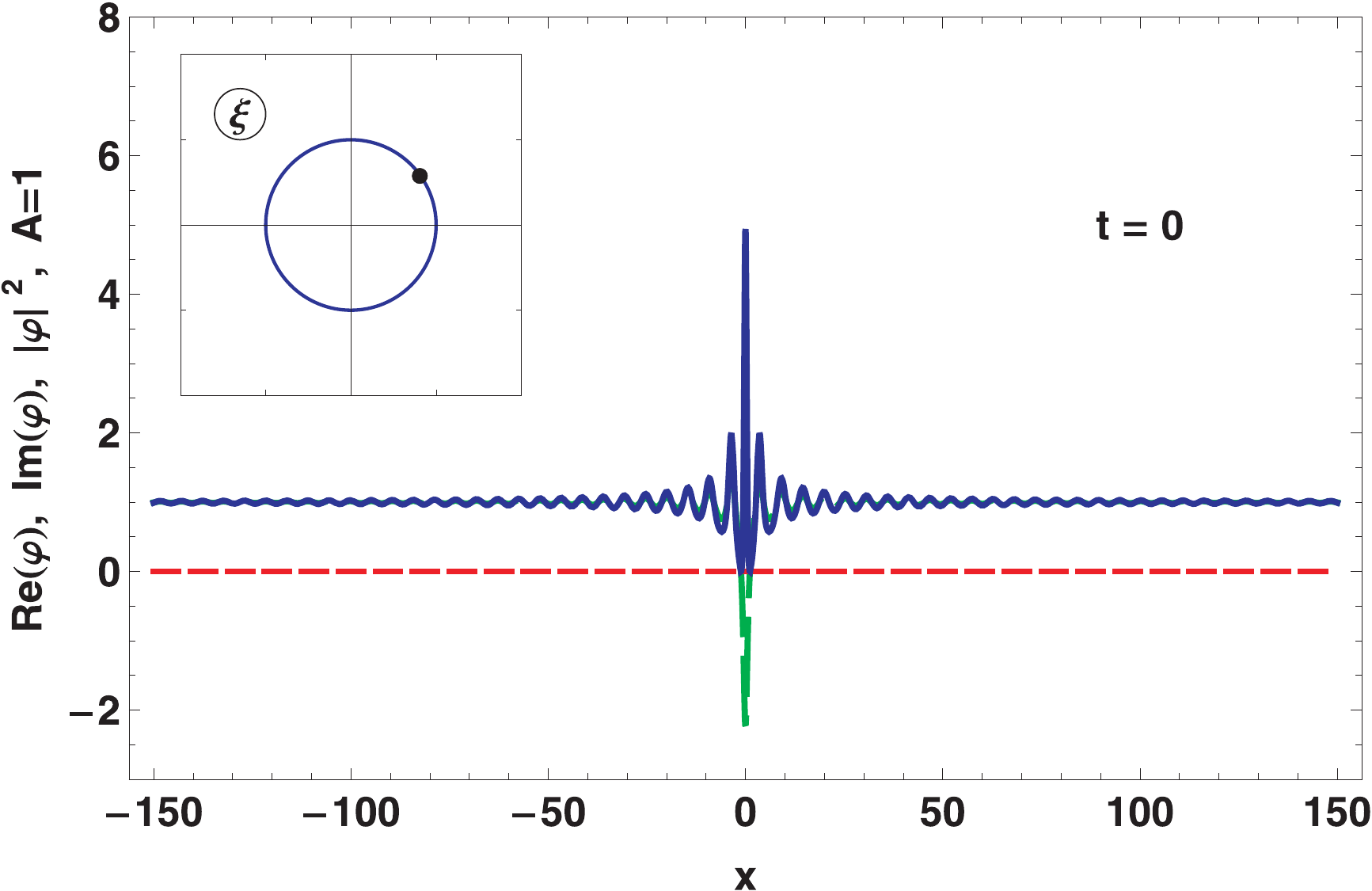}
\includegraphics[width=3in]{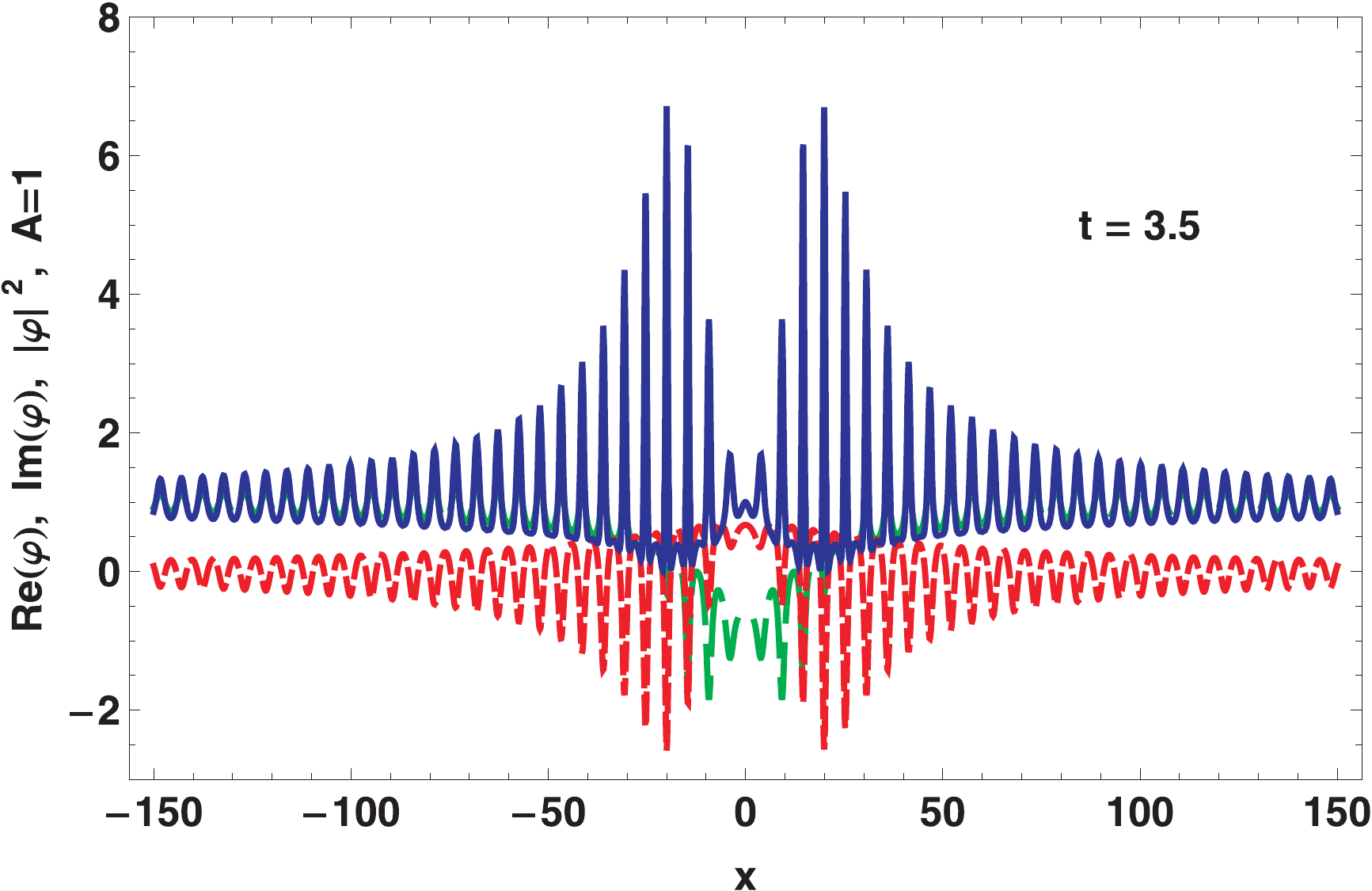}
\caption{\label{DG_AB}
Degenerate two-solitonic solution $\varphi$ at moments of time $t=0$ (left picture) and $t=3.5$ (right picture) with parameters: $R=1,\;\alpha_1=\pi/5,\;\mu=0,\;\theta=0$. Green dashed lines - real part of $\varphi$, red short dashed lines - imaginary part of $\varphi$ and blue solid lines - absolute squared value of $\varphi$.
}
\end{figure}

We start this paragraph by discussing the degenerate case $\theta^+=0$ mentioned previously. Now the numerator and the denominator of the two-solitonic solution are both zero. This indeterminate form can be resolved using l'H\^{o}pital's rule. We choose $\alpha_1=\alpha_2=\alpha$, $R_{1,2}=1 + \varepsilon, \;\;\; \varepsilon \ra 0$ and $\theta_1= -\theta_2 = \theta$, $\mu_1= -\mu_2 = \mu$ in (\ref{2S_symmetric}) and expand the numerator and the denominator to the second order by $\varepsilon$. It is convenient to use explicit expressions (\ref{f(q)_a},\ref{f(q)_b},\ref{f(q)_c}) for quadratic functions of vectors $\bi{q_n}$ and its components. If we put $\theta = 0,\;\mu=0$ the solution is centered in space and time and given in the following form
\begin{eqnarray}
\varphi=A-2A\sin2\alpha\frac{M-iK}{H}.
\end{eqnarray}
Here
\begin{eqnarray}
\fl M=\sin\alpha(\cosh 2\gamma t+\cos\alpha\cos 2kx)(\cos\alpha\cosh 2\gamma t+\cos 2kx)+
\nonumber\\
\cos\alpha\sin\alpha\sinh 2\gamma t(\sinh 2\gamma t-2\Omega t\sin\alpha\cos2\alpha\cos 2kx)+
\nonumber\\
\cos\alpha\sin\alpha\sin 2kx(2\xi x\sin\alpha\cosh 2\gamma t+\sin 2kx),
\nonumber\\
\fl K=\sin^2\alpha\sinh 2\gamma t(\cosh 2\gamma t+\cos\alpha\cos 2kx)-
\nonumber\\
\cos\alpha(\sinh 2\gamma t-2\Omega t\sin\alpha\cos 2kx)(\cos\alpha\cosh 2\gamma t+\cos 2kx)-
\nonumber\\
2\cos\alpha\sin\alpha\sin 2kx(\xi x\cos\alpha\sinh 2\gamma t-\Omega t \sin 2kx),
\nonumber\\
\fl H=\cos^2\alpha
\bigl[(2\xi x \sin\alpha+\sin 2kx\cosh 2\gamma t)^2+(\cos 2kx\sinh 2\gamma t-2\Omega t\sin\alpha)^2\bigr]+
\nonumber\\
\sin^2\alpha(\cosh2\gamma t+\cos\alpha\cos 2kx)^2.
\end{eqnarray}
and
\begin{eqnarray}
k=A\sin\alpha, &\qquad& \gamma=-\frac{A^2}{2}\sin2\alpha,
\nonumber\\
\xi=A\cos\alpha, &\qquad& \Omega=A^2\cos2\alpha.
\end{eqnarray}
The solution is a combination of trigonometric, hyperbolic and polynomial functions. The typical behavior is illustrated in  figure \ref{DG_AB}.

The second question which should be discussed here is the following. The numerator vanishes not only when poles are complex conjugate but also when they coincide at an arbitrary point $\lambda_1 = \lambda_2$. Indeed, now
\begin{eqnarray}
R_1=R_2=R, &\qquad& \alpha_1=\alpha_2=\alpha.
\end{eqnarray}
\begin{equation}
\lambda_1=\lambda_2=A\cosh z\cos\alpha+\rmi A\sinh z\sin\alpha.
\end{equation}
Then
\begin{eqnarray}
\fl N_{\lambda}=\frac{|\bi{q}_1|^2 q^*_{21}q_{22}-(\bi{q}^*_1\cdot\bi{q}_2) q^*_{21}q_{12}-(\bi{q}_1\cdot\bi{q}^*_2) q^*_{11}q_{22}+|\bi{q}_2|^2 q^*_{11}q_{12}}{2A\cosh z\cos\alpha+2\rmi A\sinh z\sin\alpha} \equiv 0,
\nonumber\\
\fl \Delta_{\lambda}=\frac{|q_{11}q_{22}-q_{12}q_{21}|^2}{4A^2\cosh^2 z\cos^2 \alpha},
\end{eqnarray}
as well
\begin{eqnarray}
\fl \ae_1=\ae_2=\ae=A\sinh z \cos\alpha, &\qquad& \omega_1=\omega_2=\omega=\frac{A^2}{2}\sinh2z\cos 2\alpha,
\nonumber\\
\fl k_1=k_2=k=A\cosh z \sin\alpha, &\qquad& \gamma_1=\gamma_2=\gamma=-\frac{A^2}{2}\cosh2z\sin 2\alpha.
\end{eqnarray}
Now
\begin{eqnarray}
u_1=\ae x-\gamma t+\mu_1/2, &\qquad& u_2=\ae x-\gamma t+\mu_2/2,
\nonumber\\
v_1=k x - \omega t-\theta_1/2, &\qquad& v_2=k x-\omega t-\theta_2/2.
\end{eqnarray}
Again, the vectors $\bi{q}_1$,$\bi{q}_2$ are periodic functions of $x$ and exponential functions of time. But now
\begin{eqnarray}
\fl q_{11}q_{22}-q_{12}q_{21}=
-4\sin\alpha\biggl(\cos\frac{\theta^-}{2}\sinh\frac{\mu^-}{2}-
\rmi \sin\frac{\theta^-}{2}\cosh\frac{\mu^-}{2}\biggr),
\end{eqnarray}
and
\begin{eqnarray}
\fl \Delta_{\lambda}=\frac{4\sin^2\alpha}{A^2\cosh^2 z\cos^2\alpha}
\biggl(\cos^2\frac{\theta^-}{2}\sinh^2\frac{\mu^-}{2}+
\sin^2\frac{\theta^-}{2}\cosh^2\frac{\mu^-}{2}\biggr).
\label{denominator leading order}
\end{eqnarray}
Again the denominator $\Delta$ does not depend on $x$ and $t$. But now $\Delta \ne 0$ if $\theta^- \ne 0$ or $\mu^-\ne 0$. The degenerate case appears when $\theta_1=\theta_2 = \theta$ and $\mu_1=\mu_2 = \mu$. However, as we will see later, there is no cancelation of solitons and small perturbations do not appear in the limit $\lambda_1 \ra \lambda_2$. Roughly this can be explained as follows. By the results of \S 4 the phase difference for this case is $4\alpha$. A small perturbation cannot dramatically change the phase of the solution, thus they cannot appear in an arbitrary point of $\xi$ plane. But how we approach this solution to the condensate in the limit $\lambda_1 \ra \lambda_2$?

\begin{figure}[h]
\centering
\includegraphics[width=3in]{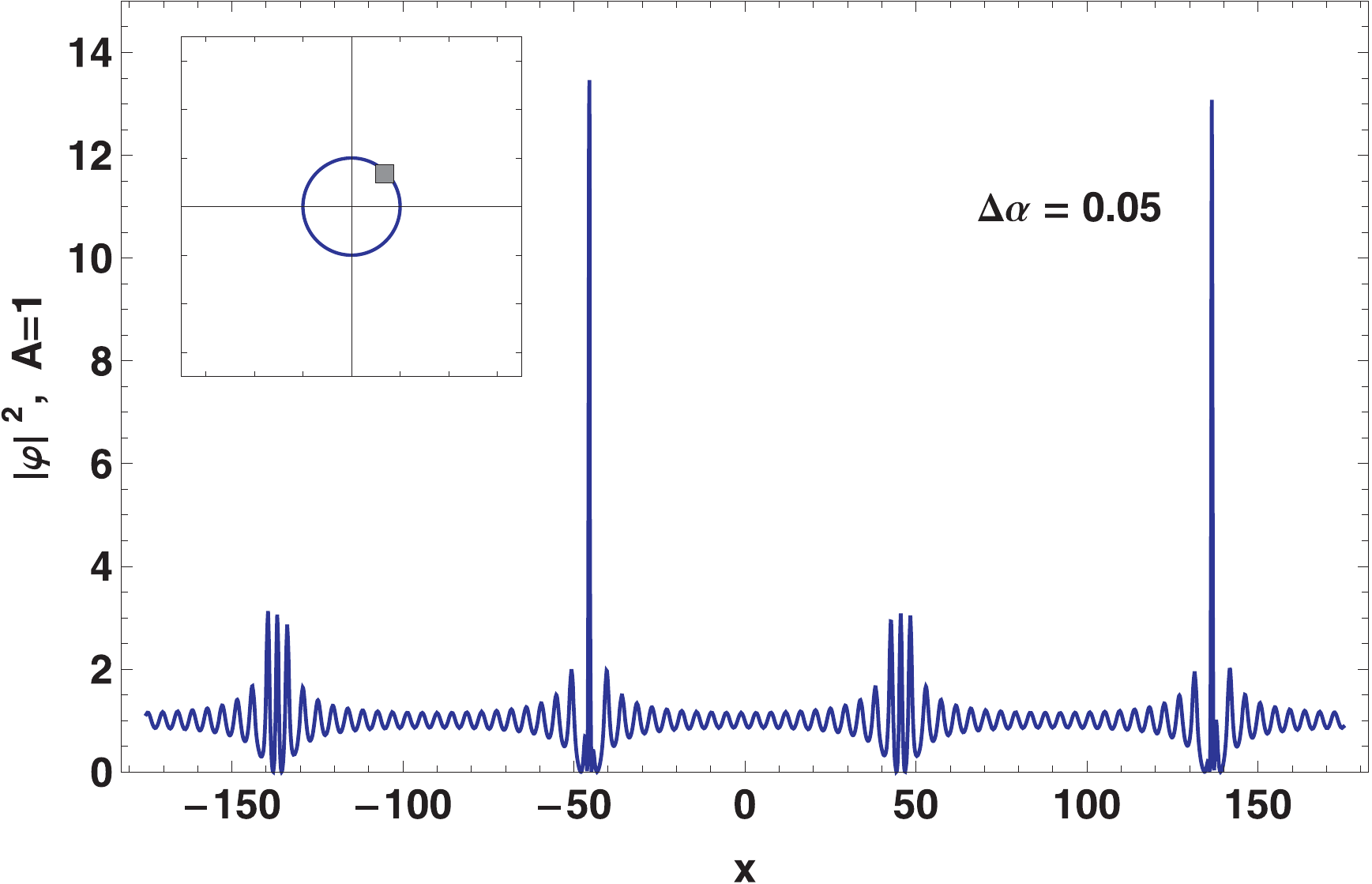}
\includegraphics[width=3in]{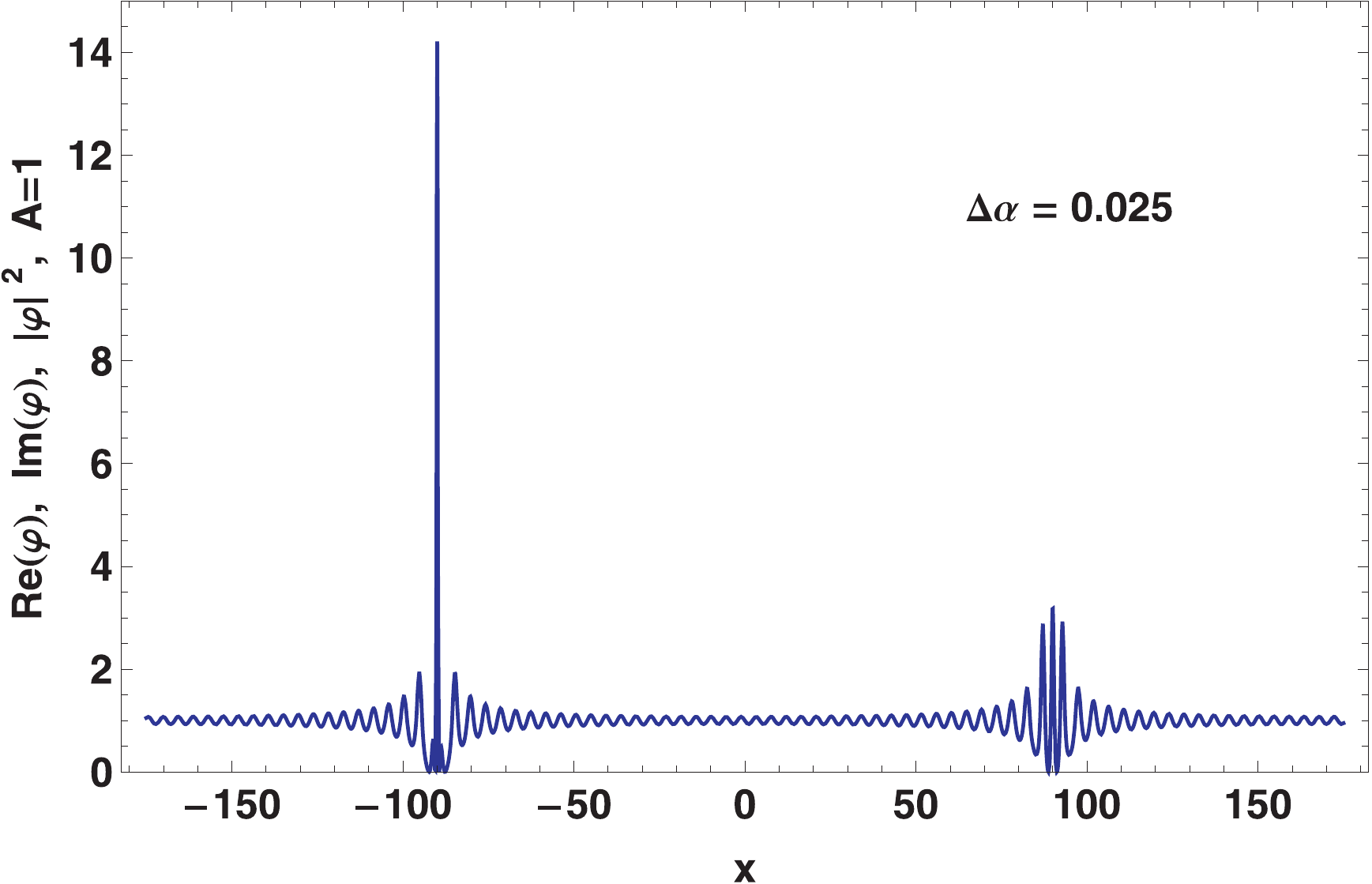}
\caption{\label{close_poles_AB}
Absolute squared value of two-solitonic solution $\varphi$ with close poles located on the unit circle (the area of poles location is illustrated in subfigure) at the moment of time $t=0$. The difference in angular variable is $0.05$ (left picture) and $0.025$ (right picture). Parameters: $R_1=R_2=1,\;\alpha_1=3\pi/4+0.05$ (left picture) $;3\pi/4+0.025$ (right picture) $,\alpha_2=3\pi/4,\;\mu_1=\mu_2=0,\;\theta_1=\pi/2.\;\theta_2=-\pi/2$.
}
\end{figure}

We first discuss the case when the poles are close on the unit circle. As we know this is a particular case of a double Akhmediev breather, thus the solution is periodic in space. When $\alpha_1 \ra \alpha_2$ periodic beats occur at the difference frequency. The amplitude of the beats tends to a constant, while the distance between the beats $L$ increases in inverse proportion to the difference $\alpha_1 - \alpha_2$. This situation is illustrated in figure \ref{close_poles_AB}. Thus the limit is reached by tending the distance between beats to infinity: $L \ra \infty$.

\begin{figure}[h]
\centering
\includegraphics[width=3in]{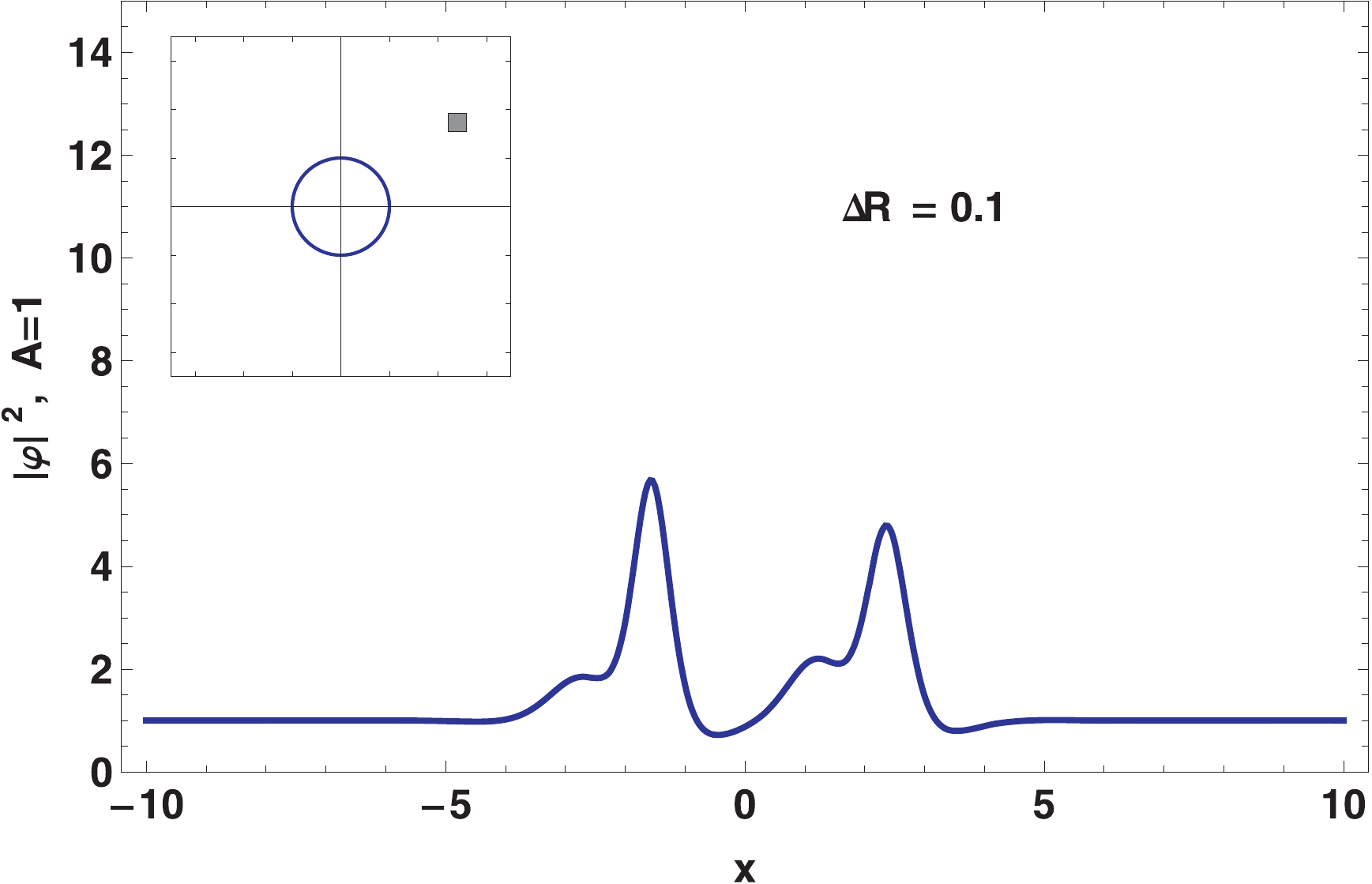}
\includegraphics[width=3in]{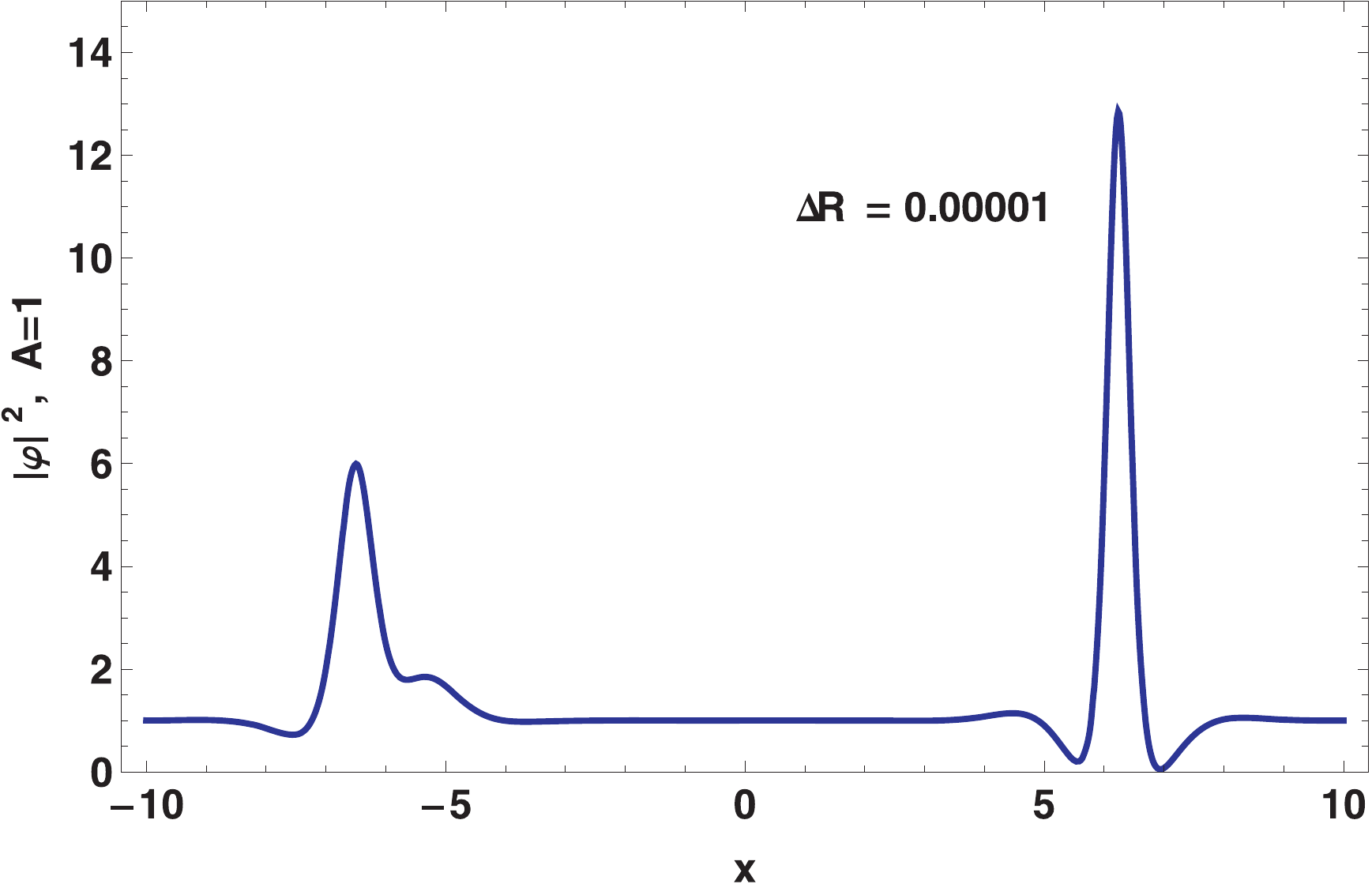}
\caption{\label{close_poles_arb_point}
Absolute squared value $\varphi$ of two-solitonic solution $\varphi$ with close poles located in arbitrary point (the area of poles location is illustrated in subfigure) at the moment of time $t=0$. The distance between poles is $0.1$ (left picture) and $0.00001$ (right picture). Parameters: $R_1=2.9$ (left picture) $;2.99999$ (right picture) $,\alpha_1=\pi/5,\;R_2=3,\;\alpha_2=\pi/5,\;\mu_1=\mu_2=0,\;\theta_1=\theta_2=0$.
}
\end{figure}

\begin{figure}[h]
\centering
\includegraphics[width=3in]{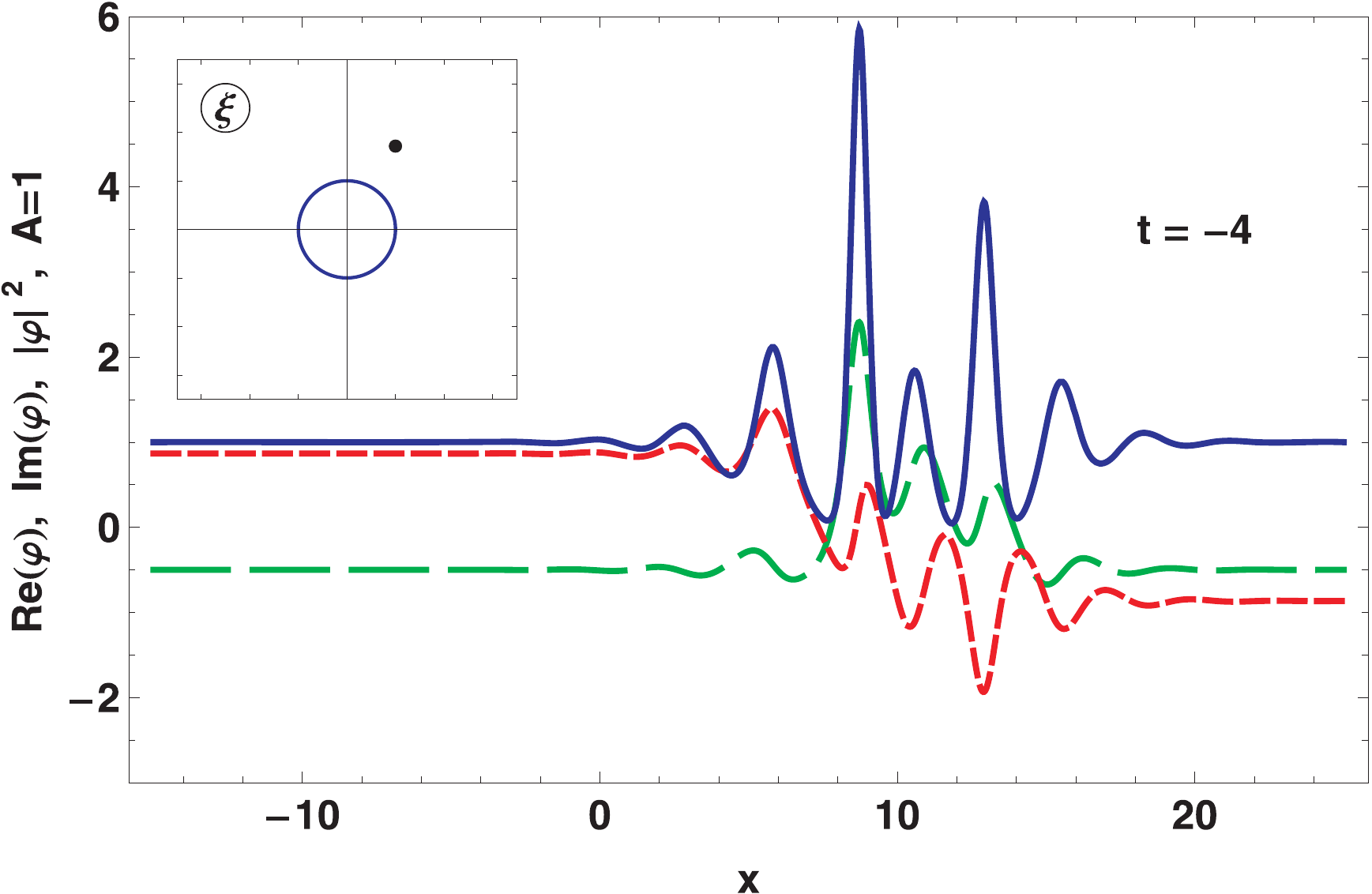}
\includegraphics[width=3in]{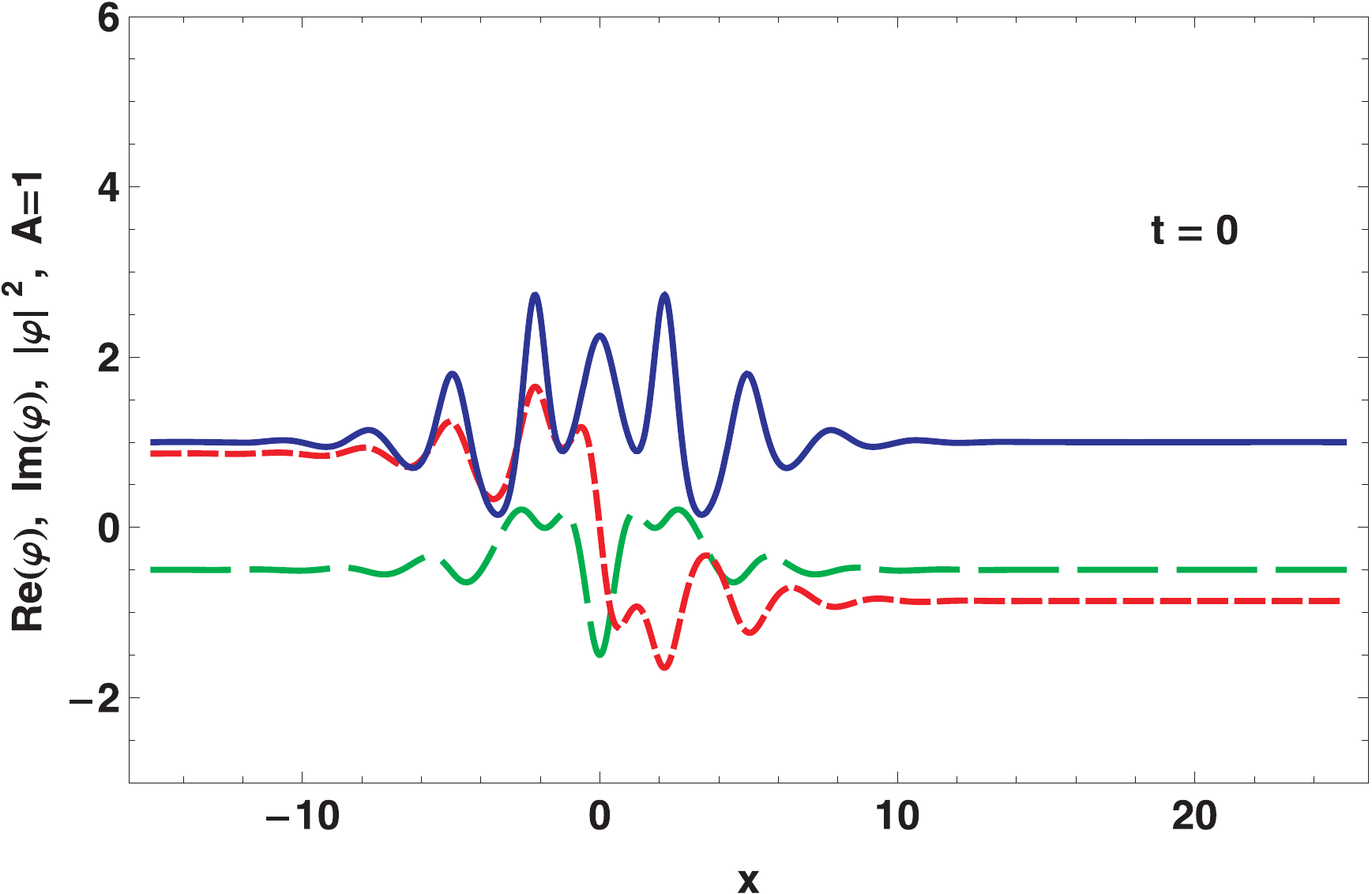}
\caption{\label{DG_arb_point}
Degenerate two-solitonic solution $\varphi$ at moments of time $t=-4$ (left picture) and $t=0$ (right picture) with parameters: $R=2,\alpha=\pi/3,\mu=0,\theta=0$. Green dashed lines - real part of $\varphi$, red short dashed lines - imaginary part of $\varphi$ and blue solid lines - absolute squared value of $\varphi$.
}
\end{figure}

\begin{figure}[h]
\centering
\includegraphics[width=3in]{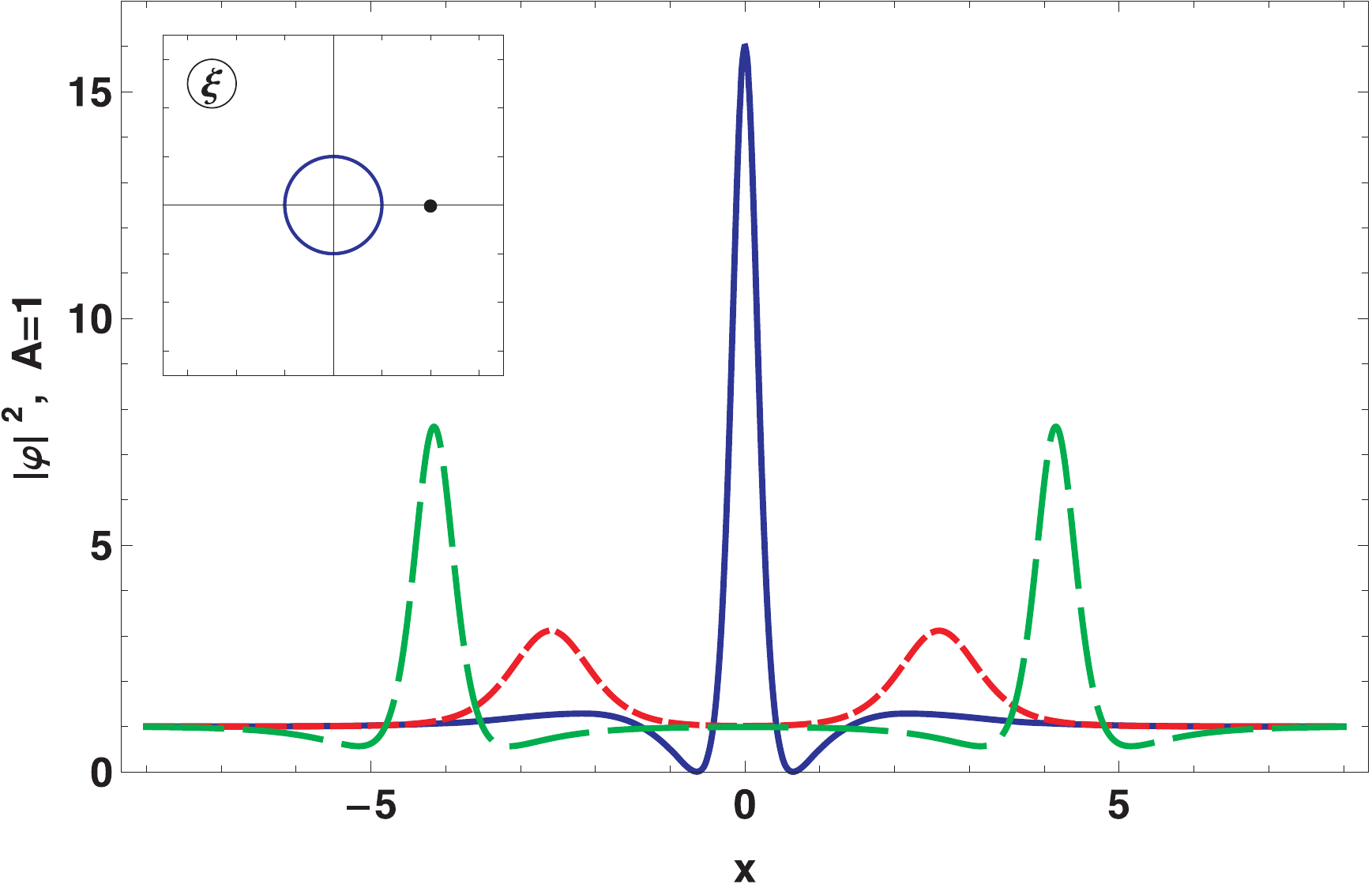}
\caption{\label{DG_bounded}
Absolute squared value of degenerate two-solitonic solution $\varphi$ at moments of time $t=0,10,100$. Poles located on the real axis out of the cut: $R=2,\;\alpha=0,\;\mu=0,\;\theta=0$.
}
\end{figure}

When one pole tends to another at an arbitrary point, the solitons move in one direction. As we mentioned in \S 6 if we put $\mu_1=\mu_2=0$ the solitons collide at $x=0,\;t=0$. But the distance of interaction becomes nonzero. Indeed, on  figure \ref{close_poles_arb_point} solitons at the moment of collision are presented. We see that they interact with nonzero distance which is in inverse proportion to the logarithm of the difference between poles location. It means that now the limit is reached by tending to infinity the distance between solitons at the moment of collision.
To obtain the degenerate solution in the general case we must put $\mu_1=\mu_2=\mu$ and $\theta_1=\theta_2=\theta$.    Now l'H\^{o}pital's rule should be applied to general two-solitonic solution (\ref{2S_general}). The result is a solution which moves and oscillates in time. It is again a combination of trigonometric, hyperbolic and polynomial functions. The expression for the degenerate solution at an arbitrary point is rather cumbersome, so we only give a picture for this case (see figure \ref{DG_arb_point}).
When $\alpha_1=\alpha_2=0$ (the bounded state) the situation is similar - with $R_1 \ra R_2$ the distance between solitons at the moment of their collision increases logarithmically. The degenerate solution can be written in following explicit form.
\begin{eqnarray}
\varphi=A-2A\sinh 2z\frac{M-\rmi K}{H}.
\end{eqnarray}
Here
\begin{eqnarray}
\fl M=\sinh z (\cosh 2\ae x + \cosh z\cos 2\omega t)(\cos 2\omega t + \cosh z \cosh 2\ae x) -
\nonumber\\
\sinh z \cosh z \sinh 2\ae x \cos 2\omega t (\cos 2\omega t\sinh 2\ae x + 2\xi x\sinh z)-
\nonumber\\
\sinh z \cosh z\cosh 2\ae x \sin 2\omega t (\sin 2\omega t\cosh 2\ae x + 2\Omega t\sinh z),
\nonumber\\
\fl K=\sinh^2 z \sin 2\omega t(\cos 2\omega t + \cosh z \cosh 2\ae x) -
\nonumber\\
\cosh^2 z \sinh 2\ae x \sin 2\omega t (\cos 2\omega t\sinh 2\ae x + 2\xi x\sinh z)+
\nonumber\\
\cosh z(\cosh 2\ae x \cos 2\omega t \cosh z +1) (\sin 2\omega t\cosh 2\ae x + 2\Omega t\sinh z),
\nonumber\\
\fl H=
\cosh^2 z \bigl[
(\cos 2\omega t\sinh 2\ae x + 2\xi x\sinh z)^2+(\sin 2\omega t\cosh 2\ae x + 2\Omega t\sinh z)^2
\bigr]
+
\nonumber\\
\sinh^2 z (\cos2\omega t + \cosh z\cosh 2\ae x)^2
\end{eqnarray}
and
\begin{eqnarray}
\ae=A\sinh z,
&\qquad&
\omega=\frac{A^2}{2}\sinh 2z,
\nonumber\\
\xi=A\cosh z,
&\qquad&
\Omega=A^2\cosh 2z.
\end{eqnarray}
This is two solitons with the distance increasing logarithmically in time. The example is given in figure \ref{DG_bounded}. Note that a particular case on a zero background is well known and was mentioned in the first work on focusing NLSE of Zakharov and Shabat ~\cite{Zakharov-Shabat1972}.
\section{Superregular $2N$-solitonic solutions}
Let us consider the case of N superregular pairs of poles. Now
\begin{equation}
\alpha_n=-\alpha_{n+N}, $\qquad$ R_n=1+\varepsilon, $\qquad$ R_{n+N}=1+a_n\varepsilon.
\end{equation}
Here $n=1,..N$ is the number of the pair. A small parameter expansion of the general formula (\ref{N-solitoniic solution}) is a quite tedious task. Fortunately, the scheme of the dressing method avoids this difficulty. As was mentioned at the end of the \S 2 we can start from an arbitrary initial solution $\varphi_0$. In this way, we can consistently add (dress the initial solution) complex conjugated pairs of poles to the solution.

Let us denote by $\varphi_n,\;\bchi_n,\;\bPsi_n$ the solution, the dressing function and matrix $\bPsi$ which correspond to the solution with $n$ pairs of poles. $\varphi_0 = A$ and $\bPsi_0$ (is given by (\ref{condensat lax solution})) correspond to the condensate as usual. Then $\varphi_n$ can be found by the use of the recurrence relation
\begin{equation}
\varphi_n=\varphi_{(n-1)}-2\widetilde{\chi}_{(n),12}.
\end{equation}
We know how to find the function $\bchi_n$ knowing $\bPsi_n$ from \S 6. Thus the main problem is to construct a recurrence relation for the function $\bPsi_n$. Suppose that we know the solution $\varphi_{n-1}$ and the corresponding function $\bPsi_{n-1}$, $\bchi_{n-1}$. Then $\bPsi_n$ can be found by the use of (\ref{bchi definition}) as $\bPsi_n=\bchi_n \bPsi_{n-1}$. Then
\begin{equation}
\bPsi_n = \bchi_n \bchi_{n-1} \cdots \bchi_1 \bPsi_0
\label{Psi_n}
\end{equation}
by virtue of (\ref{bchi_alpha beta})
\begin{eqnarray}
\fl \chi_{n,\mu\nu}=
\delta_{\mu\nu}+\frac{p_{1,\mu}q_{1,\nu}}{\lambda-\lambda_n}+\frac{p_{2,\mu}q_{2,\nu}}{\lambda-\lambda_{n+N}}
=
\nonumber\\
\fl \delta_{\mu\nu}+
\frac{(\lambda-\cos\alpha_n)(p_{1,\mu}q_{1,\nu}+p_{2,\mu}q_{2,\nu})
-\rmi\varepsilon\sin\alpha_n(p_{1,\mu}q_{1,\nu}-p_{2,\mu}q_{2,\nu})}{(\lambda-\lambda_n)(\lambda-\lambda_{n+N})}.
\label{bchi 2n solitonic}
\end{eqnarray}
Now $\bi{q_n}$ and $\bi{p_n}$ are much more complicated than vectors calculated earlier, because they are result of dressing on $2(n-1)$ solitonic background. $\bchi_n$ can be presented in the following form
\begin{equation}
\bchi_n = \bchi_n^{(0)}E + \bchi_n^{(1)} + \bchi_n^{(2)} + \cdots + \bchi_n^{(n)}
\end{equation}
Here $\bchi_n^{(0)}$ is a leading order. This is a result of dressing on the condensate background. $\bchi_n^{(1)}$ is proportional to $\varepsilon$, $\bchi_n^{(2)}$ is proportional to $\varepsilon^2$ etc. - the results of previous superregular dressings. $\bchi_n^{(0)}$ can be obtained by the limit $\varepsilon \ra 0$ in (\ref{bchi 2n solitonic}) as
\begin{equation}
\bchi_n^{(0)}=
\biggl(
1+\frac{2A\cos\alpha_n}{\lambda-A\cos\alpha_n}
\biggr)
\left(
  \begin{array}{cc}
    1 & 0 \\
    0 & 1 \\
  \end{array}
\right).
\end{equation}
Note that this is exactly formula (\ref{bchi 2 solitonic}) with $\alpha = \alpha_n$. $\bchi_n^{(0)}$ satisfy to the condition
\begin{equation}
\bchi_n^{(0)^*}(-\lambda^*) = \bchi_n^{(-1)}(\lambda)
\end{equation}
Now if we expand (\ref{Psi_n}) and keep only the first order terms, we will get such sum:
\begin{equation}
\sum_{k=1}^n (\bchi_n^{(0)}\cdots \bchi_{k+1}^{(0)} \bchi_k^{(1)} \bchi_{k-1}^{(0)}\cdots \bchi_{1}^{(0)})\bPsi_0
\label{first order sum}
\end{equation}
However, according to (\ref{q_p_invariance}), multiplying of dressing matrix on scalar factors such as $\bchi_n^{(0)}$ does not change the result of dressing. Then in the first order we write (\ref{Psi_n}) as
\begin{equation}
\bPsi_n = (E + \bchi_n^{(1)} + \bchi_{n-1}^{(1)} + \cdots + \bchi_1^{(1)}) \bPsi_0
\label{Psi_n}
\end{equation}
It proves that $\varphi_n$ is a linear sum of perturbations $\delta\varphi_n$ corresponding to the $n$-th pair of poles:
\begin{equation}
\varphi_n=A+\sum_{m=1}^N \delta\varphi_n.
\end{equation}
Each $\varphi_n$ is calculated in the first order by (\ref{delta_varphi}) with its own parameters.

\begin{figure}[h]
\centering
\includegraphics[width=3in]{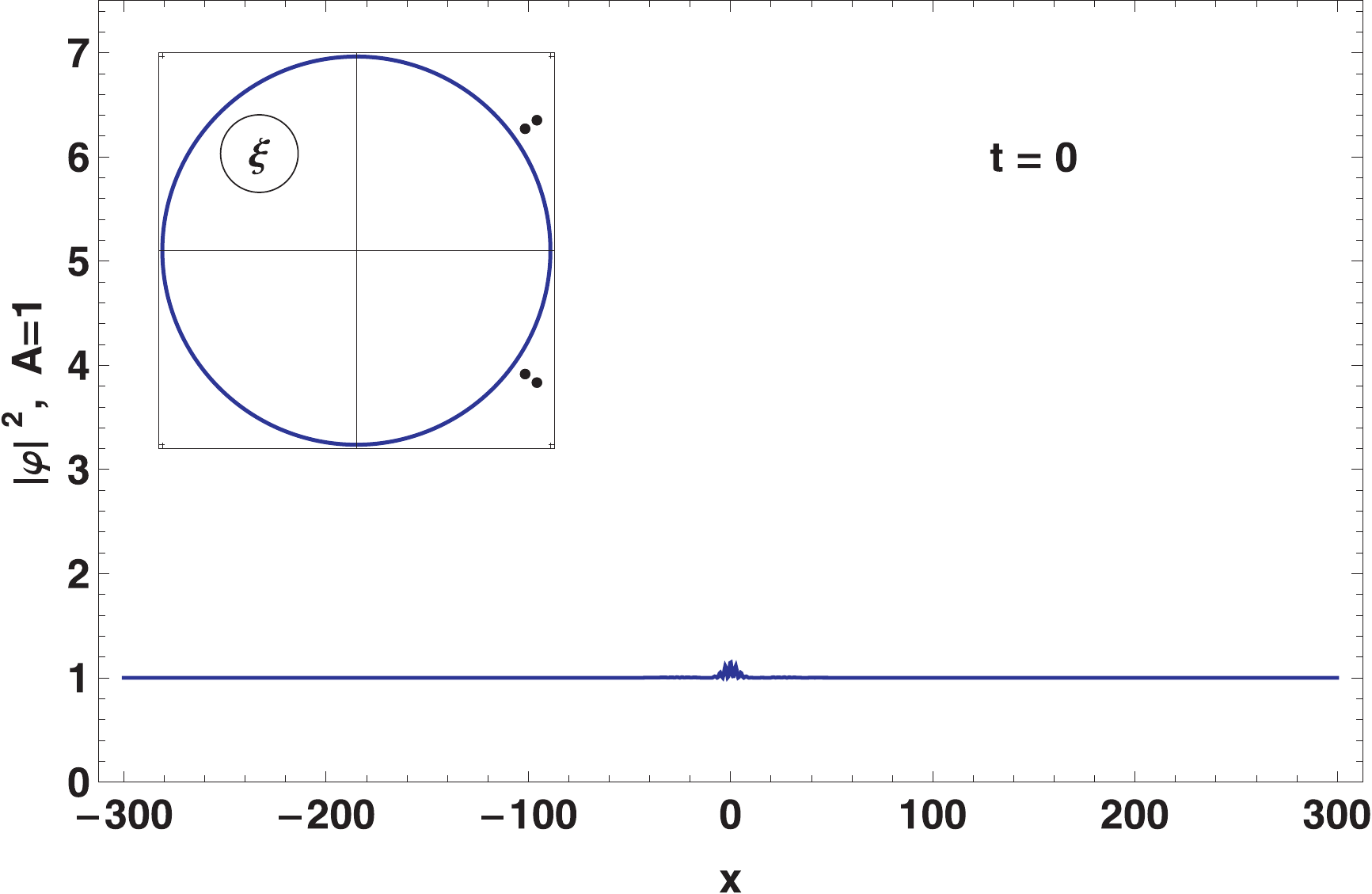}
\includegraphics[width=3in]{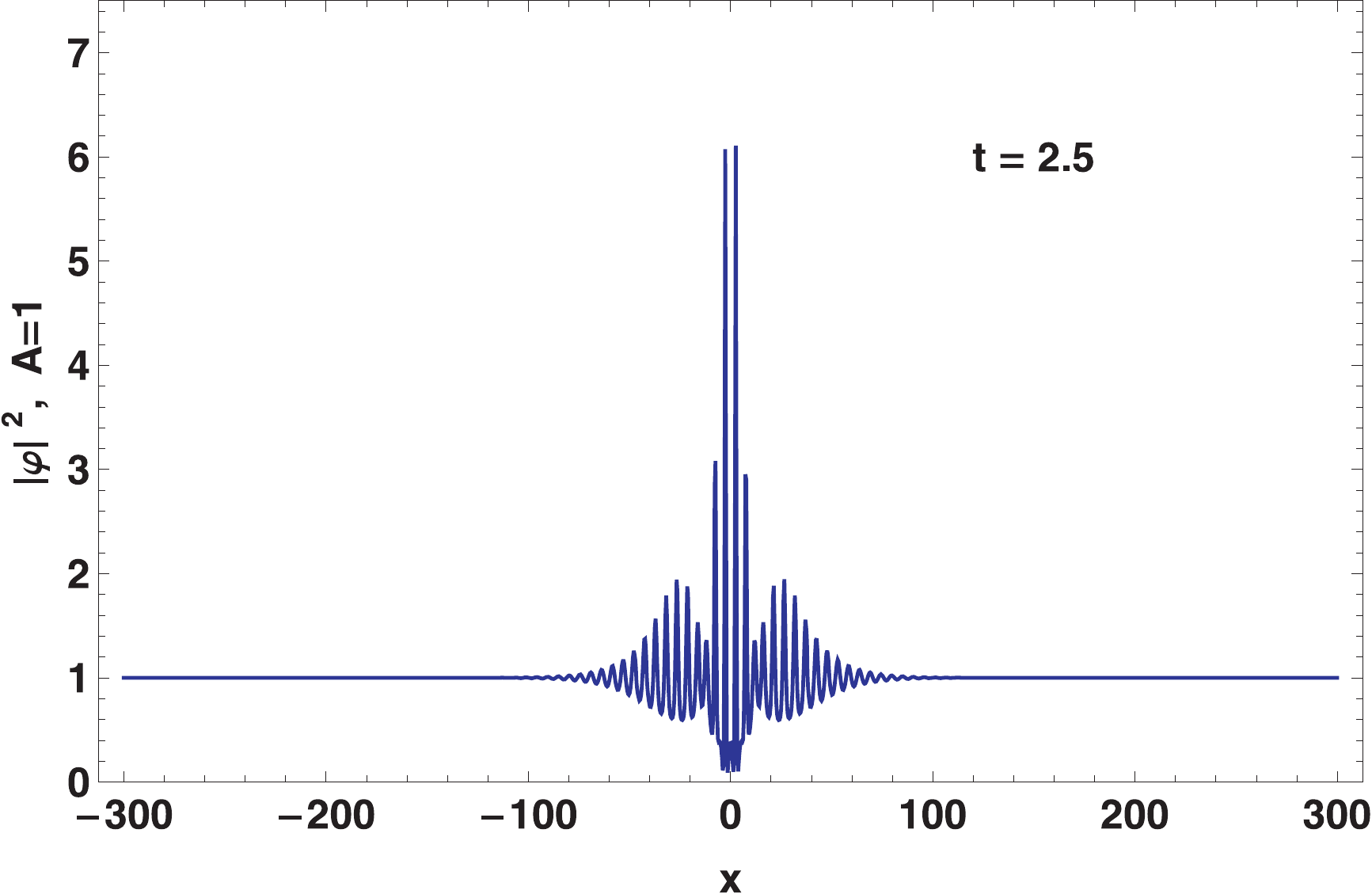}
\includegraphics[width=3in]{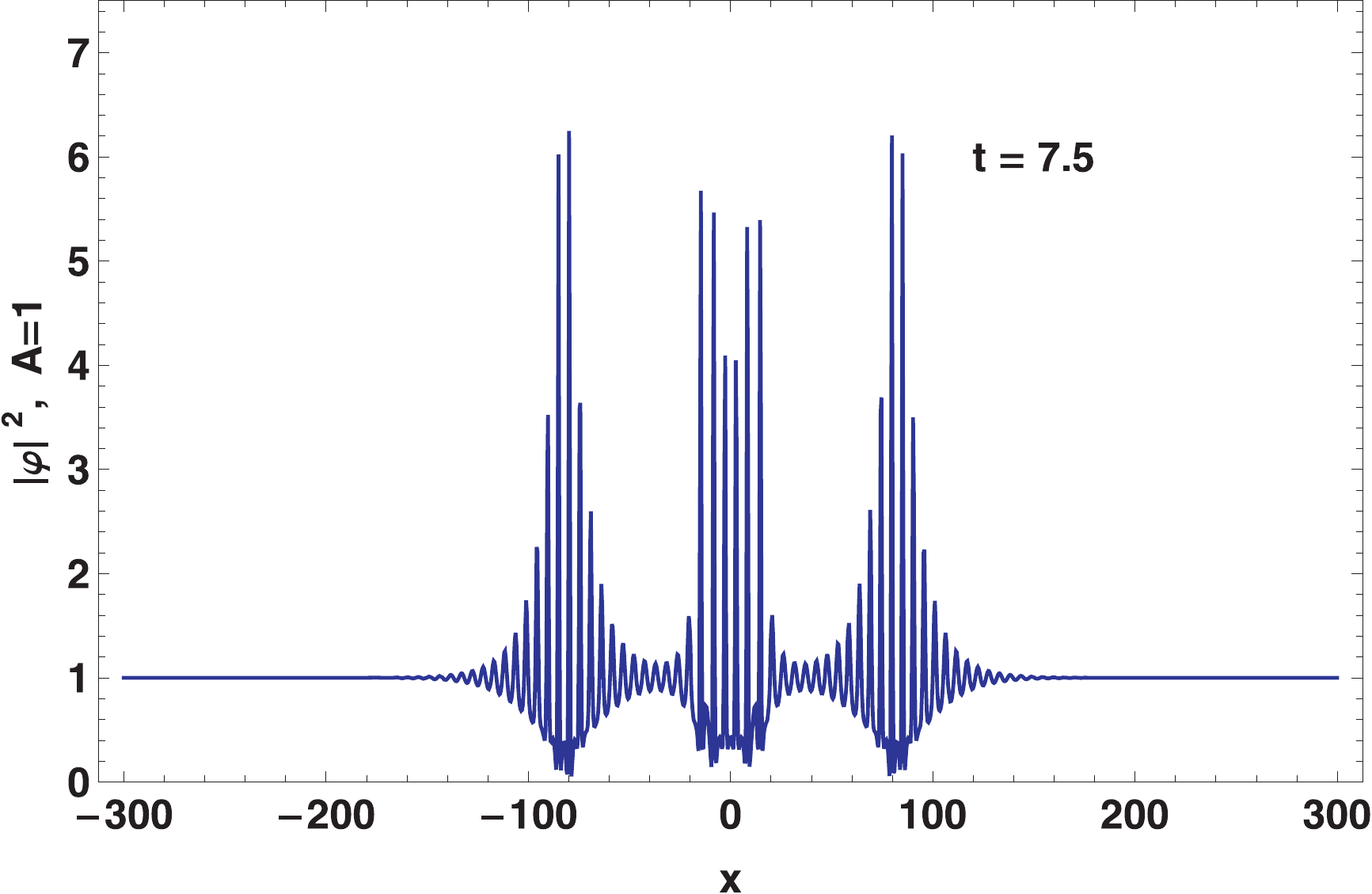}
\includegraphics[width=3in]{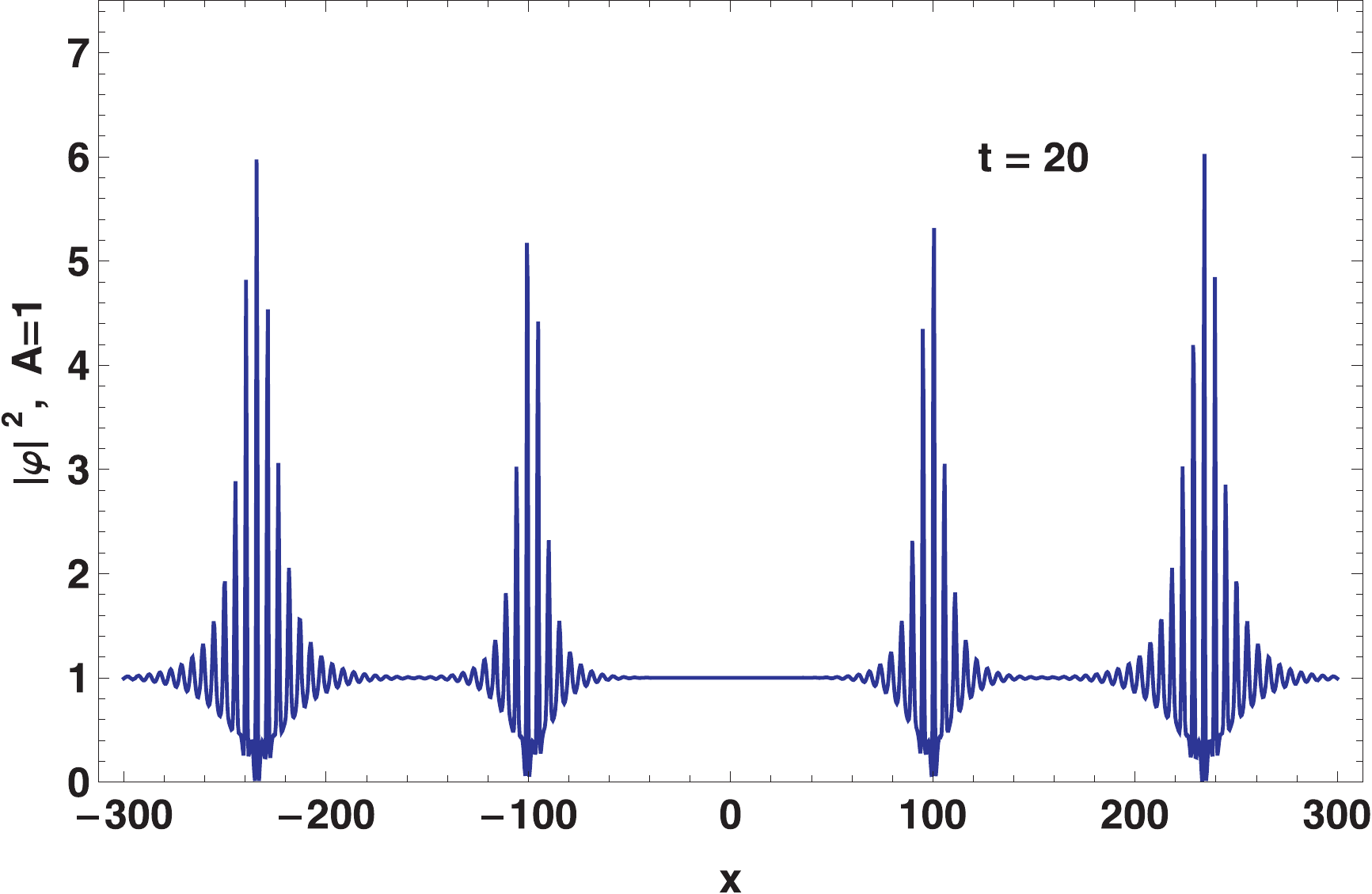}
\caption{\label{4S}
The development of superregular four-solitonic solution. Absolute squared value of solution $\varphi$ at different moments of time with parameters: $R_1=1.05,\;R_3=1.05,\;\alpha_1=\pi/5,\;\alpha_3=-\pi/5,\;\mu_1=\mu_3=0,\;\theta_1=\theta_3=\pi/2$; $R_2=1.075,\;R_4=1.075,\;\alpha_2=\pi/5,\;\alpha_4=-\pi/5,\;\mu_2=\mu_4=0,\;\theta_2=\theta_4=\pi/2$.
}
\end{figure}

\begin{figure}[h]
\centering
\includegraphics[width=3in]{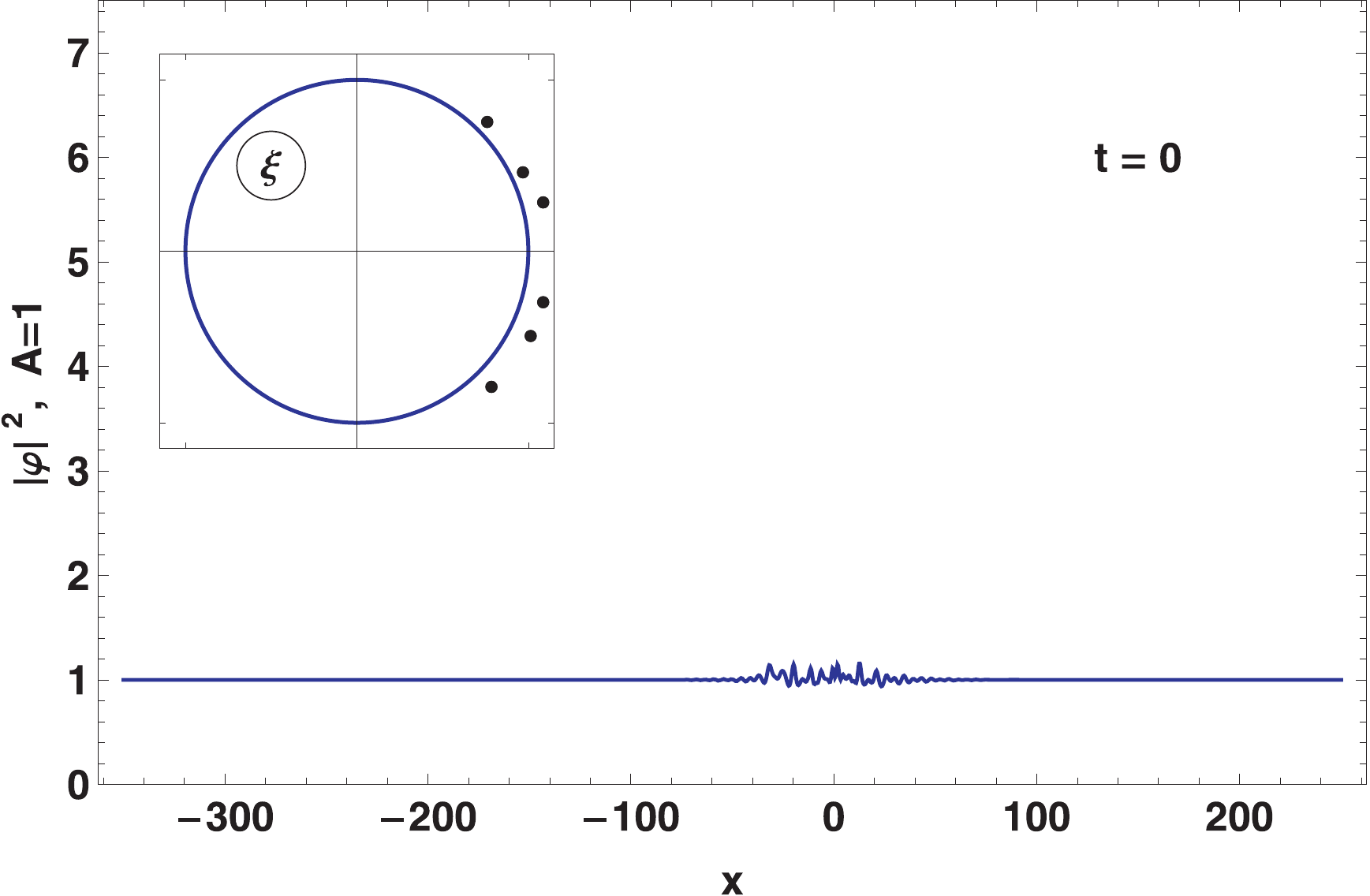}
\includegraphics[width=3in]{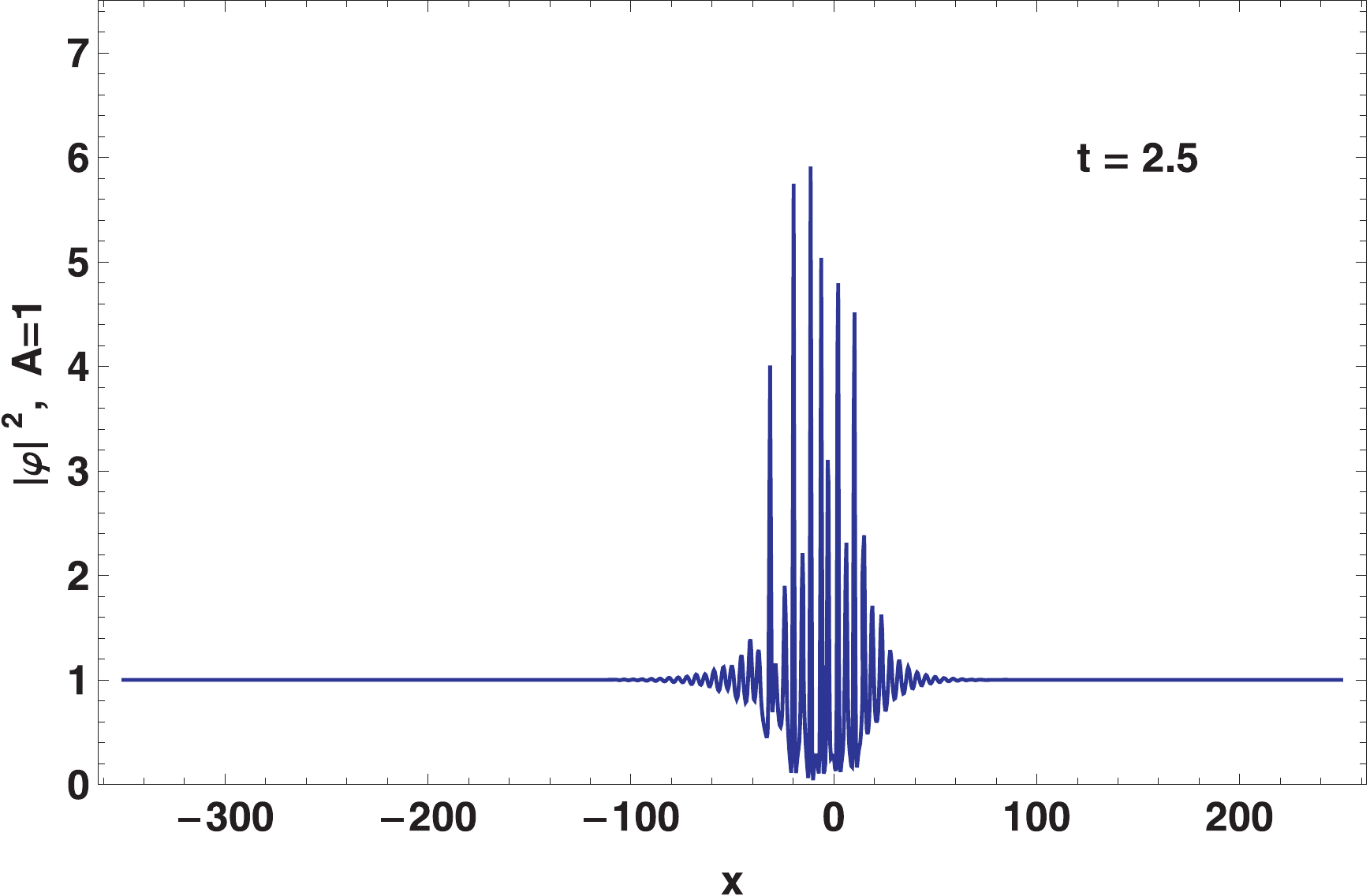}
\includegraphics[width=3in]{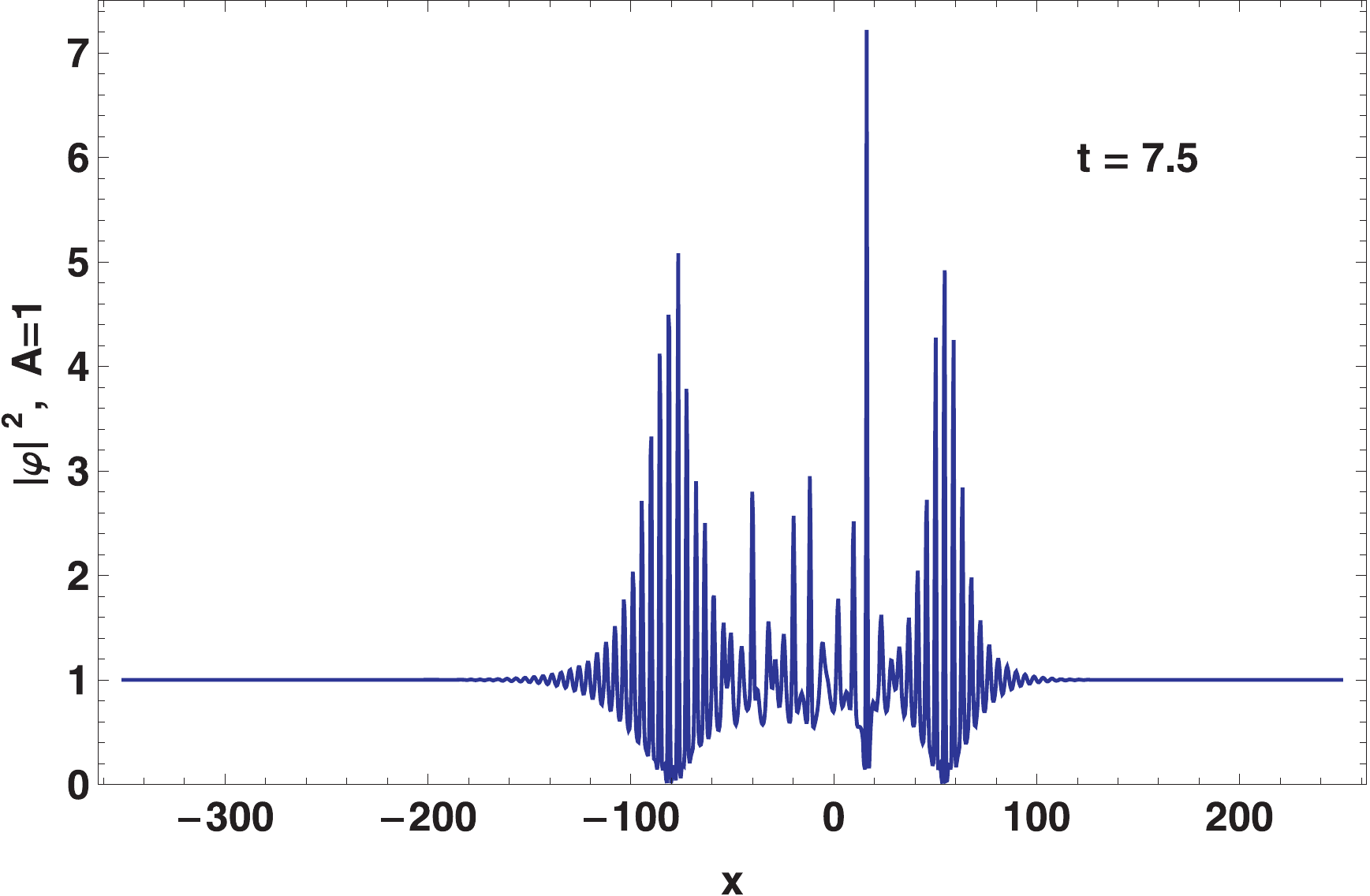}
\includegraphics[width=3in]{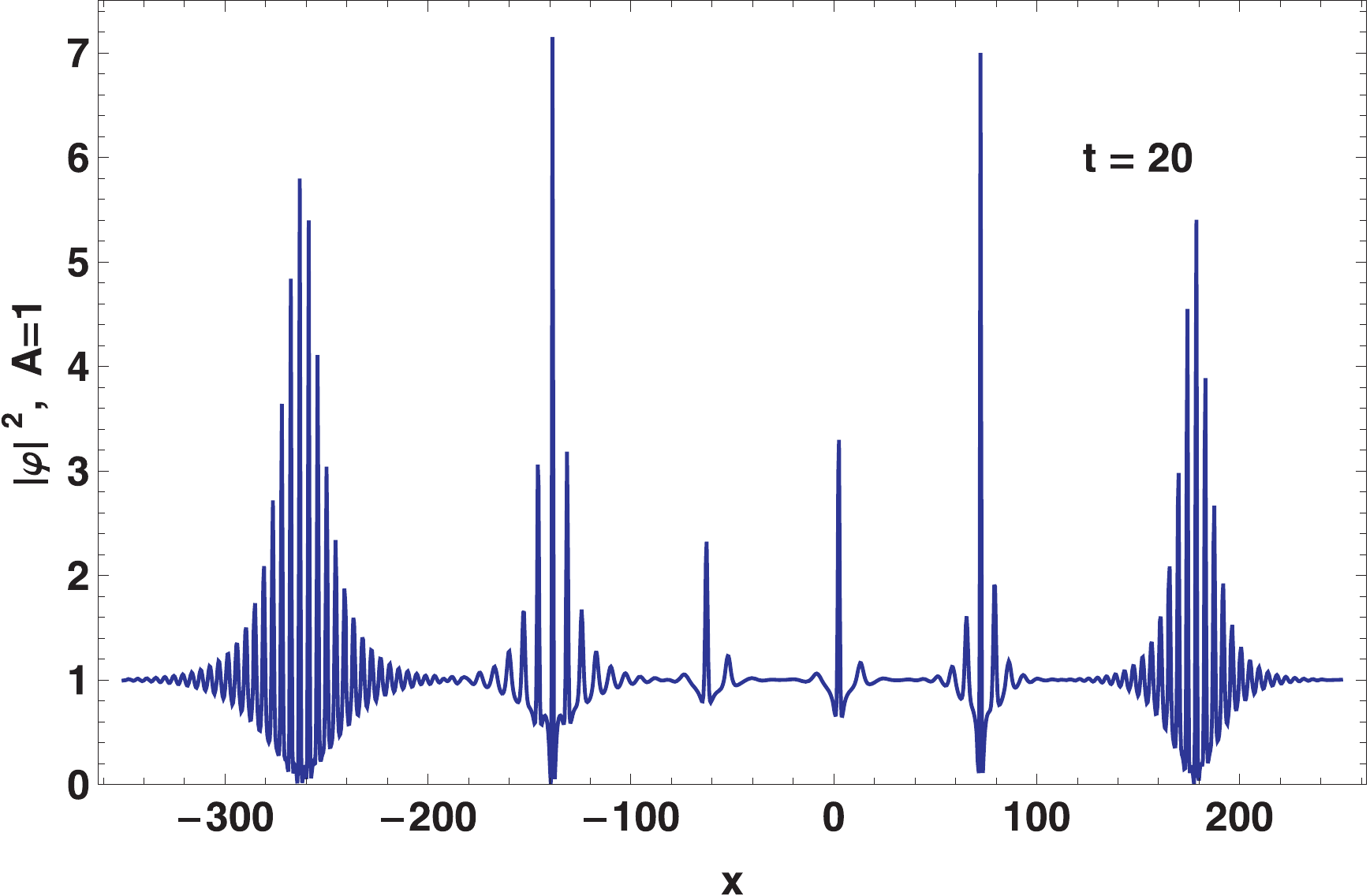}
\caption{\label{6S}
The development of superregular six-solitonic solution. Absolute squared value of solution $\varphi$ at different moments of time with parameters: $R_1=1.05,\;R_4=1.075,\;\alpha_1=\pi/4,\;\alpha_4=-\pi/4,\;\mu_1=\mu_4=0,\;\theta_1=\theta_4=\pi/2$; $R_2=1.05,\;R_5=1.1,\;\alpha_2=\pi/7,\;\alpha_5=-\pi/7,\;\mu_1=\mu_5=0,\;\theta_1=\theta_5=\pi/2$;
$R_3=1.1,\;R_6=1.1,\;\alpha_3=\pi/12,\;\alpha_6=-\pi/12,\;\mu_3=\mu_6=5,\;\theta_3=\theta_6=\pi/2$
}
\end{figure}

An example of a small localized perturbation of the condensate at the moment $t=0$ for four- and six-solitonic solutions are given in figures \ref{4S} and \ref{6S}. Note that in figure \ref{4S} superregular pairs have the same angle.

\begin{figure}[h]
\centering
\includegraphics[width=3in]{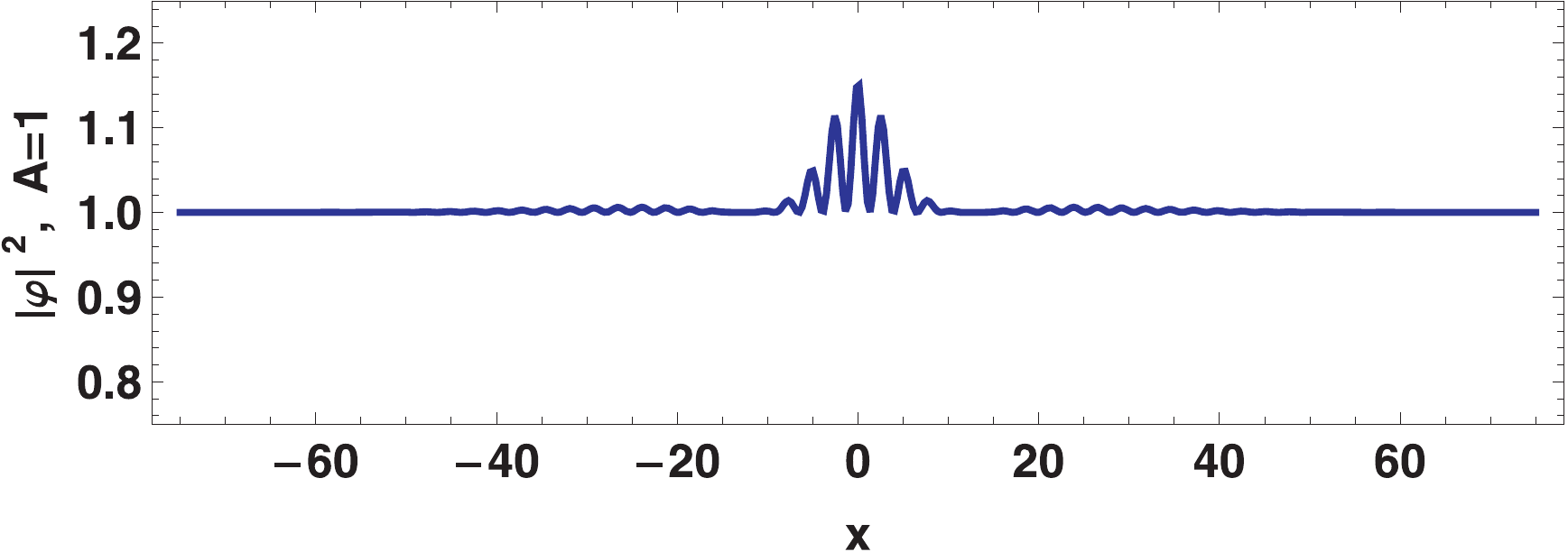}
\includegraphics[width=3in]{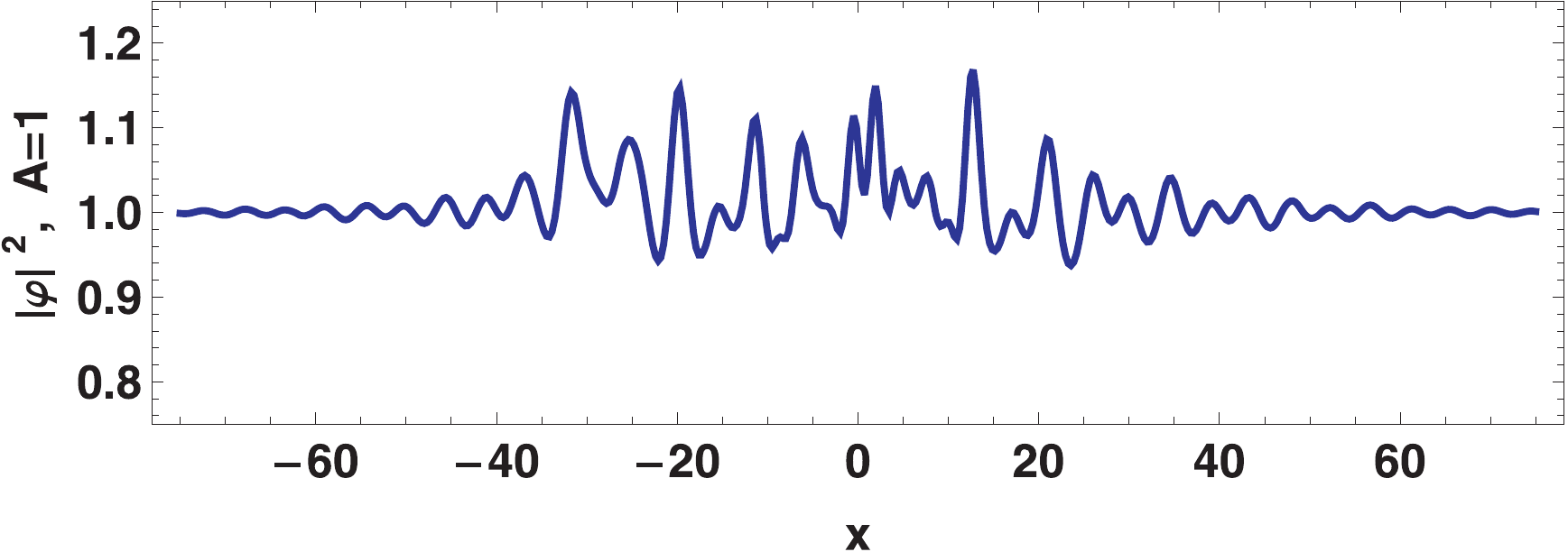}
\caption{\label{2S_SmallPert3}
Enlarged small perturbations at the moment of time $t=0$ presented on the figure \ref{4S} (left picture) and on the figure \ref{6S}(right figure).
}
\end{figure}

\subsection{Conclusion}
We apply the dressing method to the focusing NLSE in the presence of a condensate. We describe the general $N$ - solitonic solution. We apply uniformizing variables to simplify the analysis. We introduce regular solitonic solutions, which do not disturb the phases of the condensate at infinity by coordinate. The criterion of regularity is particularly simple in the uniformazing variables.

We obtain explicit expressions for one- and two-solitonic solution. We carefully describe different species of these solutions. We study regular two-solitonic solutions in particular detail. We describe a pure homoclinic double Akhmediev breather. We also describe solitonic atoms as well as degenerate solutions.

The central result of our work is the following. We find a broad class of $2N$-solitonic solutions which are small localized perturbations of the condensate at $t=0$. They develop into a nonlinear superposition of $N$ pairs of quasi-Akhmediev breathers. This can be treated as a sort of "integrable turbulence" appearing as a result of nonlinear development of the modulation instability. When $\varepsilon \ra 0$ this superposition is linear, so that at $t=0$ superregular solutions generated by separate regular pairs of soliton forms an $N$ - dimensional linear space. This remarkable fact will be discussed in another article.

\subsection{Acknowledgments}
The authors express deep gratitude to Dr. E.A. Kuznetsov and for helpful discussions.
This work was supported by: Russian Federation  Government  Grant (No. 11.G34.31.0035 with Ministry of Education and Science of RF, November 25, 2010), the RFBR (grant no. 12-01-00943), and by the Grant "Leading Scientific Schools of Russia" (No. 6170.2012.2).

\end{document}